\documentclass{jaa}
\usepackage{natbib}
\usepackage{longtable}
\usepackage{subfloat}
\usepackage{caption}
\usepackage{subcaption}
\setcitestyle{numbers}

\usepackage{graphicx}

\begin{document}\sloppy

\title{Understanding the Inner Structure of Accretion disk in GX 17+2: AstroSat's Outlook}


\author{K. Sriram\textsuperscript{1, *}, P. Chiranjeevi\textsuperscript{1},  S. Malu\textsuperscript{1} and V. K. Agrawal\textsuperscript{2}}
\affilOne{\textsuperscript{1}Department of Astronomy, Osmania University , Hyderabad 500007, India.\\}
\affilTwo{\textsuperscript{2}Space Astronomy Group, ISITE Campus, U R Rao Satellite Center, Bangalore, 560037, India.}


\twocolumn[{

\maketitle

\corres{astrosriram@yahoo.co.in}


\begin{abstract}
We performed the timing and spectral studies of a Z source GX 17+2 observed from Astrosat LAXPC instrument. 
Cross-Correlation function (CCF) was performed using soft  (3-5 keV) and hard (16-40 keV) X -ray bands across the hardness intensity 
diagram and found correlated/anti-correlated hard and soft lags which seems to be a common feature in these sources. 
 We performed spectral analysis for few of these observations and found no consistent variation in the spectral parameters 
during the lags, however 10-40\% change was noticed in diskbb and power-law components in few of observations. 
For the first time, we report the detection of HBOs around $\sim$25 Hz and $\sim$ 33 Hz along with their harmonics using AstroSat LAXPC data. 
On comparison with spectral results of HB and other branches, we found that inner disk front 
is close to the last stable orbit and as such no systematic variations are observed.  
We suggest that the detected lags are readjustment time scales of corona close to the NS and 
constrained its height to be around few tens to hundreds of km. The detected lags and no significant variation 
of inner disk front across the HID strongly indicate that structural variation in corona is the most possible cause of Z track in HID.

\end{abstract}


}]


\doinum{12.3456/s78910-011-012-3}
\artcitid{\#\#\#\#}
\volnum{000}
\year{0000}
\pgrange{1--}
\setcounter{page}{1}
\lp{1}

\section{INTRODUCTION}
Neutron Star (NS) Low Mass X-Ray Binaries (LMXBs) that are highly luminous (close to or more than Eddington luminosity) exhibiting
a Z shaped track on the Hardness Intensity Diagram (HID)/ Colour-Colour Diagram (CCD) are termed as Z sources (Hasinger \& Van Der Klis 1989). 
These are further classified into Sco X-1 like and Cyg X-2 like sources based on the particular Z track traced by them (Kuulkers et al. 1994, 1997).
The various branches of this track are characterized by varying temporal and spectral features along them. The three main branches are 
the horizontal, normal and flaring branch (HB, NB and FB). The two existing contrary pictures to explain the motion of the source along the HID
are the varying mass accretion ($\dot{m}$) rate hypothesis (Hasinger et al. 1990; Vrtilek et al. 1990) and the constant mass accretion rate hypothesis where
other physical or radial instabilities cause the source to trace out the Z shaped track (Homan et al. 2002, Lin et al. 2012).

One of the fundamental characteristics of each branch is the signature of Quasi Periodic Oscillations (QPOs) exhibited in each of them (Hasinger \& van der Klis 1989).
QPOs in the horizontal branch (HB) termed as horizontal branch oscillations (HBOs) are found in the 15--60 Hz frequency range, while those in the
normal branch (NB) termed as normal branch oscillations (NBOs) are found in the 5--8 Hz frequency range (van der Klis 2006).
Flaring branch Oscillations, so far detected only in two sources Sco X-1 and GX 17+2 (Priedhorsky et al. 1986; Penninx et al. 1990, 
Homan et al. 2002), are found in the frequency range of 10--25 Hz. Each of these oscillations are considered to have different physical origins.
Alpar \& Shaham (1985) proposed a model where HBOs could be associated with the keplerian orbital frequency of the inner edge of 
the disk and the spin frequency of the neutron star. Stella \& Vietri (1998a, 1999) proposed the Relativistic Precession Model (RPM) that associates
HBOs to the nodal precession of tilted orbits near the neutron star. NBOs and FBOs could possibly be oscillations in the optical depth of accretion flow 
in the inner disk region as proposed by Lamb (1989), Fortner, Lamb \& Miller (1989) or they could be oscillations associated with the sound 
waves in a thick disk (Alpar et al. 1992). Another model for NBOs were proposed by Titarchuk et al. (2001), where these oscillations are considered to
be acoustic oscillations of a spherical viscous shell around the NS. 

Spectral modeling has been ambiguous in these sources, especially when it comes to the origin of soft and hard energy photons.
As per the eastern spectral model, soft photons emanating from the disk are modeled using a multi-temperature black body emission (MCD, Mitsuda et al. 1984) 
and these photons are the seed for inverse comptonization in the compact corona producing hard X-rays.
The western model on the other hand uses a single temperature blackbody emission from the NS surface (or the immediate surrounding region)
and a high energy power-law model for describing the hard emission (White et al. 1986).
Another model proposed by Popham \& Sunyaev (2001) considers a hot, low density boundary layer around the NS surface to be responsible for the hot comptonized spectrum
and the optically thick accretion disk to be responsible for the black body spectrum. Though both the models fairly describe the X-ray spectrum of Z 
sources, the picture is not clear in terms of understanding the location and size of corona and whether the accretion disk is truncated or not. 

A vital timing tool to resolve some of these ambiguities in understanding the accretion disk corona geometry 
is a Cross Correlation function (CCF) study between soft and hard X-ray energy bands (eg. Vaughan et al. 1999; Sriram et al. 2007, 2012, 2019.)
Previous CCF studies of Z sources have led to the detection of lags of the order of hundred seconds in the HB and NB branch
of the Z track while FBs are mostly seen to have strong positively correlated CCFs (Lei et al. 2008, Sriram et al. 2012, 
Sriram et al. 2019). It was suggested that these lags are attributed to the readjustment of 
the coronal structure which could in effect help in constraining the size of the coronal structure (Sriram et al. 2019). 

Here we present an extensive spectro-temporal study of the source GX 17+2 using the AstroSat LAXPC archival data of the source.
GX 17+2 is a burster Z source with no confirmed optical counterparts, located at a distance of 13 kpc (Galloway et al. 2008), having a spin frequency of
293.2 Hz (Wijnands et al. 1997). GX 17+2 is a low inclination system (i $<$ 40$^{\circ}$; Cackett et al. 2010, Ludlam et al. 2017a, Malu et al. 2020). 
Cackett et al. (2010) used a relativistic diskline model on the Suzaku data of the source, which led to an estimate of 7-8 GM/c$^2$ 
for the inner disk radius. Using NuStar spectra of GX 17+2 (3-30 keV) Ludlam et al. (2017a) determined that the disk extends to 1.0-1.02 ISCO.
Sriram et al. (2019) based on the RXTE spectrum of the source estimated an inner disk radius of 20-35 km. 
Using the AstroSat spectrum of the source, Agrawal et al. (2020) found a decreasing power-law component along the Z track from HB to NB and this component was
found to be increasing from the NB to FB. They estimated a 28-42 km inner disk radius from their spectral analysis. 
Based on the AstroSat SXT+LAXPC spectra, Malu et al. (2020) found an inner disk radii $\sim$ 12--16 km (5.7--8.0 R$_{g}$), along the NB, 
which is close to ISCO. Hence the inner disk can be considered to be almost at the last stable orbit without much change in the position of the disk front.

Energy dependent CCF studies of GX 17+2 using RXTE and NuStar data (3-5 keV and 16-30 keV) performed by Sriram et al. (2019) led to the detection
of a few hundred second delays when the source was in the NB and HB. Using these lags the coronal height was constrained to be around 20-35 km. 
A similar study using the SXT (0.8-2 keV) and LAXPC (3-5 keV, 16-20 keV, 20-40 keV and 20-50 keV) data performed by Malu et al. (2020) again revealed
lags of the order few hundred seconds in the NB of the Z track and the height of the corona was constrained to few tens of km. Here the spectral analysis
revealed only a varying power-law index across the NB thus leading to the conclusion that only the hot comptonized region was varying.

Earlier CCF study of GX 17+2 using AstroSat (Malu et al. 2020) was focused only on the NB and an extensive CCF study along with a correlated spectral 
and timing study of the source across the complete HID can help in resolving the inconsistencies related to the accretion disk corona geometry.
Hence we report here, the CCF, PDS and spectral study of the source GX 17+2 using AstroSat’s LAXPC data.

\section{DATA REDUCTION AND ANALYSIS}

AstroSat Large Area X-ray Proportional Counter (LAXPC) archival data of the source was used in this study.
Datasets used for the study include the Obs.ID G05-112T01-9000000452 observed from 2016 May 11 to 2016, May 14 ($\sim$ 100 ks) ,
Obs. ID T02-087T01-9000002352 observed on 2018, September 10 ($\sim$ 49 ks) and Obs. ID G08-037T01-9000002256 observed on July 26, 2018 ($\sim$ 18 ks).

LAXPC data in the EA mode was used for this study and this has a time resolution of 10$\mu$s.
LAXPC onboard AstroSat has three proportional counter units that are co-aligned -LAXPC 10,20 and 30. The combined effective area is 6000 cm$^{2}$ at 15 keV 
and it is operational in the 3--80 keV energy range with a moderate energy resolution (Yadav et al. 2016a, Antia et al. 2017). 

The Level 1 data was analysed using LAXPC software (Format A  May 19, 2018) which is provided by the AstroSat Science Support Center (ASSC).
LAXPC\_MAKE\_EVENT, LAXPC\_MAKE\_STDGTI, LAXPC\_MAKE\_LIGHTCURVE and LAXPC\_MAKE\_SPECTRA
modules were used to generate the event file, GTI file, corresponding lightcurves and spectra
respectively. LAXPC\_MAKE\_BACKLIGHTCURVE and LAXPC\_MAKE\_BACKSPECTRUM were used to generate the corresponding background light curves and spectrum.  
The corresponding response files were generated along with it. Spectrum in the 3--50 keV band was used for the study and 1\% systematic error was 
considered (Agrawal et al. 2020). 
LAXPC10 data has been used for spectral analysis (otherwise mentioned) as it is better calibrated and have less background issues when compared to 
other LAXPC units. LAXPC 20 data has been used for timing analysis for observations during 2018, due to low gain operation of LAXPC 10. For remaining
observations LAXPC 10 and 20 has been used.

\section{TIMING ANALYSIS}
Timing analysis was performed on the background subtracted binned LAXPC10 and LAXPC20 light curves of GX 17+2.
Using the hard colour 10.5-19.7 keV/7.3- 10.5 keV and intensity in the range 7.3-19.7 keV, the HID
was obtained for sections that exhibited CCF lags (see Table 1). 
The light curves were corrected for dead-time effects (van der Klis (1988); for LAPXC the dead time is 42.5 $\mu$ s Yadav et al. (2016)). 
For all the datasets, LAXPC 20 data was used to obtain the HID (Figure 1).

\begin{figure}[!ht]
\includegraphics[width=10.0cm,height=8.0cm, angle=270]{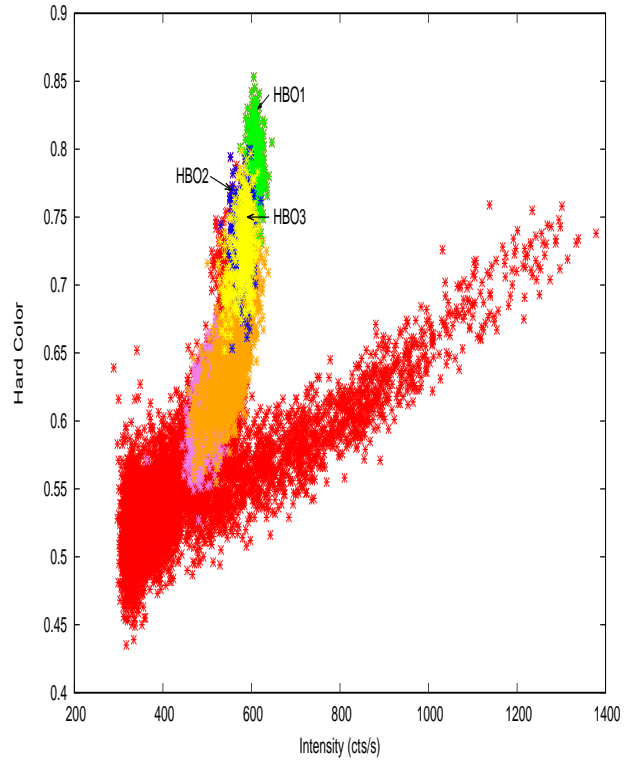} \\
\caption{HID for GX 17+2 using AstroSat LAXPC observations.  LAXPC 20 data was used. Hard colour is defined as 10.5-19.7 / 7.3-10.5 keV and Intensity is that
in the 7.3--19.7 keV range. G05-112T01-9000000452 is represented using colours red ,blue ,green and yellow, T02-087T01-9000002352 is represented using colours orange. 
G08-037T01-9000002256 is represented using colour violet.}\label{fig1}
\end{figure}

 Crosscor tool of XRONOS package was used to perform the cross correlation function (CCF) analysis (Sriram et al. 2007; Lei et al. 2008). 
We used the direct slow method (fast=no) to compute the correlation coefficients (CC) as a function of lags between the two light curves. 
The error bars of CCF are obtained by propagating the theoretical error bars of the cross correlations from the individual intervals and the 
cross correlations are normalized by dividing them with the square root of the product of newbins of the two light curves in each interval
\footnote{ https://heasarc.gsfc.nasa.gov/docs/xandau/xronos/help/crosscor.html}. CCF analysis was performed between the 3--5 keV (soft)
and 16--40 keV (hard) 32 s binned light curves (Sriram et al. 2007, 2011, 2019; Lei et al. 2008, Malu et al. 2020). Figure 2  (top panel) 
shows the soft and hard energy band light curves of each 
section and the bottom panel shows the CCF along with the Gaussian fit (inset) 
of each section where lag was observed.
  In order to estimate the CCF lags, we adopted a procedure wherein several different
segments were considered around the hypothetical centroid and Gaussian functions were fitted. The minimum $\chi$$^{2}$ fit
was considered to estimate the CCF lags and their errors were estimated with a 90\% confidence level using the 
criterion of $\Delta$ $\chi$$^2$ = 2.71 (see Table 1).
  Figure 3 shows few of the representative lag vs. $\Delta$ $\chi$$^{2}$ (= $\chi$$^{2}$-$\chi$$^{2}$$_{min}$)figures 
obtained from the above mentioned procedure.

In the Obs. ID G05-112T01-9000000452, lags were found in five different light curve segments.
CCF lags varying from $\sim$ 306 s - 571 s were found in HB and NB with a correlation coefficients (CC) varying from $\sim$ 0.4--0.6 for the lags (see Table 1). 
This is consistent with that obtained from previous studies (Sriram et al. 2019).
Out of the five segments, one was a positively correlated soft lag and remaining were anti-correlated hard lags.
For Obs. ID G08-037T01-9000002256, lag was found in one section with 
a lag of 281 $\pm$ 31 s (CC = 0.58 $\pm$ 0.03, positively correlated soft lag) in the HB. 
For Obs. ID T02-087T01-9000002352, again four segments were found to exhibit lags in the HB varying from 78 s - 604 s, with
four of them showing anti-correlated lags. One segment was associated with hard lags (see Table 1 and figure 2).
Two of the sections (G and I) show anticorrelated CCFs but with almost no significant lags. \\ \\ \\ \\ \\ \\ \\ \\ \\ \\ \\
 
\begin{table*}[htb]
\begin{minipage}[t]{\columnwidth}
\scriptsize
\caption{CCF information of LAXPC soft (3-5 keV) vs. hard (16-40 keV) light curves.
Here CC means cross-correlation coefficient. Lags refer to the CCF lags obtained and HID location gives the
position of the lightcurve segment on the HID.}\label{Tab.1} 
\begin{tabular}{cccccccccccccc}
\hline
ObsID & Start time (UT) & Stop time (UT) & Exposure time (s) & CC $\pm$ CCerr & Lags $\pm$ error  (s) & HID location & $\chi^{2}$/dof \\
&&&&&&(Hard color, Intensity)&\\
\hline

G05-112T01-9000000452 & 2016-05-11 12 27 21 & 2016-05-14 11 17 27 \\


Section A\footnote{Detected HBO} &  &  & 2400 & {\bf -0.37$\pm$0.05} & {\bf 422$\pm$85} & {\bf 550-600}, 0.81 - 0.73 &{\bf 31/40}\\

Section B &  &  & 3000 & {\bf -0.55$\pm$0.02} & {\bf 369$\pm$30} & {\bf 540-580}, 0.77 - 0.66 &{\bf 72/54}\\

Section C &  &  & 2900 & {\bf -0.57$\pm$0.03} & {\bf 306$\pm$70} & {\bf 520-550}, 0.75 - 0.67 &{\bf 27/37}\\


Section D &  &  & 3100 & {\bf 0.46$\pm$0.03} & {\bf -571$\pm$37} & {\bf 460-500}, 0.68 - 0.59 &{\bf 24/47}\\



Section E &  &  & 1700 & {\bf -0.38$\pm$0.05} & {\bf 555$\pm$43} & {\bf 400-450}, 0.62 - 0.54 & {\bf 25/32}\\

G08-037T01-9000002256 & 2018-07-26 08 54 03 & 2018-07-26 18 56 51 \\

Section F &  &  & 3100 & {\bf 0.56$\pm$0.02} &{\bf -281$\pm$31} & {\bf  470-520, 0.58 - 0.62} & {\bf 29/50}\\

T02-087T01-9000002352 & 2018-09-10 04 24 32 & 2018-09-10 23 55 00 \\




Section G &  &  & 2300 & {\bf -0.37$\pm$0.04} & {\bf -6$\pm$36} & {\bf 540-570, 0.63 - 0.72}  & {\bf 29/50}\\

Section H &  &  & 2700 & {\bf -0.4$\pm$0.04} & {\bf -78$\pm$55} & {\bf 530-560, 0.62 - 0.68} & {\bf 32/50}\\

Section I &  &  & 3200 & {\bf -0.59$\pm$0.03} & {\bf -9$\pm$26} & {\bf 520-550, 0.60 - 0.65} &{\bf 36/51}\\

Section J &  &  & 3600 & {\bf -0.27$\pm$0.03} & {\bf 604$\pm$63} & {\bf 500-540, 0.58 - 0.64} & {\bf 25/31}\\

\hline
\end{tabular}
\end{minipage}
\end{table*}

\begin{figure}[!ht]
\caption{The background subtracted LAXPC soft (3--5 keV) and hard (16 -- 40 keV)  light curves on the top panels and the corresponding
 CCF lag observed are on the bottom panels. Energy bands used are mentioned in the light curves (top panel). 
 Bottom panels show the cross correlation function (CCF) of each section of the light curve and 
 shaded regions show the standard deviation of the CCFs. Bottom panel inset figure gives the Gaussian fit of the lag portion.}

\begin{subfigure}[b]{0.6\columnwidth}
\includegraphics[width=4cm, height=7cm, angle=270]{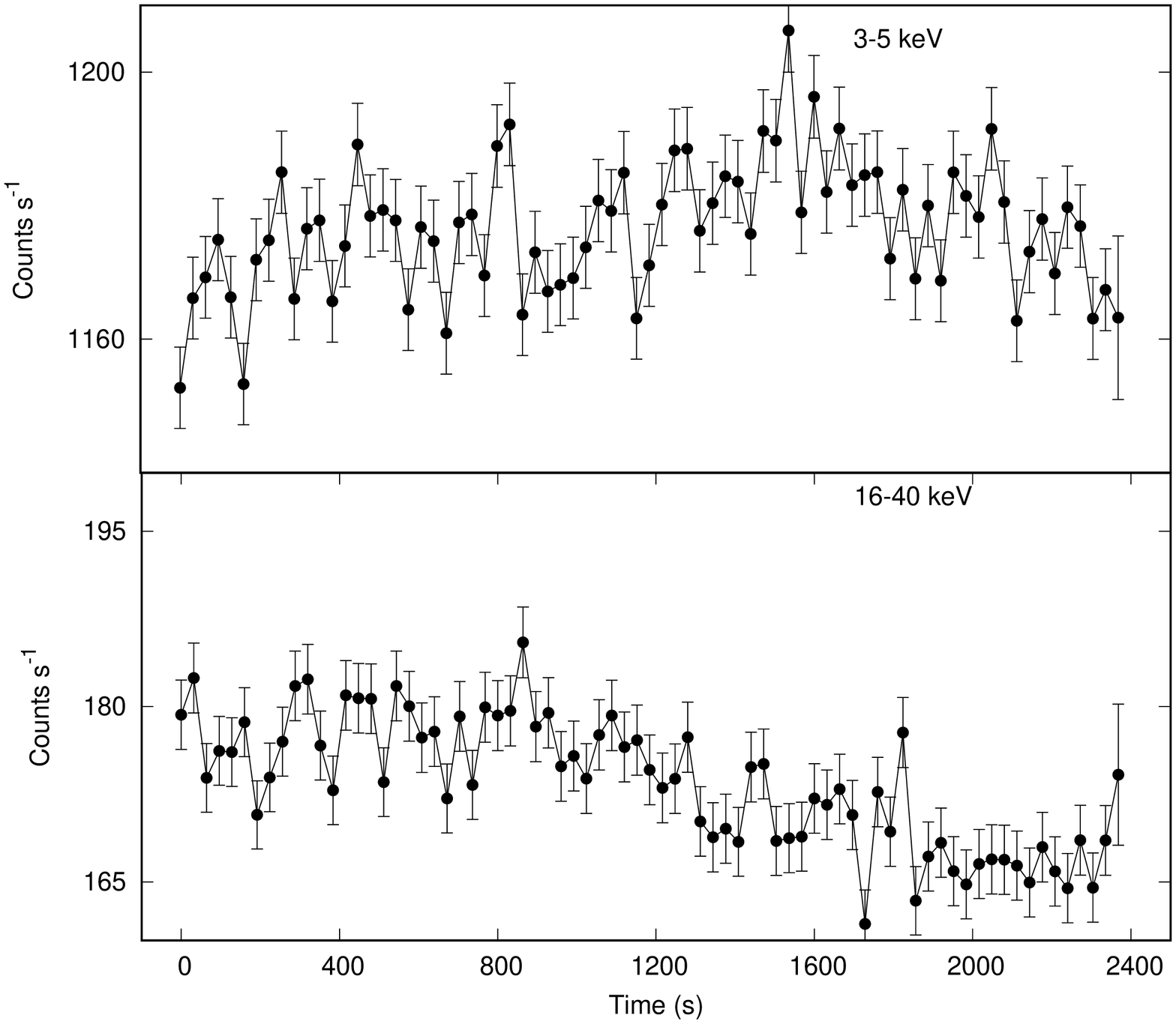}
\includegraphics[width=4cm, height=7cm, angle=270]{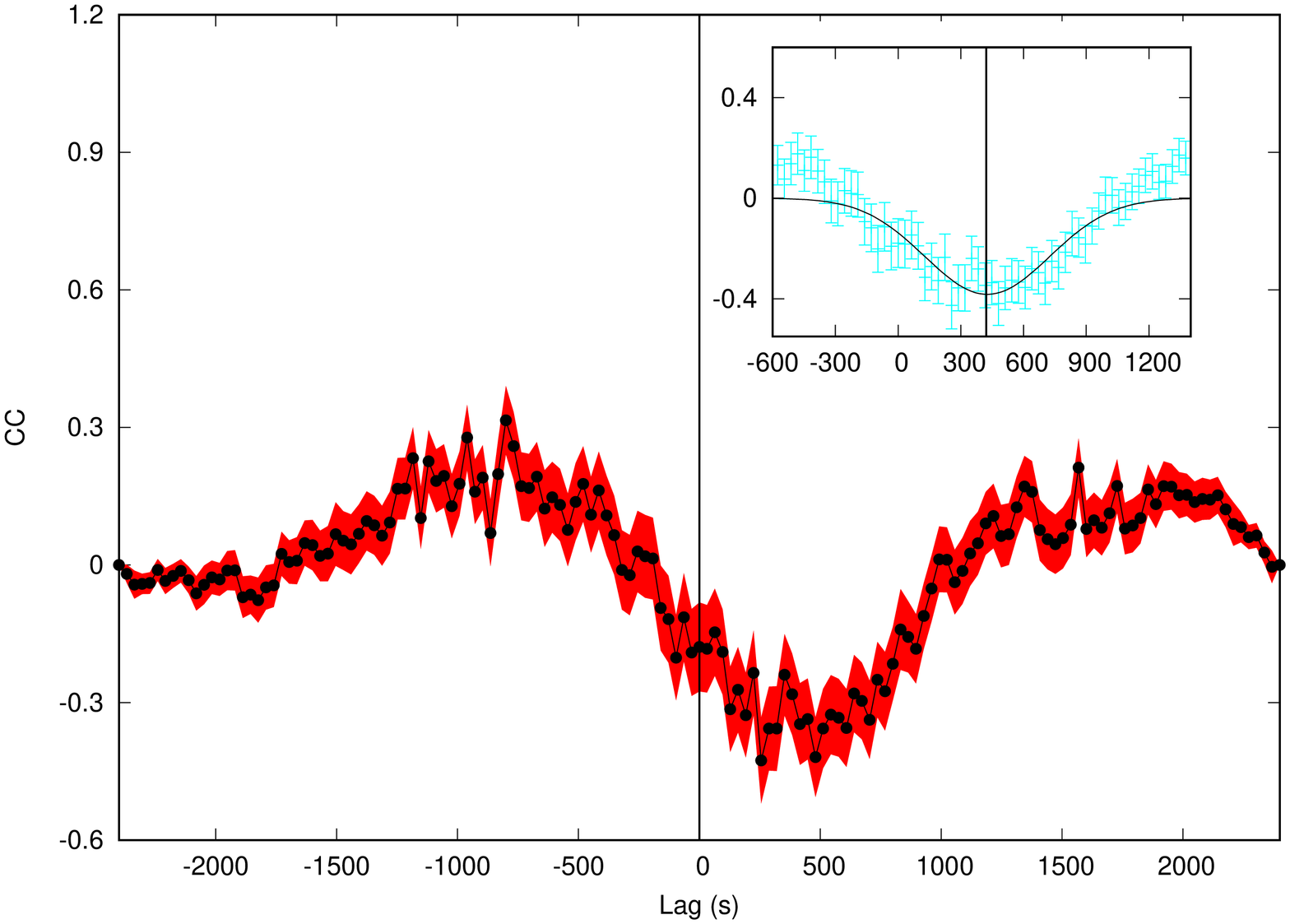}
\caption{}
\end{subfigure}
\end{figure}

\begin{figure}[!ht]\ContinuedFloat
\begin{subfigure}[b]{0.6\columnwidth}
\includegraphics[width=4cm, height=7cm, angle=270]{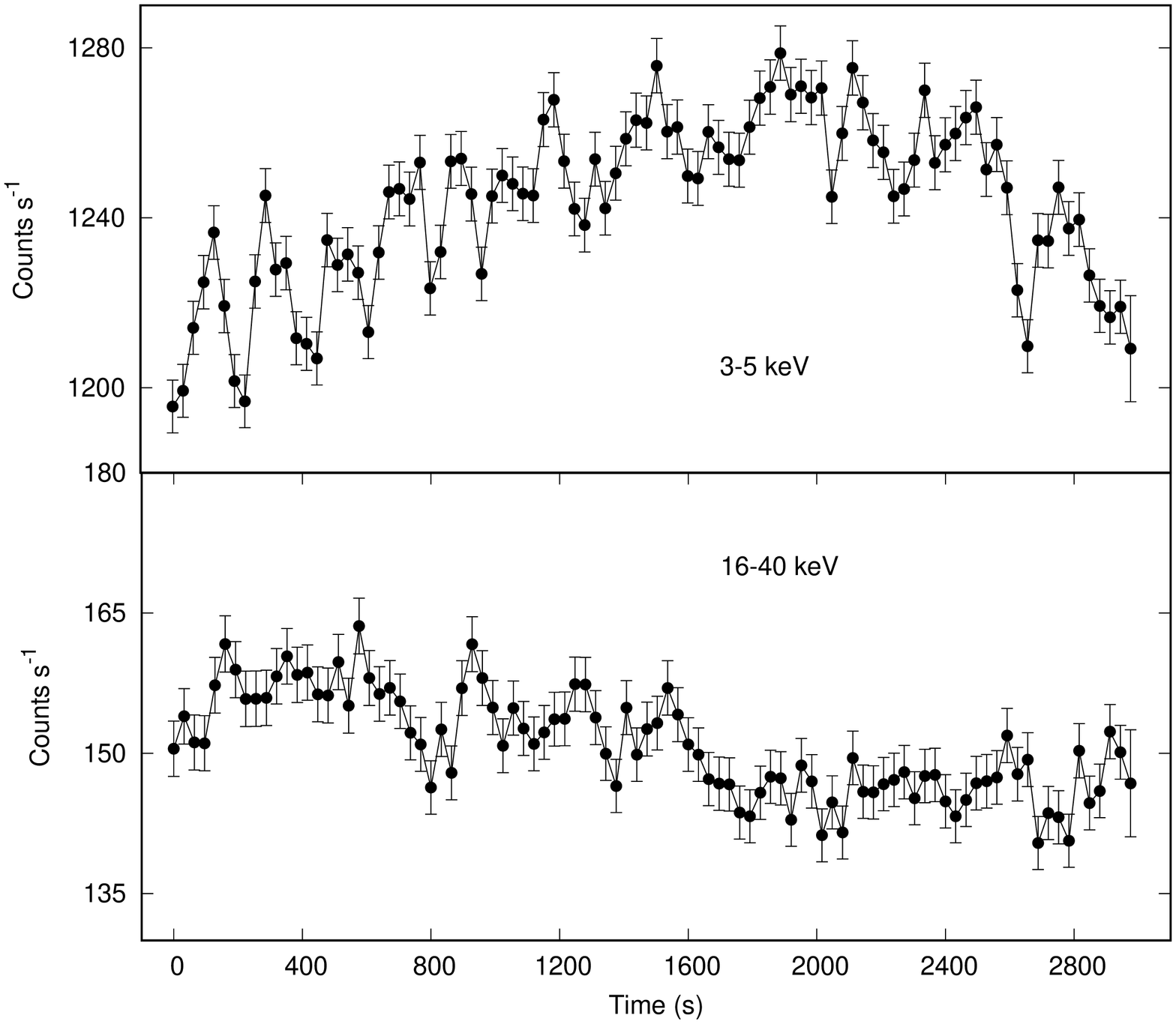}
\includegraphics[width=4cm, height=7cm, angle=270]{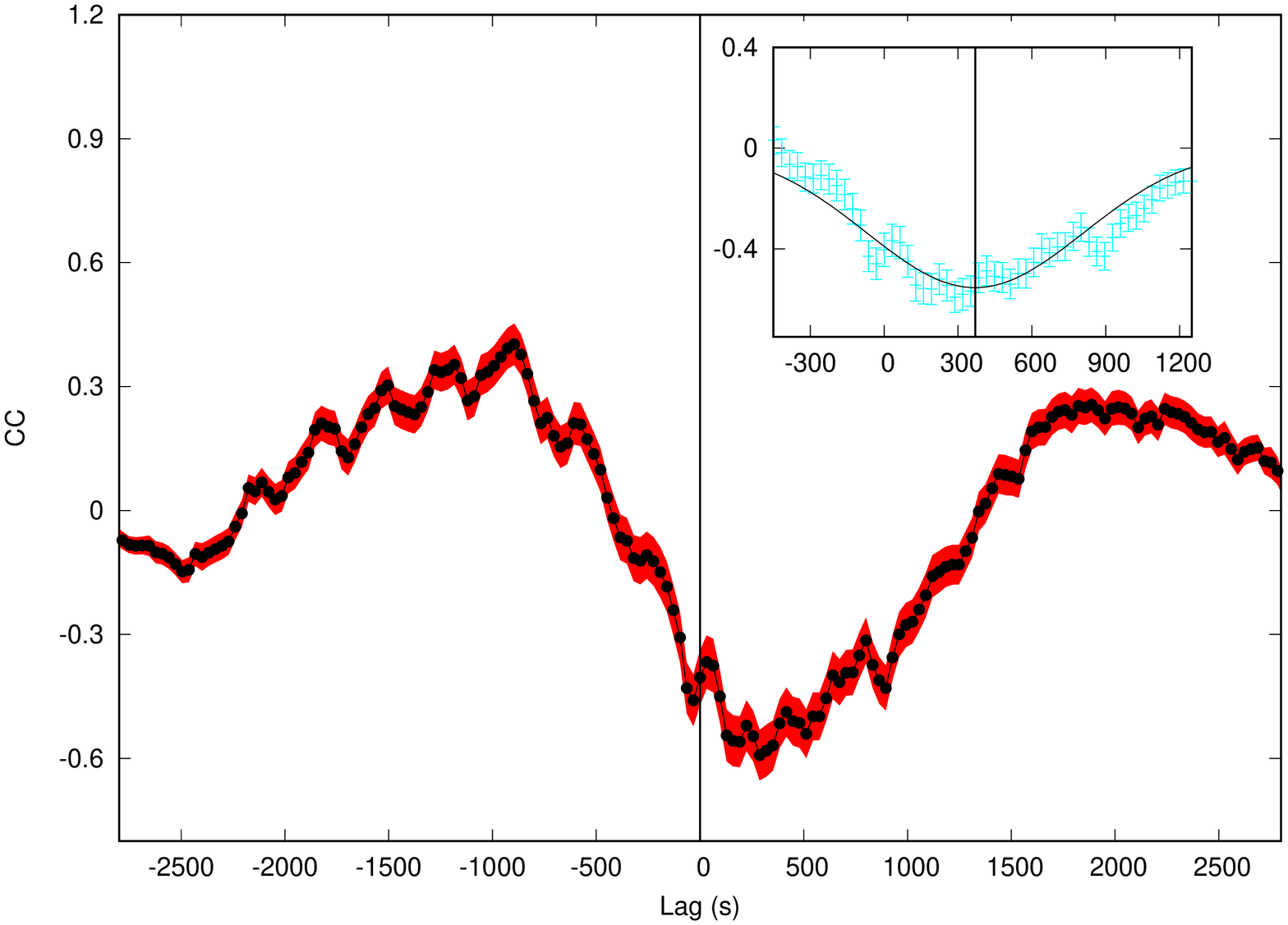}
\caption{}
\end{subfigure}

\begin{subfigure}[b]{0.6\columnwidth}
\includegraphics[width=4cm, height=7cm, angle=270]{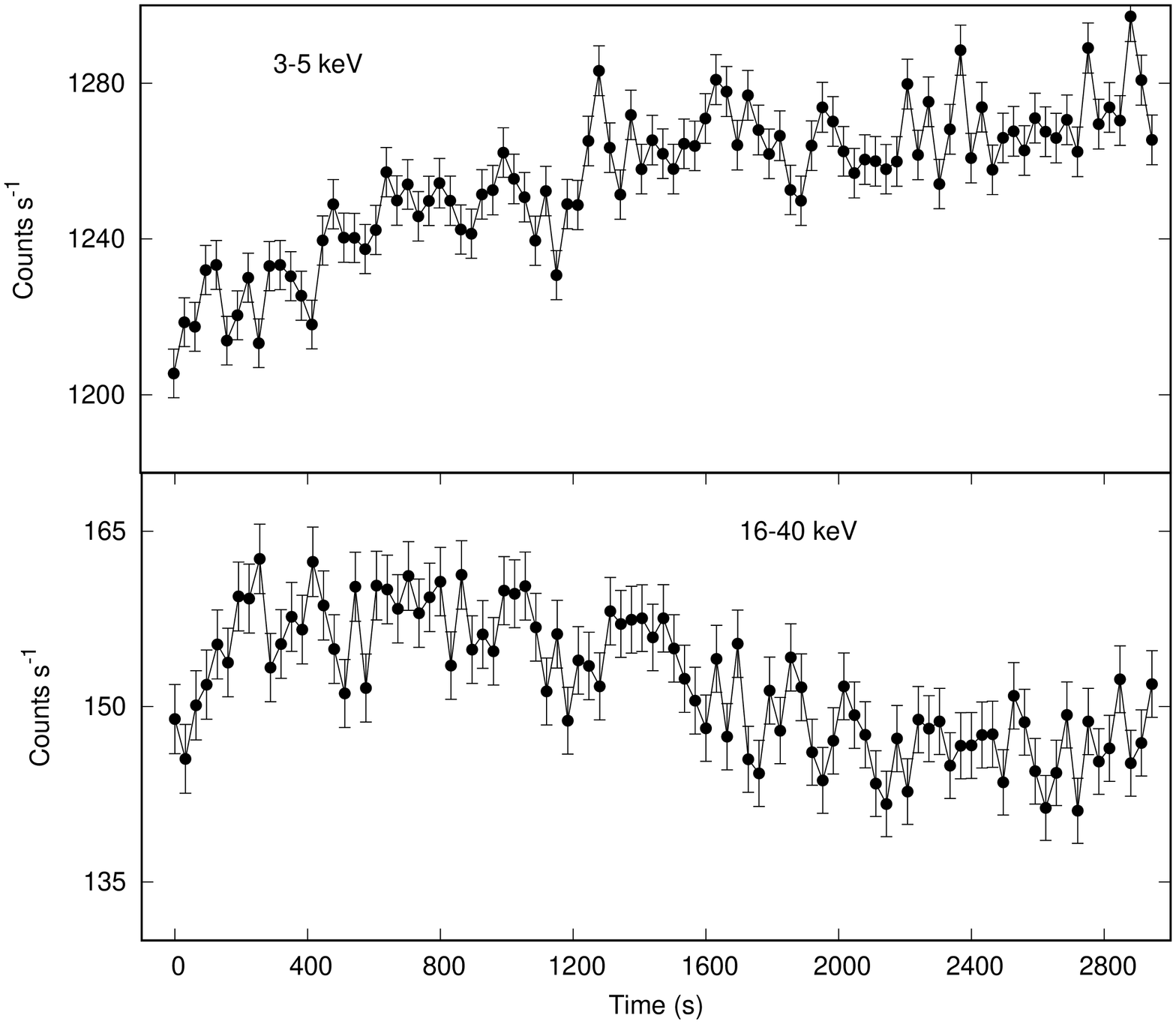}
\includegraphics[width=4cm, height=7cm, angle=270]{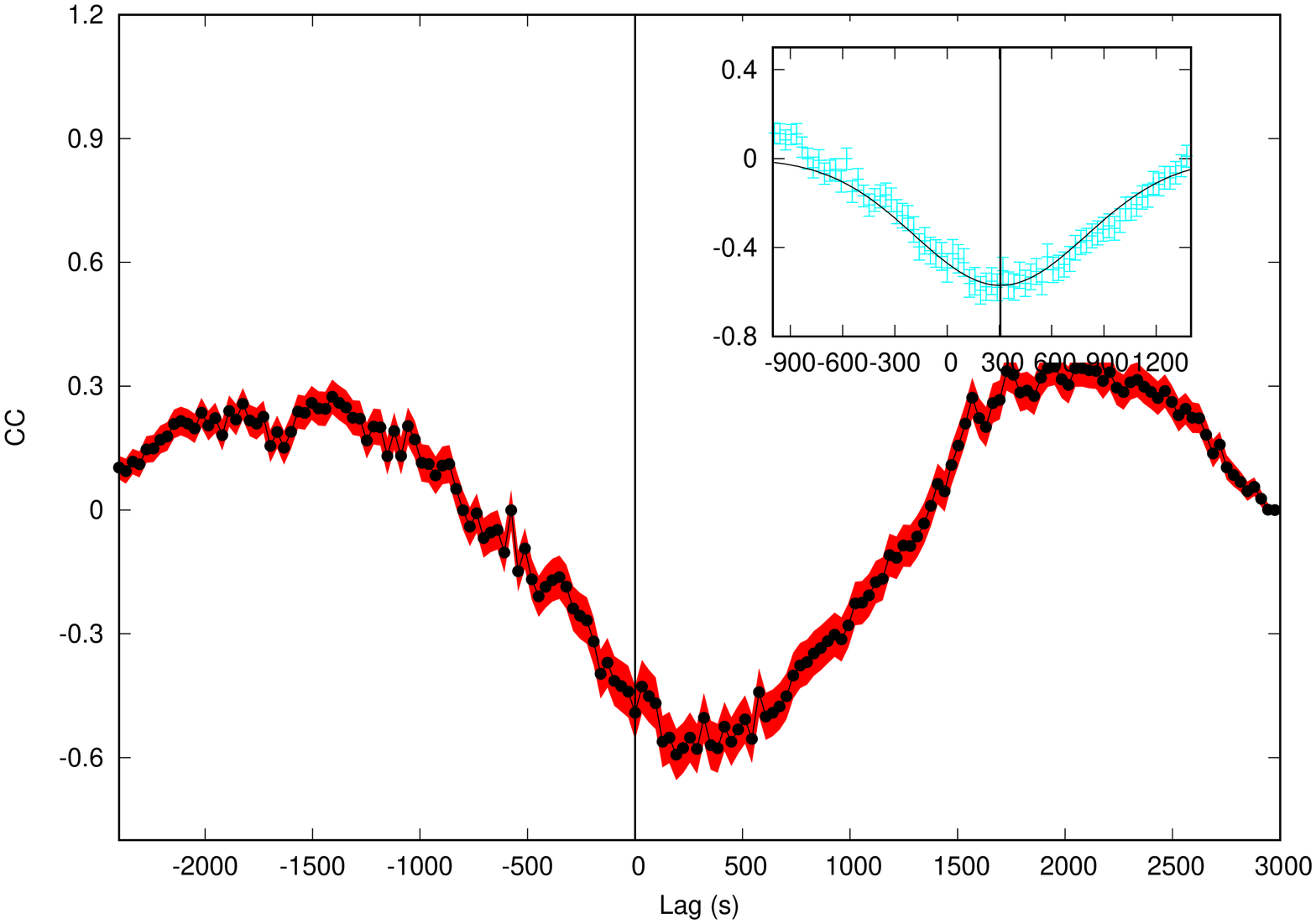}
\caption{}
\end{subfigure}
\end{figure}
\begin{figure}[!ht]\ContinuedFloat


\begin{subfigure}[b]{0.6\columnwidth}
\includegraphics[width=4cm, height=7cm, angle=270]{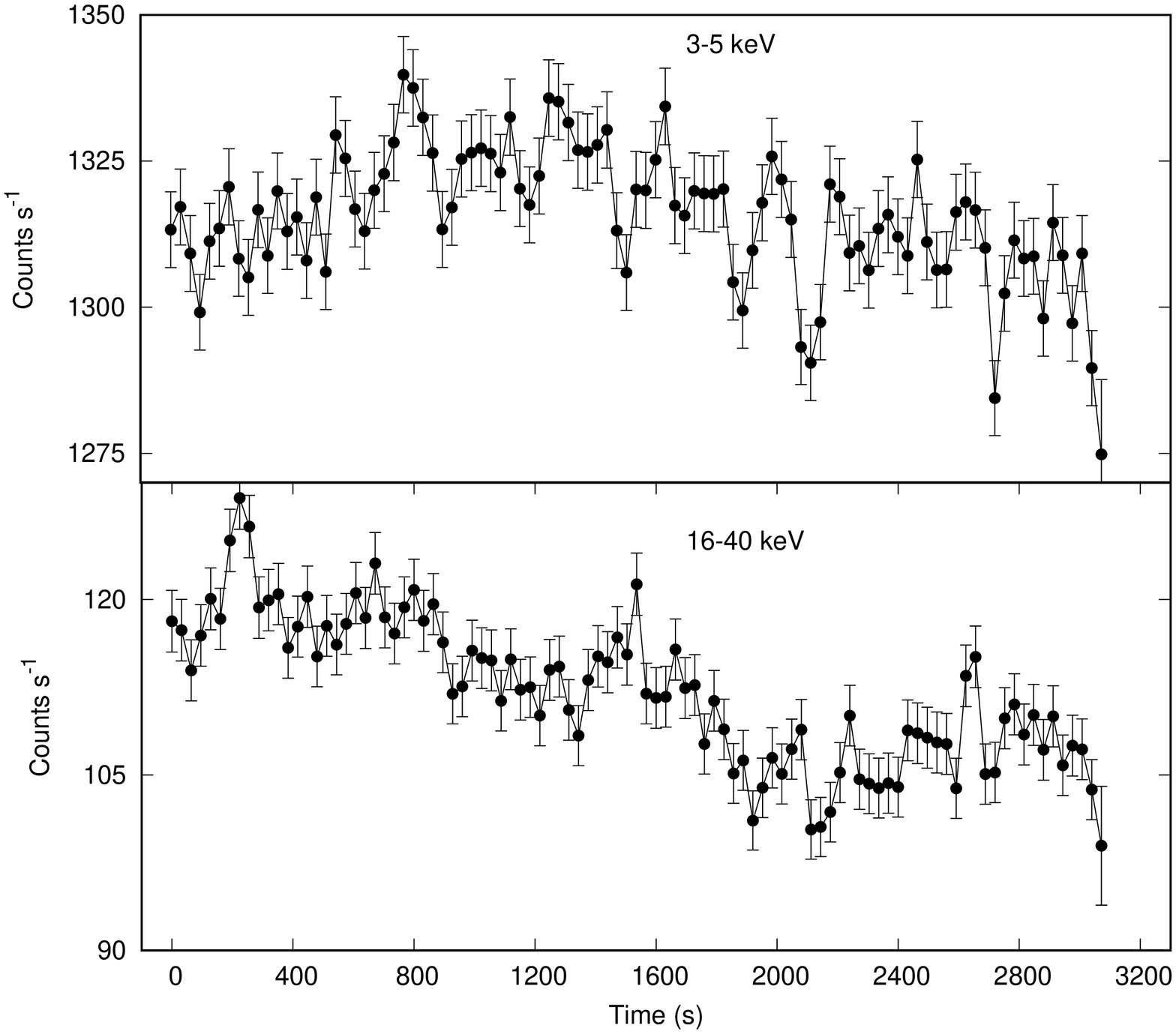}
\includegraphics[width=4cm, height=7cm, angle=270]{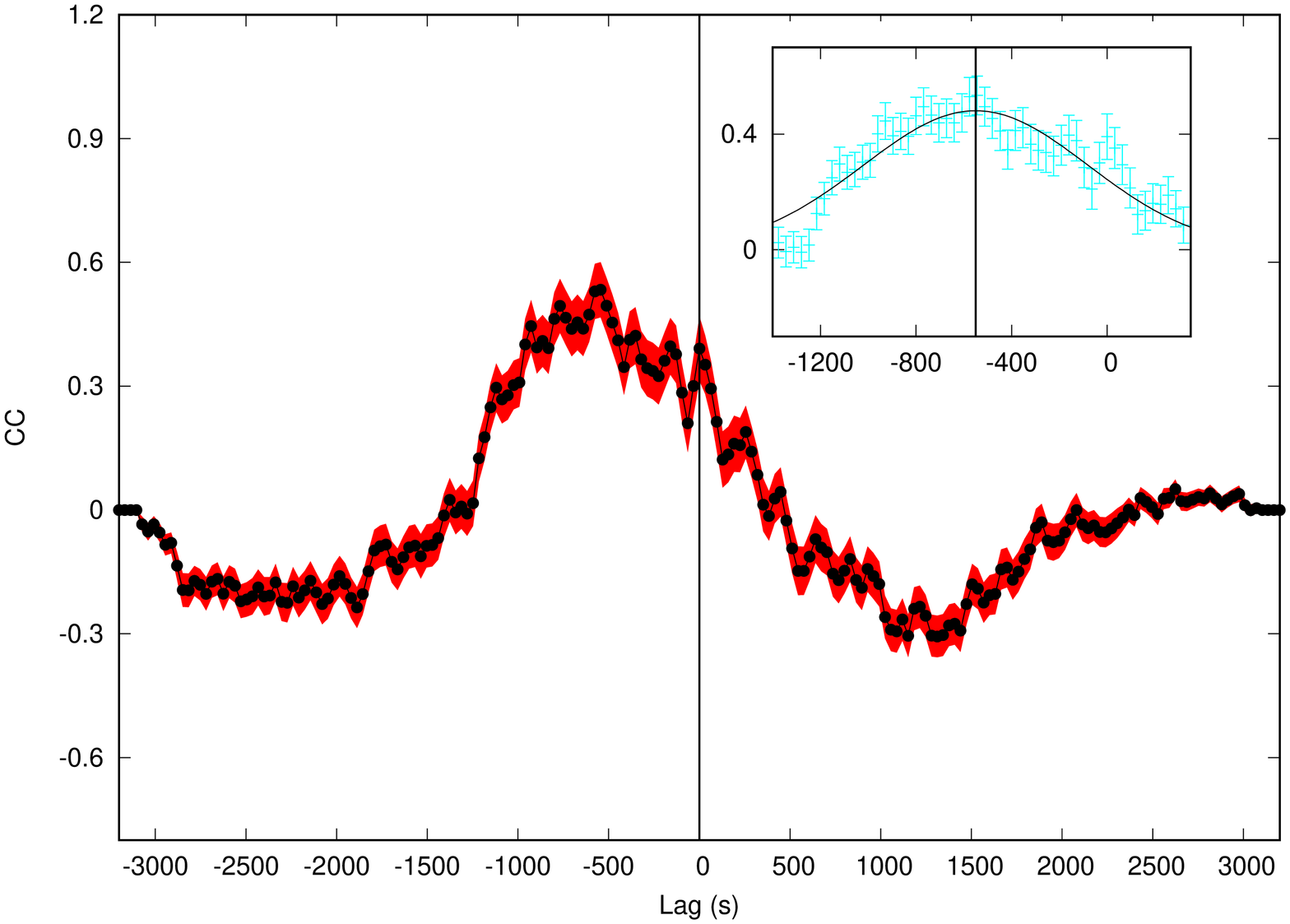}
\caption{}
\end{subfigure}



\begin{subfigure}[b]{0.6\columnwidth}
\includegraphics[width=4cm, height=7cm, angle=270]{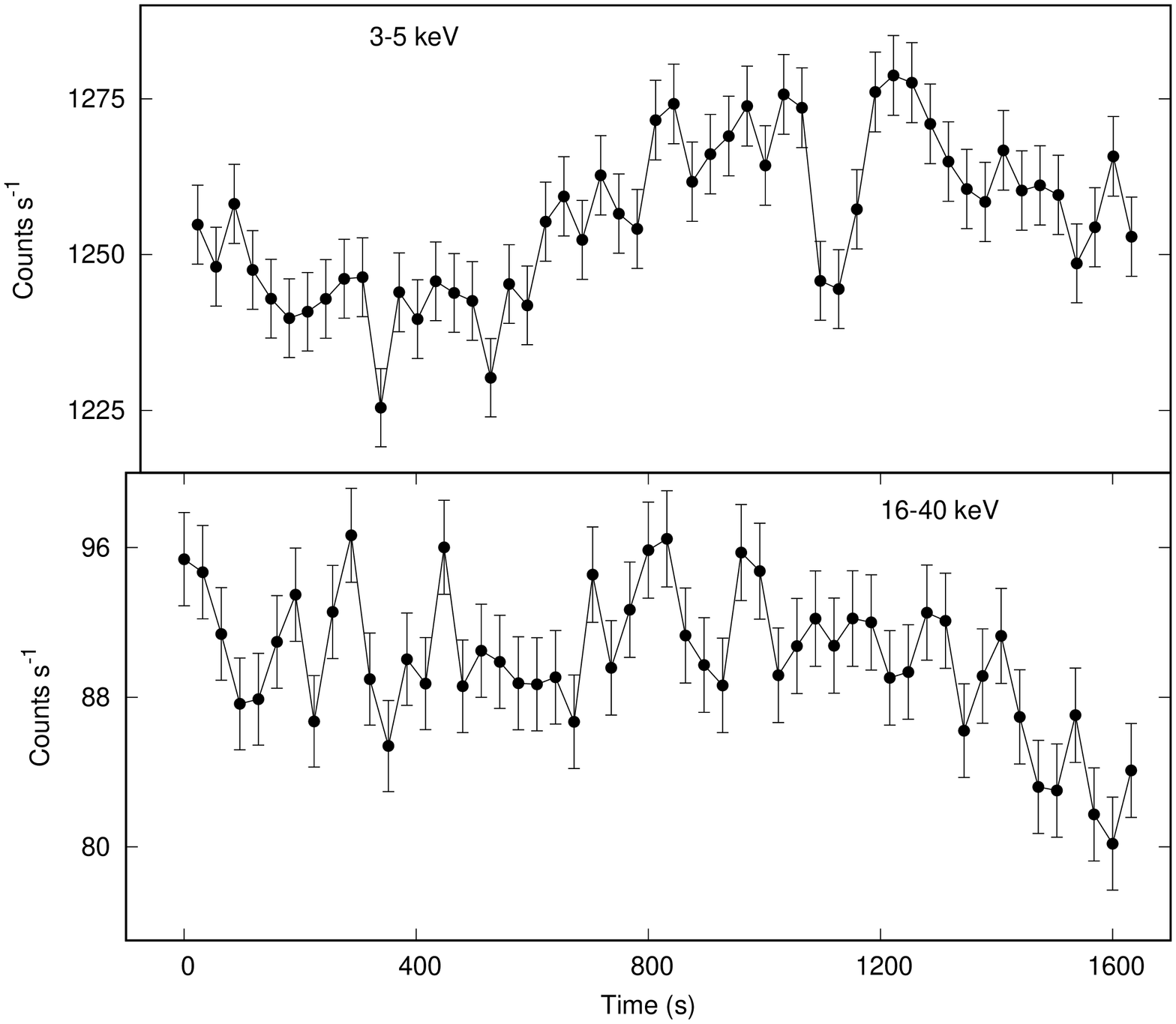}
\includegraphics[width=4cm, height=7cm, angle=270]{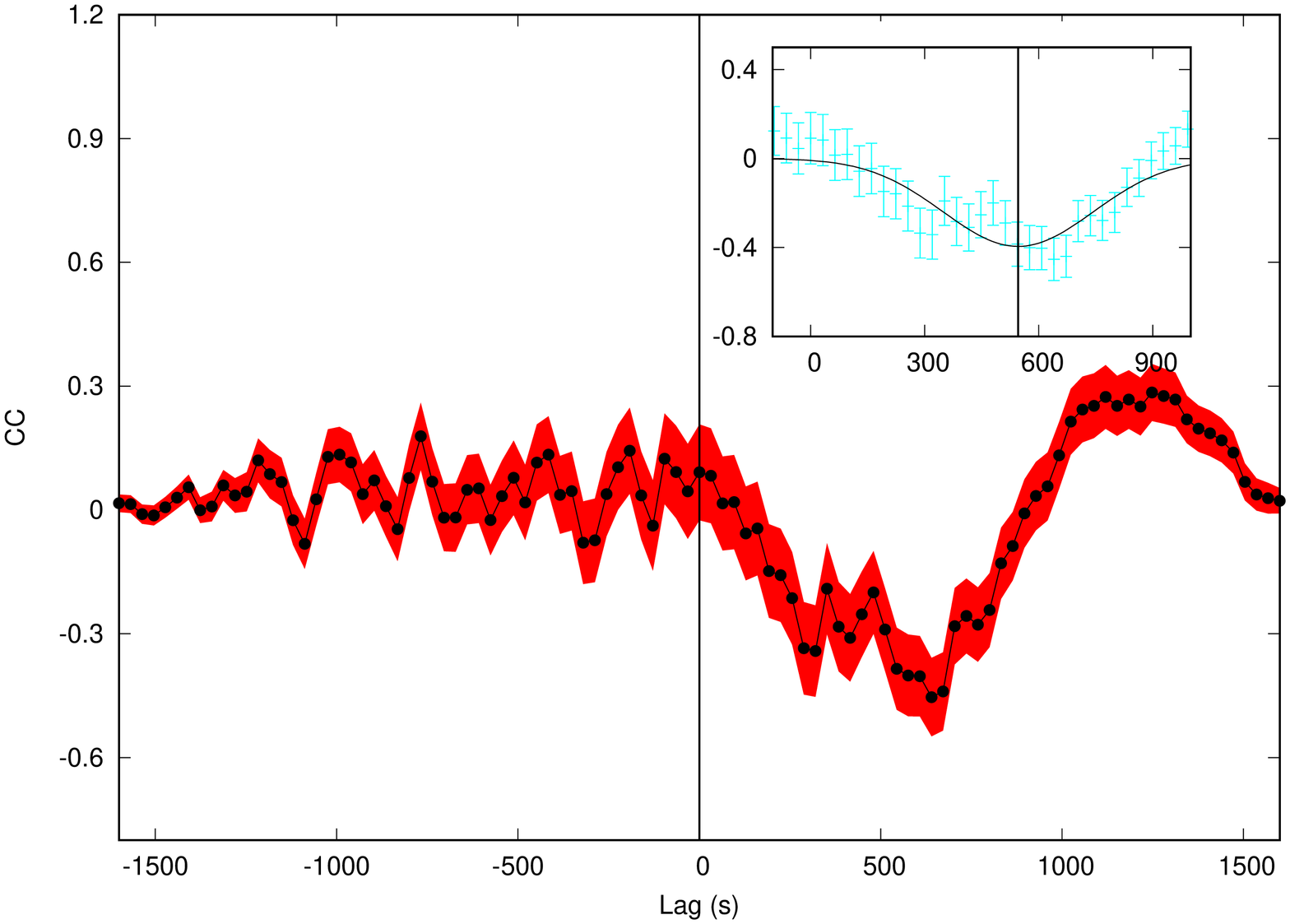}
\caption{}
\end{subfigure}
\end{figure}

\begin{figure}[!ht]\ContinuedFloat
\begin{subfigure}[b]{0.6\columnwidth}
\includegraphics[width=4cm, height=7cm, angle=270]{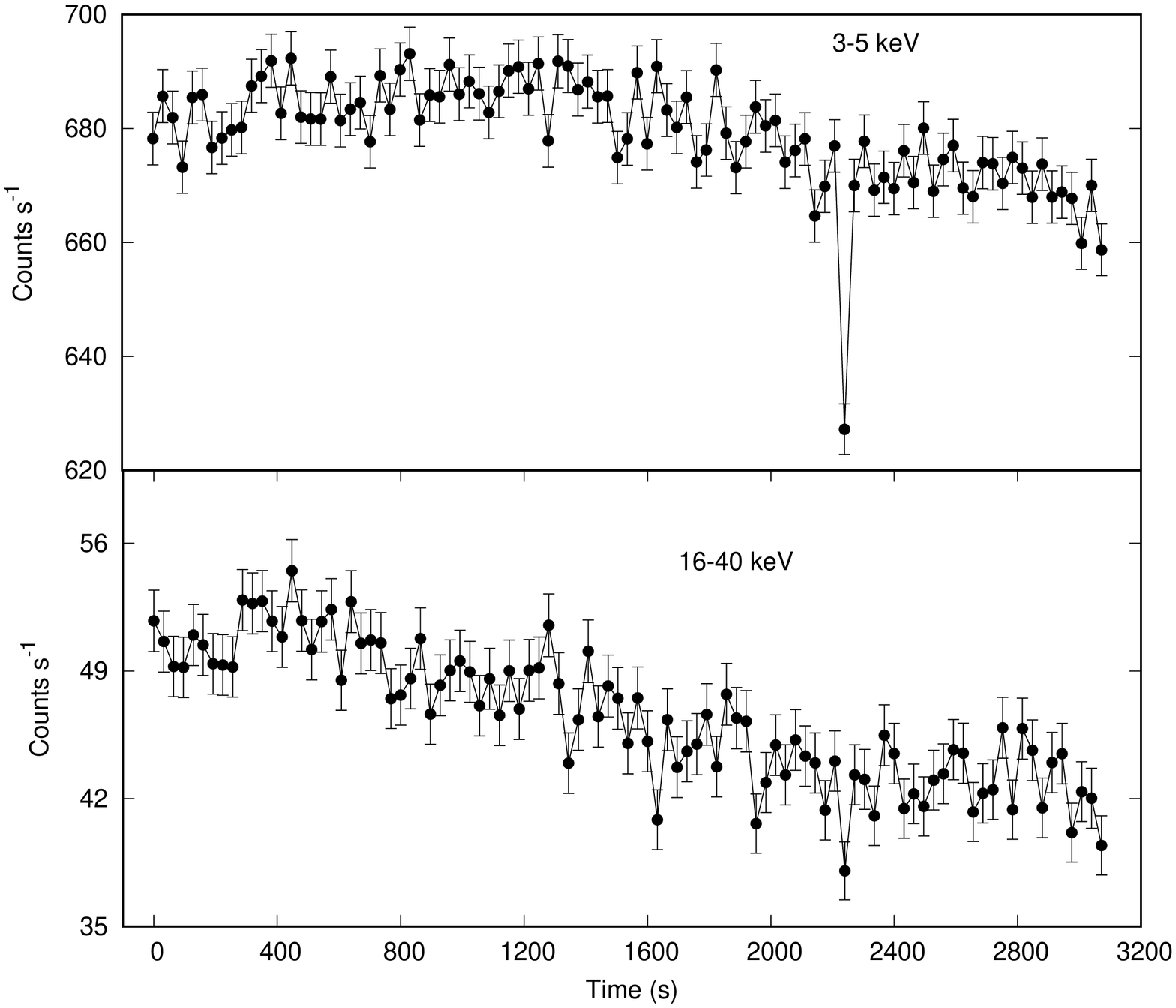}
\includegraphics[width=4cm, height=7cm, angle=270]{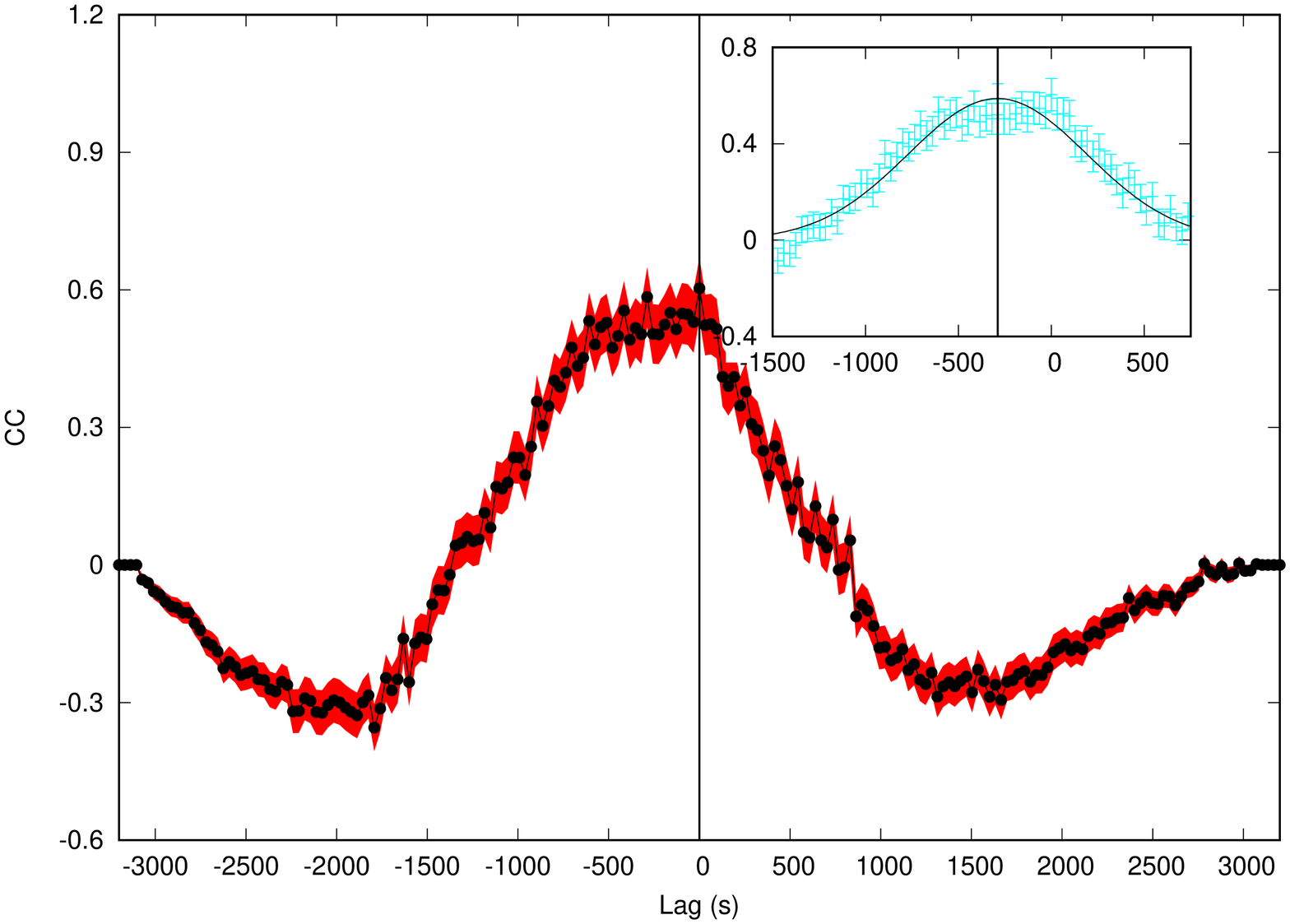}
\caption{}
\end{subfigure}





\begin{subfigure}[b]{0.6\columnwidth}
\includegraphics[width=4cm, height=7cm, angle=270]{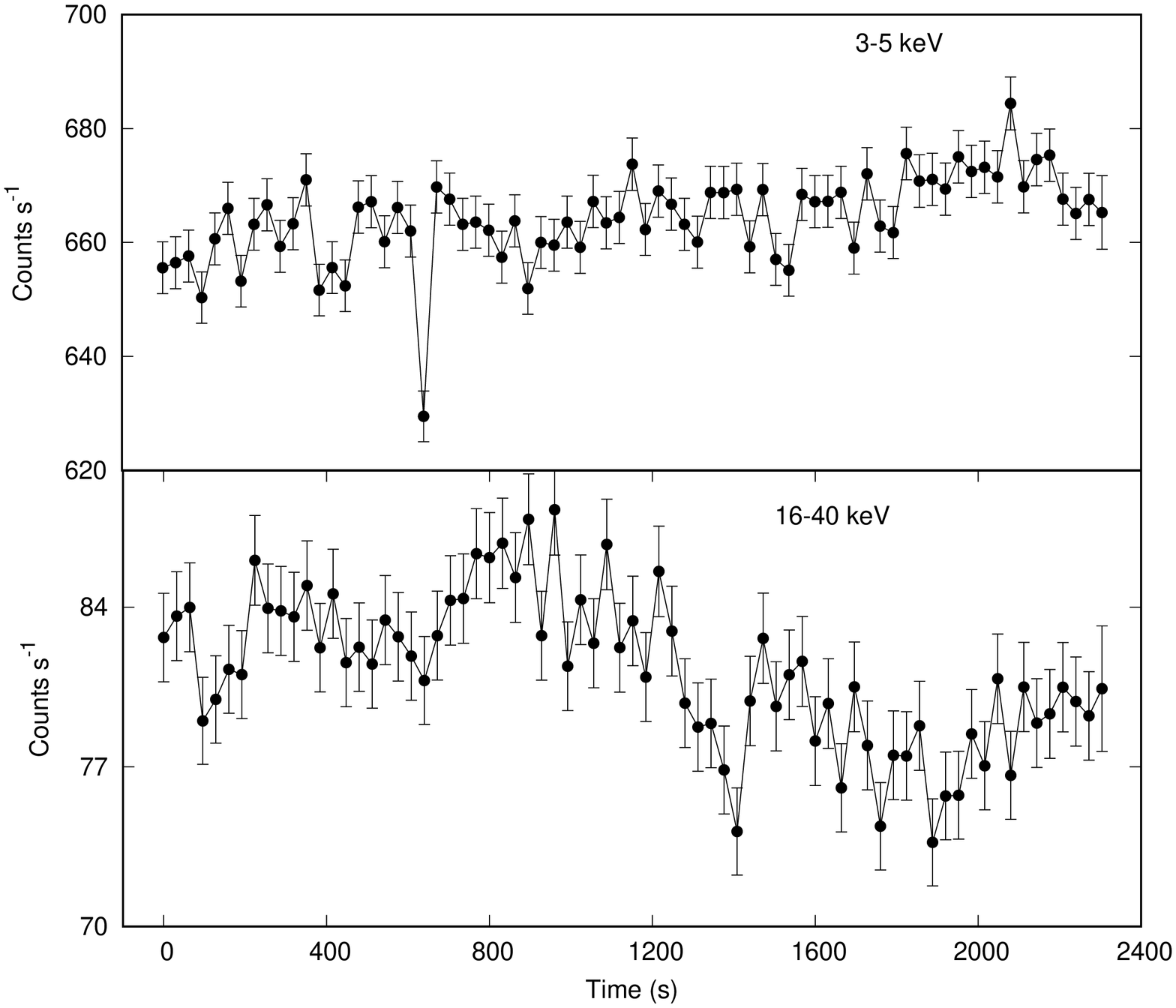}
\includegraphics[width=4cm, height=7cm, angle=270]{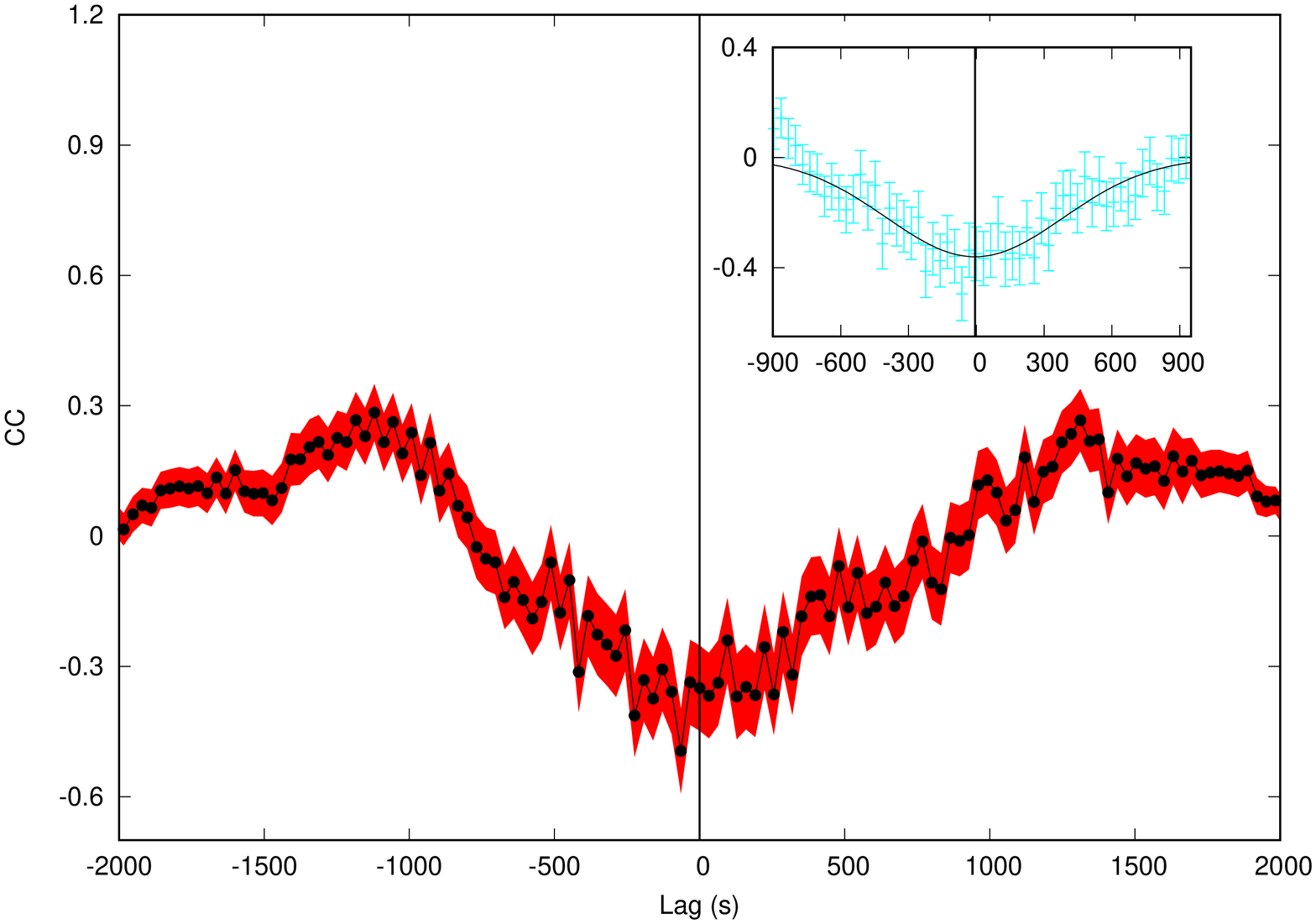}
\caption{}
\end{subfigure}
\end{figure}

\begin{figure}[!ht]\ContinuedFloat
\begin{subfigure}[b]{0.6\columnwidth}
\includegraphics[width=4cm, height=7cm, angle=270]{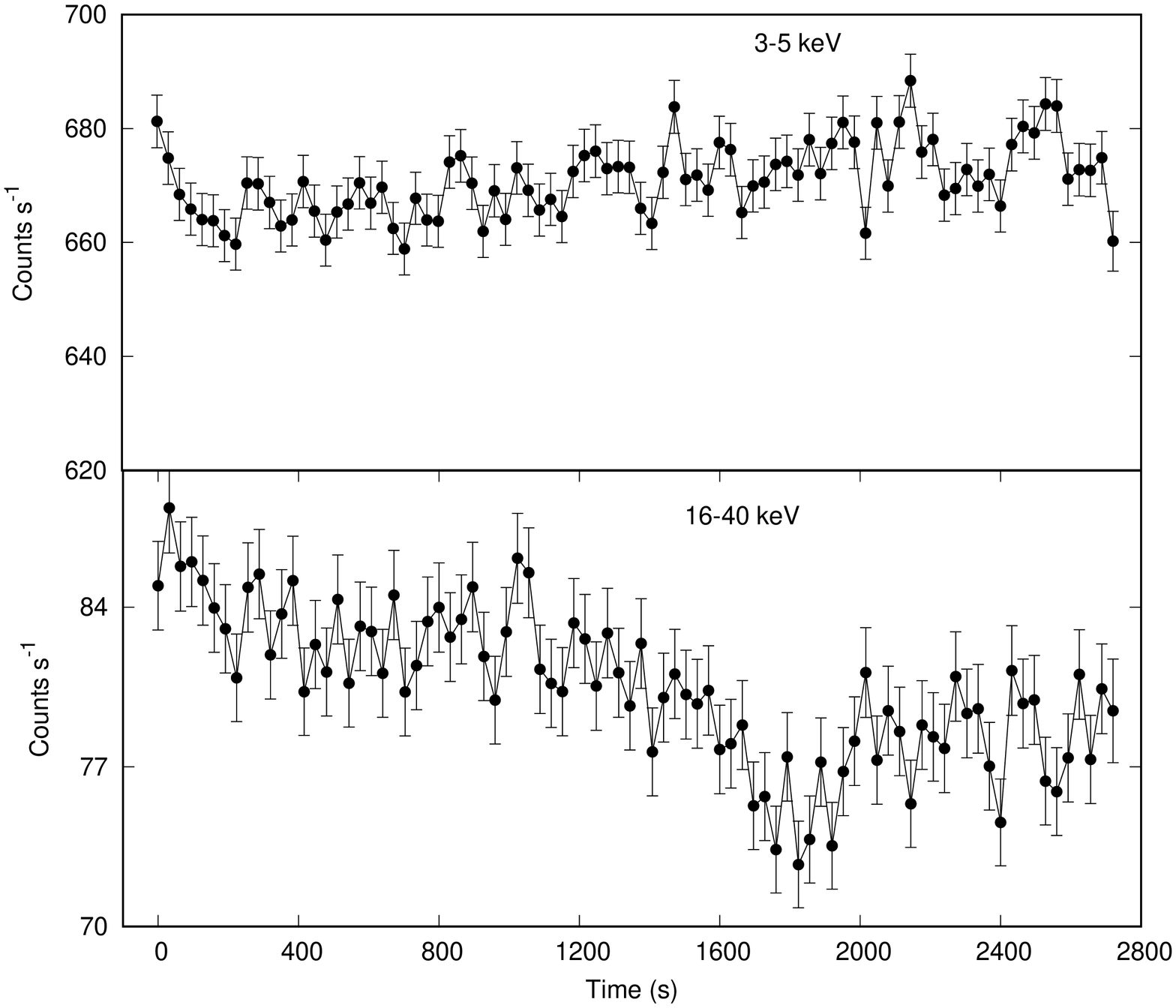}
\includegraphics[width=4cm, height=7cm, angle=270]{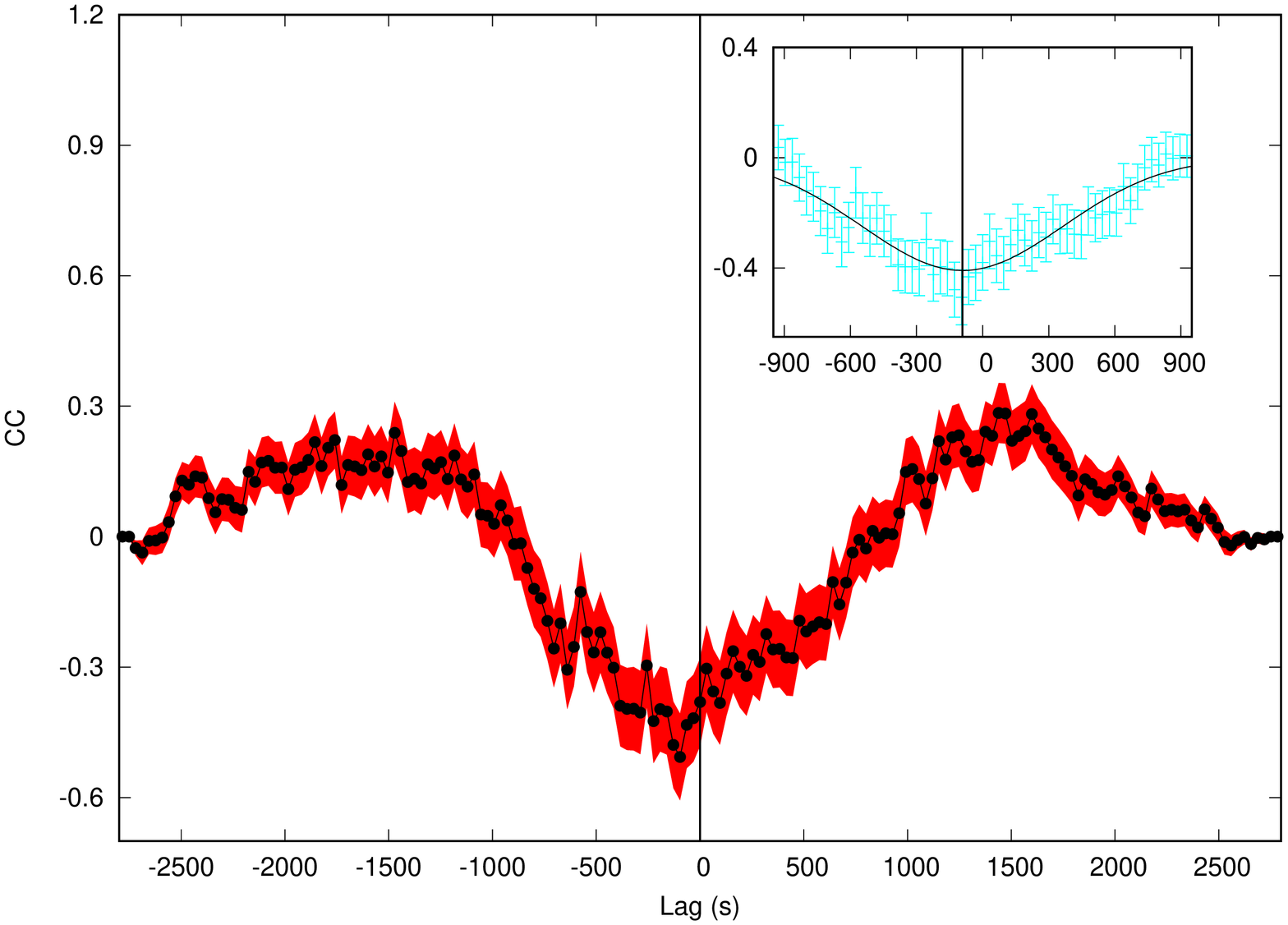}
\caption{}
\end{subfigure}

\begin{subfigure}[b]{0.6\columnwidth}
\includegraphics[width=4cm, height=7cm, angle=270]{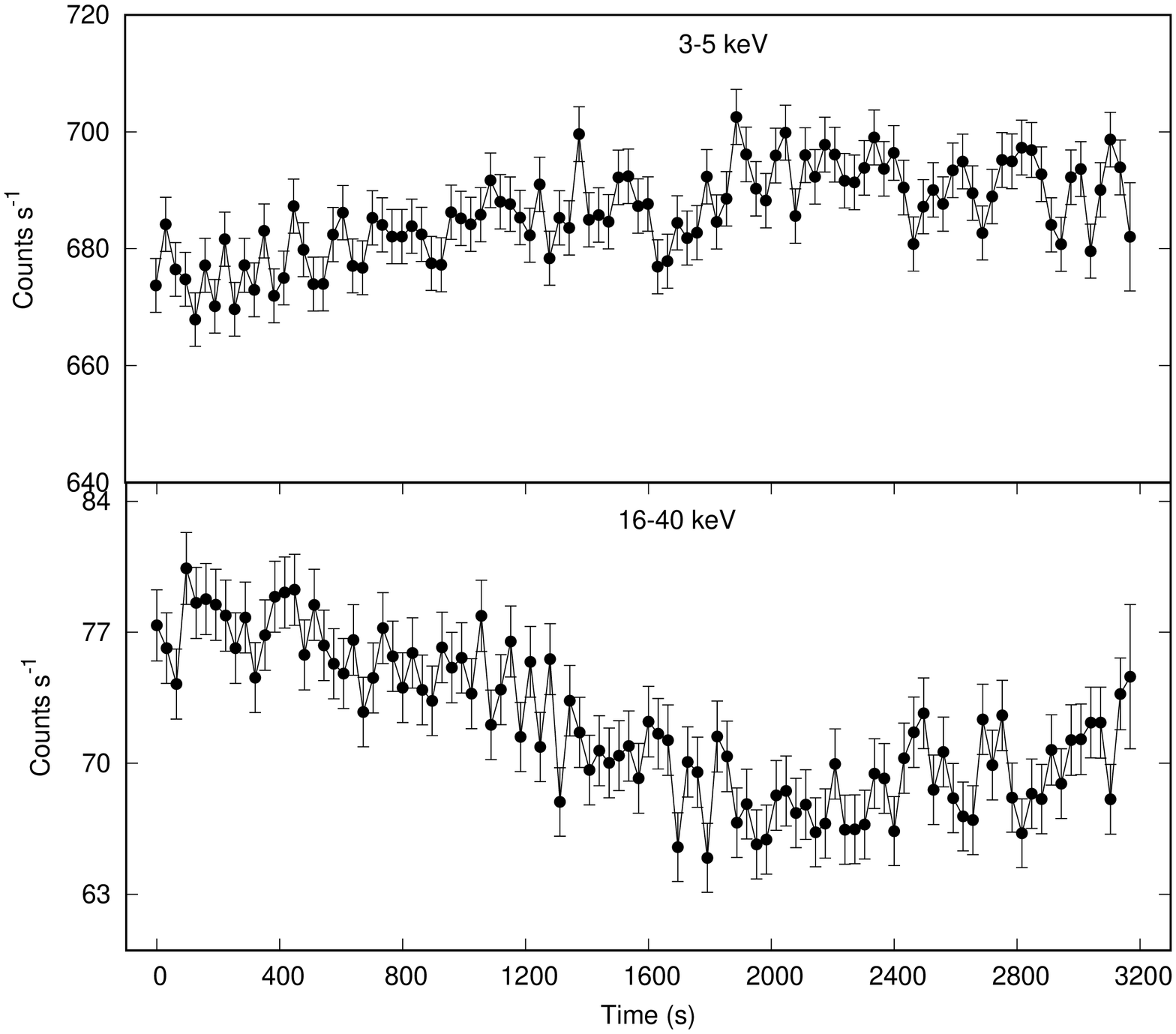}
\includegraphics[width=4cm, height=7cm, angle=270]{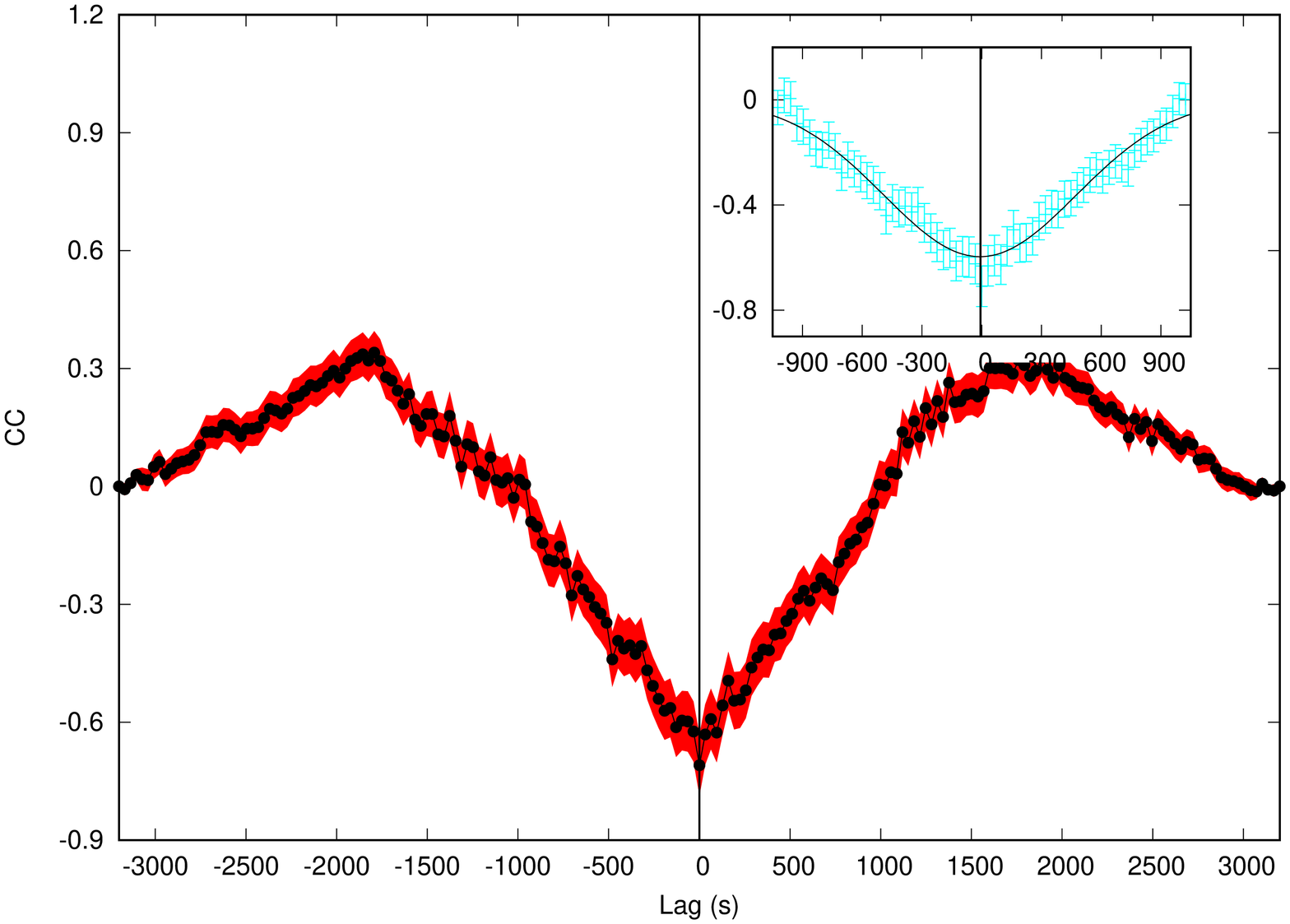}
\caption{}
\end{subfigure}
\end{figure}

\begin{figure}[!ht]\ContinuedFloat
\begin{subfigure}[b]{0.6\columnwidth}
\includegraphics[width=4cm, height=7cm, angle=270]{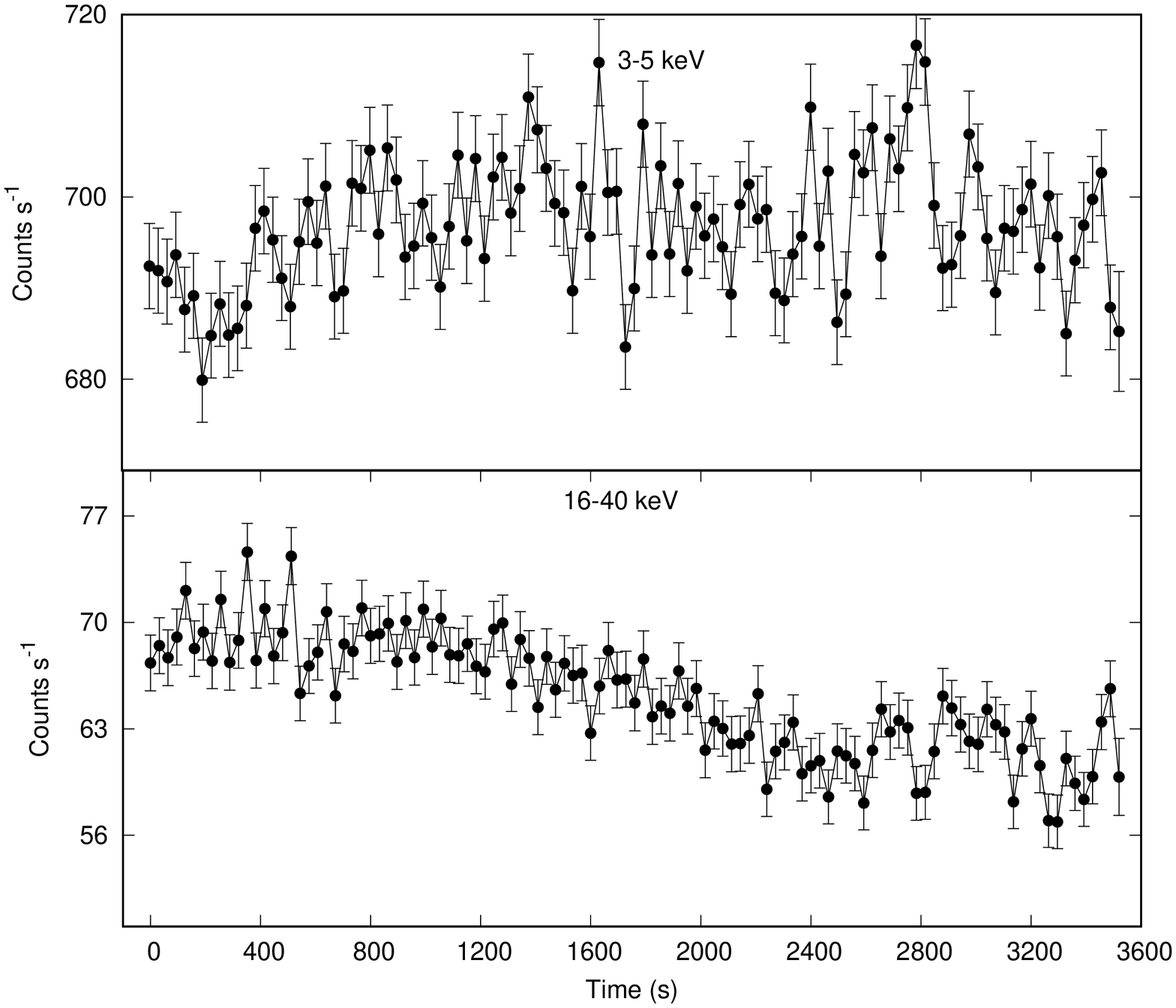}
\includegraphics[width=4cm, height=7cm, angle=270]{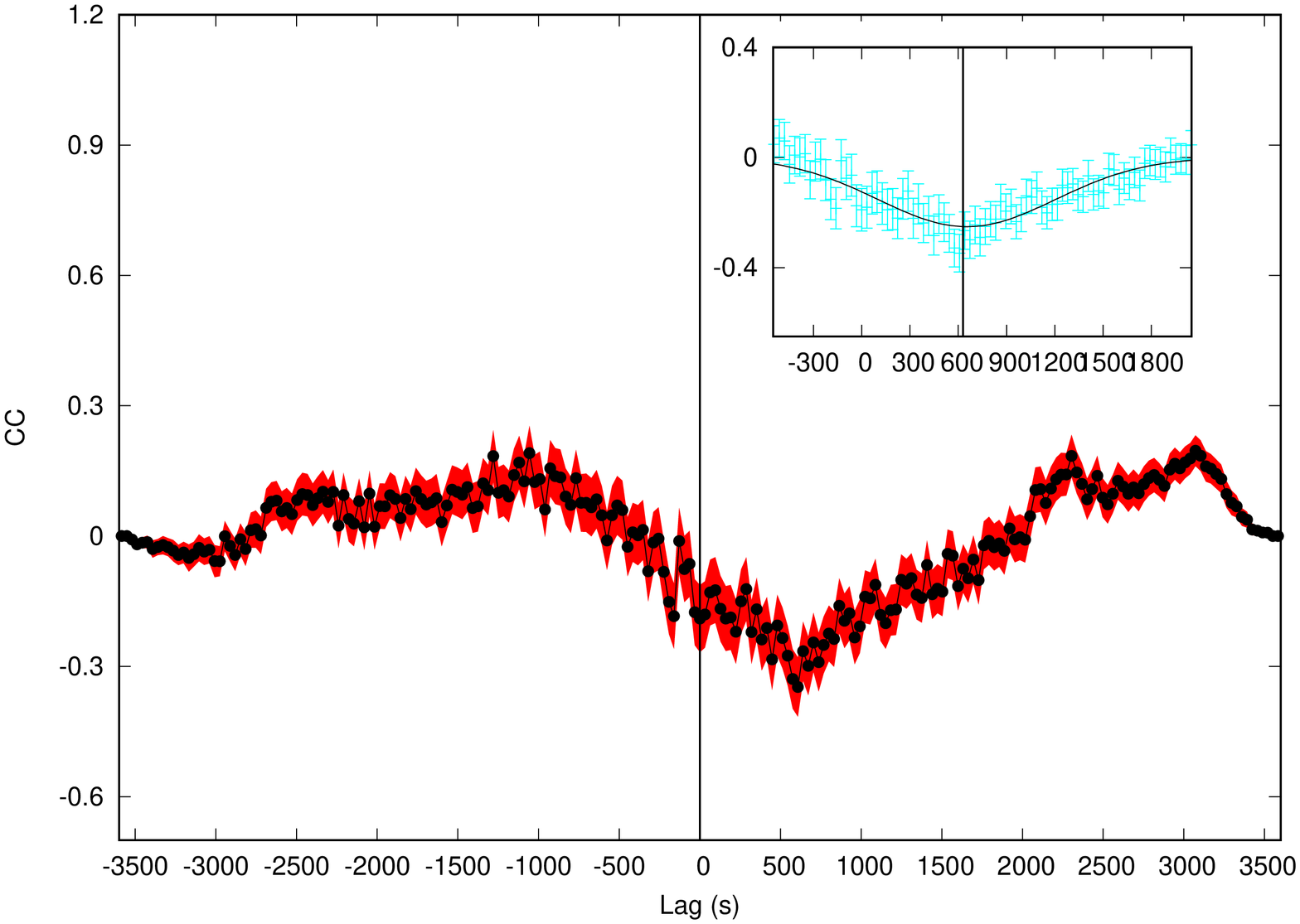}
\caption{}
\end{subfigure}
\end{figure}

\begin{figure*}[!ht]

\caption{The CCF Lag vs. $\Delta$ $\chi$$^{2}$ (= $\chi$$^{2}$ - $\chi$$^{2}$$_{min}$ ) plot obtained from the procedure mentioned in Sec. 3. }
\begin{subfigure}[b]{0.85\columnwidth}
\caption{For Section D}
\includegraphics[width=0.85\columnwidth, angle=270]{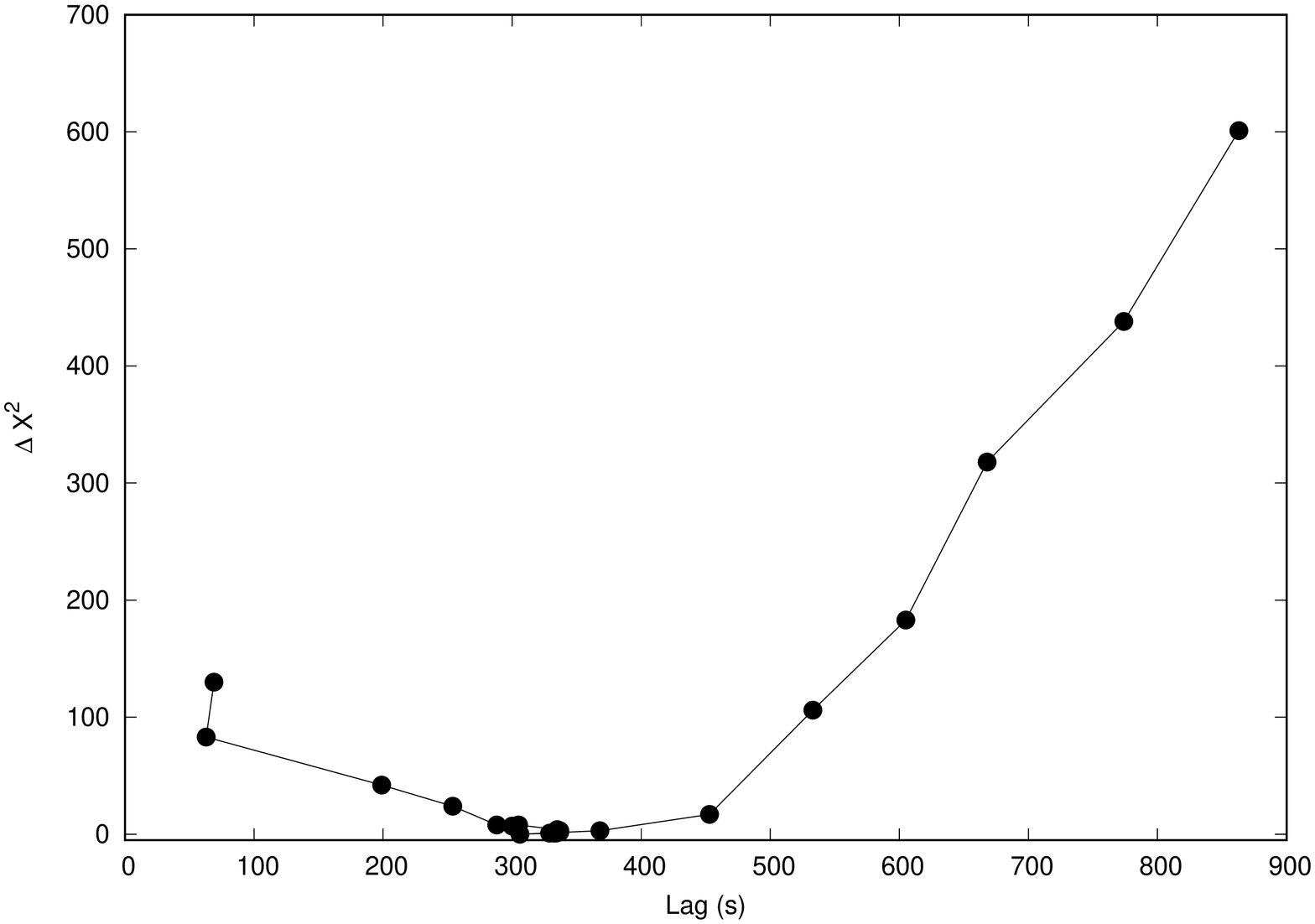}

\end{subfigure}

\begin{subfigure}[b]{0.85\columnwidth}
\caption{For Section E}
\includegraphics[width=0.85\columnwidth, angle=270]{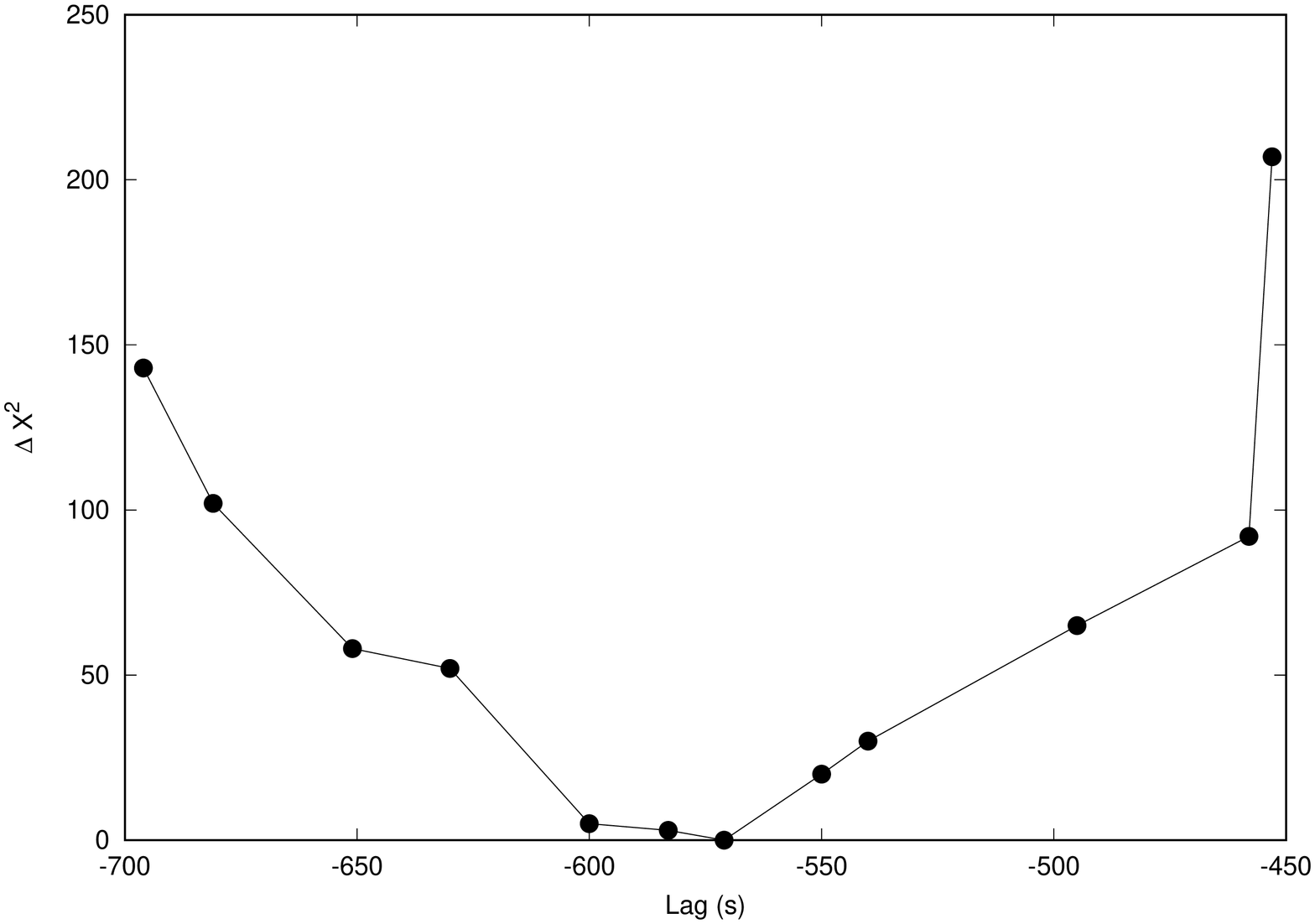}
\end{subfigure}

\begin{subfigure}[b]{0.85\columnwidth}
\caption{For Section G}
\includegraphics[width=0.85\columnwidth, angle=270]{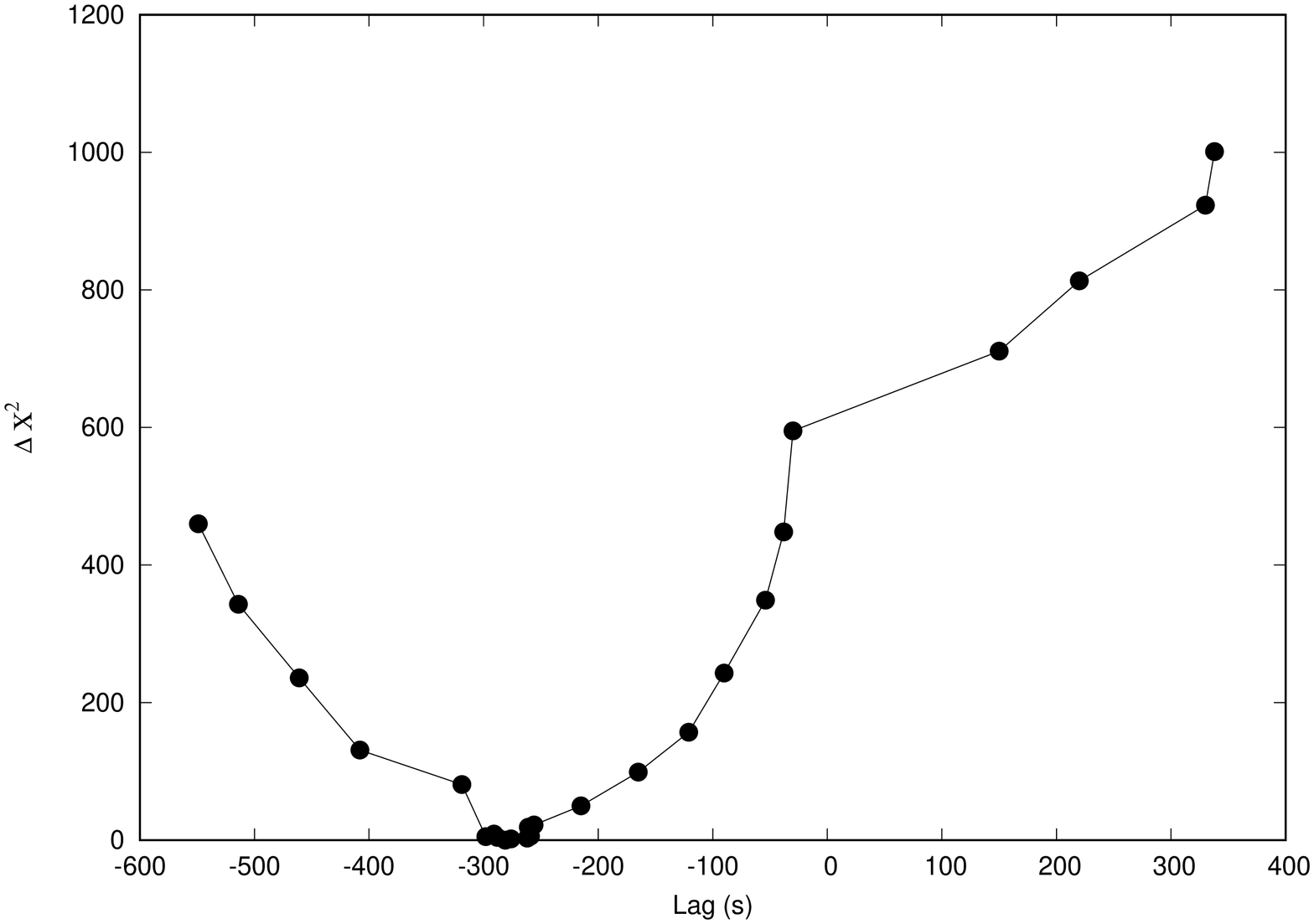}
\end{subfigure}
\end{figure*}

Power Density Spectrum (PDS) was obtained using a 1/2048 s binned light curves in the 3-20 keV energy band, using LAXPC10 and LAXPC20 in order to improve the
statistics. According to the normalization method by Miyamoto et al. (1991), power density spectra (PDS) were normalized in units of 
(rms/mean)$^2$/Hz. PDS were fitted with a one or two Lorentzian and power-law model (see Table 2). PDS has revealed a $\sim$ 25 Hz and $\sim$ 33 Hz HBO in the G05-112T01-9000000452 
(Orbit 03352, 03353,03355), when the source was in the HB (see Table 2). For one of the light curve segments (orbit 03352), PDS revealed a 25.04 $\pm$ 0.44 Hz HBO with a 50.37 $\pm$ 0.43 Hz harmonic. For orbit 03353, a 32.34 $\pm$ 0.23 Hz HBO with a harmonic
of 61.74 $\pm$ 4.75 Hz was observed. This section also exhibits an anti-correlated hard lag of 426 $\pm$ 38 s. Another segment (orbit 03355) revealed a 35.24 $\pm$ 0.33 Hz HBO with no harmonic (see Figure 4 and Table 2).

\begin{figure*}[!ht]

\caption{HBOs seen in Observation Id G05-112T01-9000000452 for the light curve in 3-20 keV energy range
 using AstroSat LAXPC observations fitted using a Lorentzian and powerlaw model to the PDS.}
\begin{subfigure}[b]{0.85\columnwidth}
\includegraphics[width=0.85\columnwidth, angle=270]{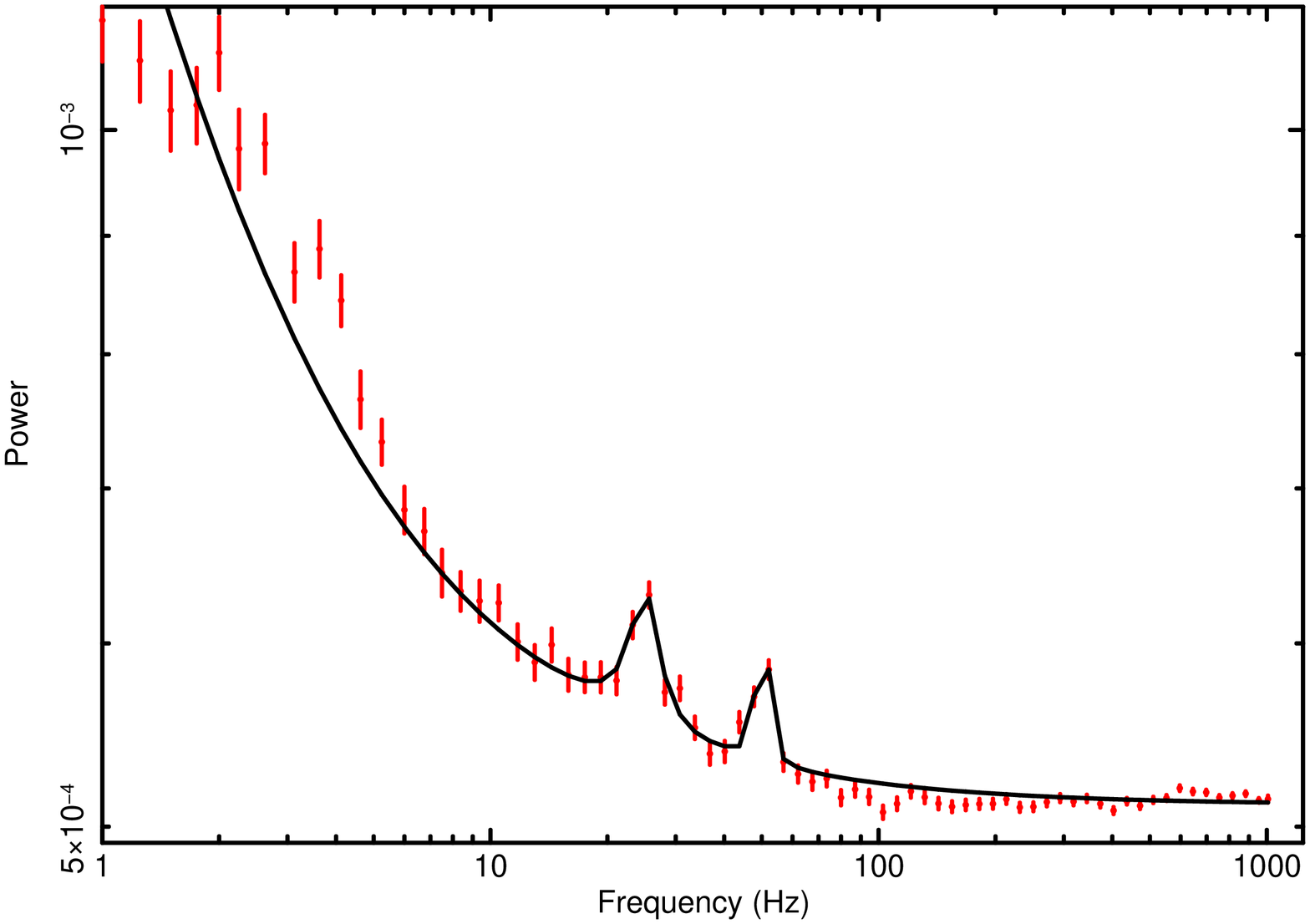}

\end{subfigure}

\begin{subfigure}[b]{0.85\columnwidth}
\includegraphics[width=0.85\columnwidth, angle=270]{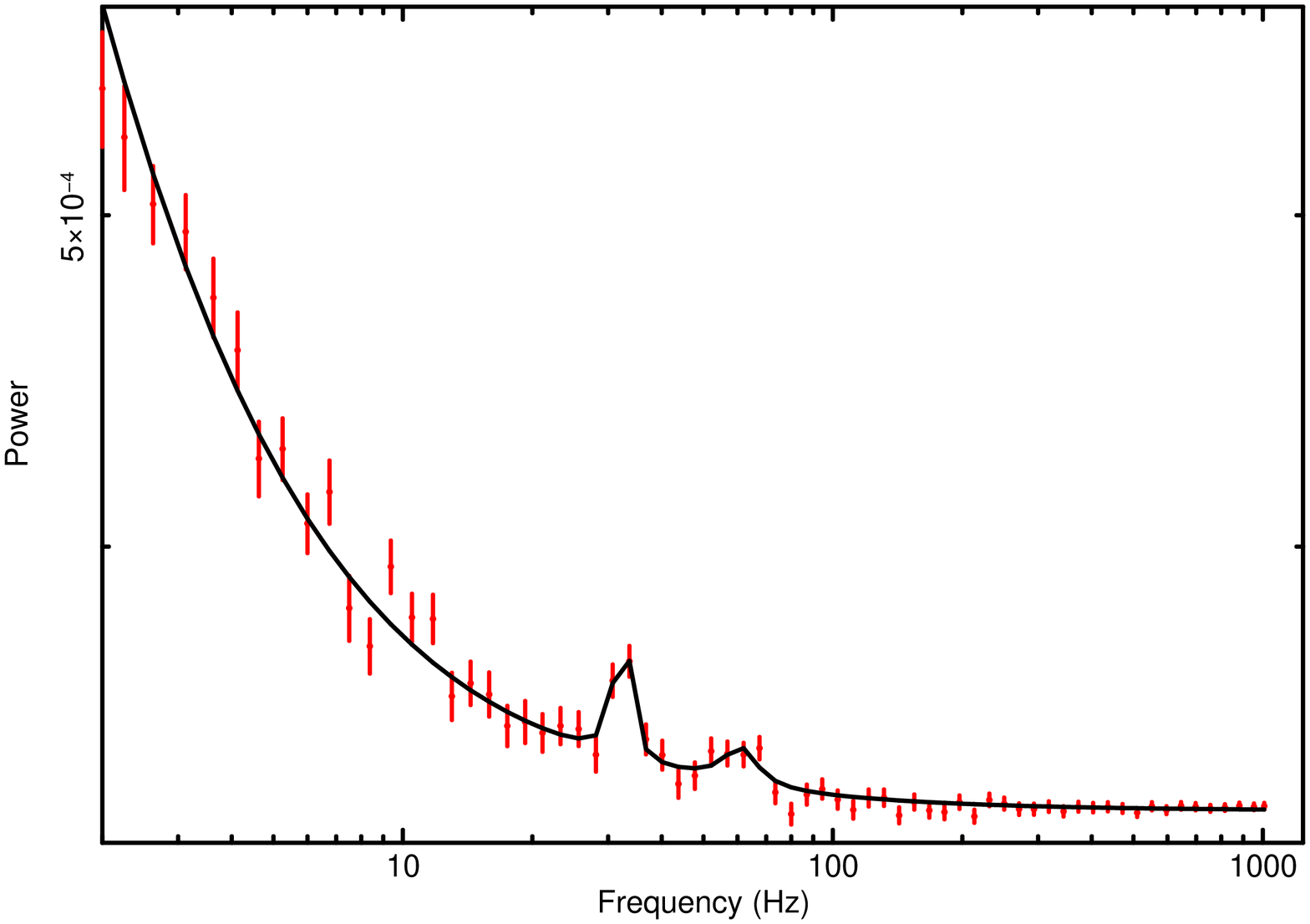}
\end{subfigure}

\begin{subfigure}[b]{0.85\columnwidth}
\includegraphics[width=0.85\columnwidth, angle=270]{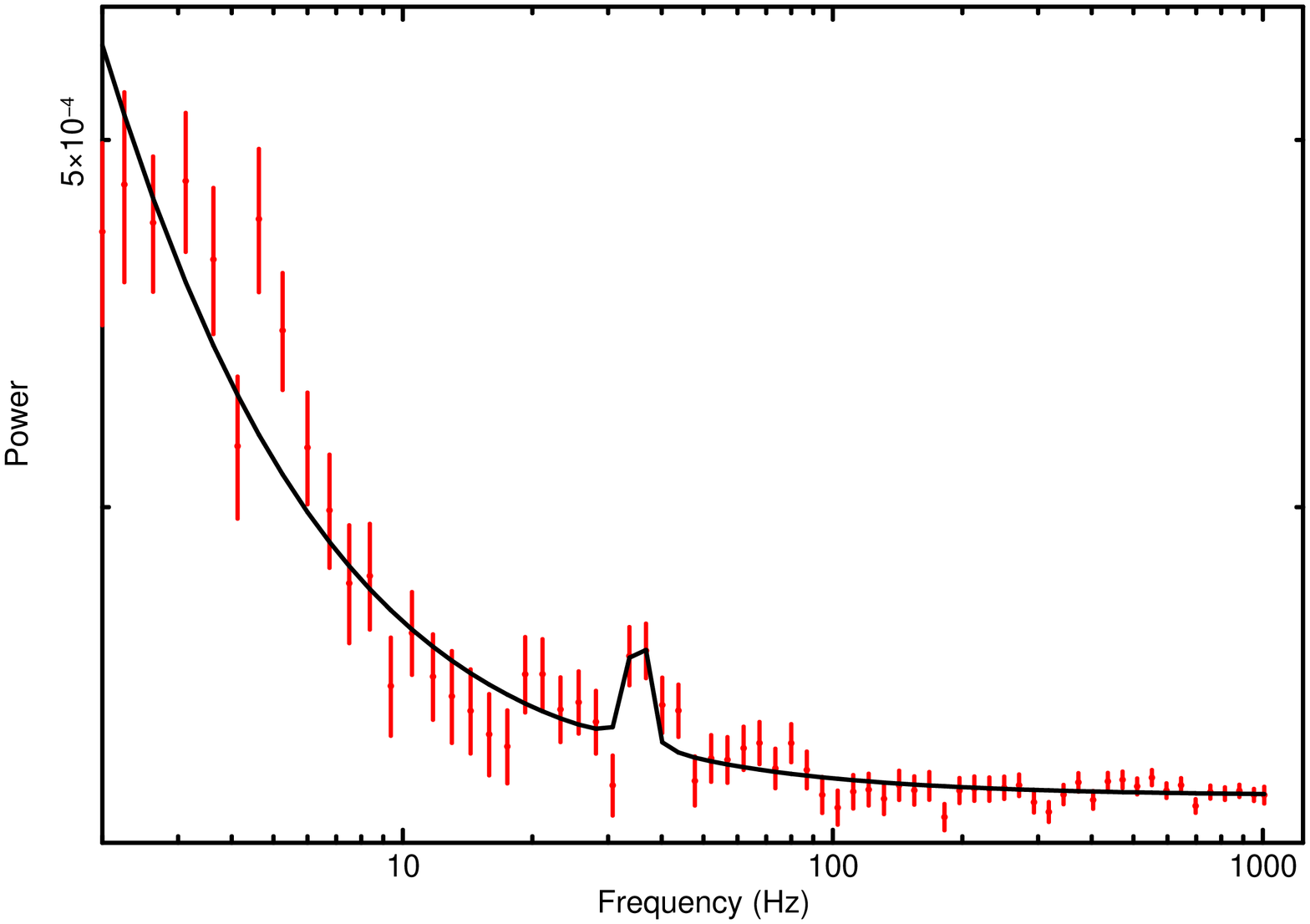}
\end{subfigure}
\end{figure*}

\begin{table}[htb]
\begin{minipage}[t]{\columnwidth}

\scriptsize
\caption{HBO parameter values}\label{tab2} 
\centering
\begin{tabular}{ccccccccccccccccc}
\topline
Observation Id & G05-112T01-9000000452 \\
\hline
Orbit & HBO $\nu$ (Hz) & $\Delta$ $\nu$ (Hz) & RMS \% \\
\hline
03352       & 25.04$\pm$0.44 &   5.97$\pm$1.91 & 1.21$\pm$0.11\\
"       & 50.37$\pm$0.43 &   5.67$\pm$2.75 & 1.60$\pm$ 0.14\\
03353       & 32.34$\pm$0.23 &   1.77$\pm$0.28 &  0.96$\pm$0.08\\
"       & 61.74$\pm$4.75 &  13.74$\pm$6.13 & 1.53$\pm$0.14 \\
03355       & 35.24$\pm$0.33 &   4.41$\pm$1.18 & 1.06 $\pm$0.10\\

\hline
\end{tabular}
\end{minipage}
\end{table}
	
\clearpage
\section{SPECTRAL ANALYSIS}
In order to check if any associated spectral variations are present, spectral analysis was performed for the first and 
last 1000 s of light curve segments (marked a, b in Table 3, figure 2) which exhibited significant lags and separately
for those segments which exhibited HBOs (Table 4, figure 4 ).
Spectra were fitted in the 3--50 keV energy range using XSPEC v12.10.1f (Arnaud 1996) and uncertainties were estimated with a 90\% confidence level 
($\Delta$ $\chi$$^2$ = 2.71). Figures 5, 6, 7 and 8 show the top panel with the unfolded spectra (thick line) 
along with the model components (dashed line) and the bottom panel with the residuals to the fit. 
Absorption column density was modeled using the Tbabs model (Wilms et al. 2000) and was fixed
at $\sim$ 2.2 $\times$ 10$^{22}$ cm$^{-2}$ (Cackett et al. 2010).
Using the continuum model {\it DiskBB + bbody + power-law} (Cackett et al. 2008, 2010; Lin et al. 2007),
the spectra were fitted which led to residuals around 6.7 keV which was then modeled using a Gaussian function 
centered at 6.7 keV (fixed) resulting in lower reduced $\chi^{2}$/dof values (see Table 3, Figure 4). 
Gaussian and power-law component models are needed to unfold the spectra as indicated by the values of F-test probability (Table 3). 
$\chi^{2}$/dof values before adding the power-law component was noted to be 93.69/39, 112.21/39 for section A(a,b) which is significantly 
higher compared to that obtained after incorporating the power-law component, i. e. 29.72/37, 47.54/37.
Spectral parameters are found to be non-varying (within error bars) for the first and last 1000 s sections of each segment that exhibited
lags (see Table 3). During these observations, the inner disk temperature (kT$_{in}$) was different, ranging from 
1.66 keV -- 2.02 keV.
It should be noted that kT$_{in}$ and kT$_{BB}$ are affected by the spectral hardening factor (f$_{col}$; for more details see Davis et al. 2005). 
The effective temperature is given by T$_{eff}$ = kT$_{in}$ / f$_{col}$ and T$_{eff}$ was found to be in the range of 1.03--1.26 keV for 
f$_{col}$ = 1.6 (Davis et al. 2005). 
We did not find any significant variation in the spectral parameters between the final and last sections of each observation. 
There is a slight change in total flux in  B  and  D  sections (Table 3).
The spectral analysis of sections exhibiting HBOs (Table 5, Figure 7) suggest that the disk is close to the 
last stable orbit during the HB and we did not find any significant variation in the
 spectral parameters when the source varied from $\sim$ 25 Hz to $\sim$ 33 Hz HBOs.

We have also unfolded the spectra of various sections using a thermal comptonization model viz. 
{\it nthComp + Gaussian + power-law)} (Agrawal et al. 2020). In the {\it nthComp model}, soft seed photons are 
assumed to be a black body emission (Zdziarski et al. 1996). Based on the F-test probability in all the sections, 
we find that the power-law model is necessary for the data (Table 4 \& 6). Unfolded spectra for various sections 
are shown in Figure 6 and 8. For the fits, we fixed the power-law index, $\Gamma$ at 3.0 in order to constrain the other parameters. 
The reason to fix $\Gamma$ came from the study of Agrawal et al. (2020), as in most of the branches, spectra exhibit $\Gamma$ around 3. 
We noted that even if we vary $\Gamma$ from 2.9 to 3.1, there is no noticeable variation in other spectral parameters. 
We found that for eg. section  A, $\Gamma_{nthcomp}$ varied slightly from 1.80 to 1.94 with in 90\% confidence level. Similar small 
changes were observed in kT$_{e}$ for sections A and  D. Total fluxes were found to be slightly varying in subsections i.e. a and b of 
sections B  and D (Table 4).

\begin{table*}[htb]
\begin{minipage}[t]{\columnwidth}
\renewcommand{\footnotesize}{\tiny}
\tiny
\caption{Best-fit spectral parameters for sections which exhibited lags using
 DiskBB + Gaussian + bbody + Power-law model. The subscript BB represents the bbody model and dBB represents DiskBB model.
The flux in units of 10$^{-8}$ ergs cm$^{-2}$ s$^{-1}$ is calculated in the energy band 3--50 keV. Errors are quoted at a 90\% confidence level.
Luminosity is in units of 10$^{38}$ erg s$^{-1}$ assuming the distance 13 kpc for GX 17+2. Mass accretion rate is in units of 10$^{18}$ g s$^{-1}$.} 
\centering
\begin{tabular}{ccccccccccccccccccccccccccccccc}
\hline
Parameter && A &&&& B &&&& C  \\
          & a && b && a && b&& a && b  \\
\hline
$k$T$_{in}$ (keV) \footnote{Temperature of the DiskBB model.} &1.74$\pm$0.16 && 2.02$\pm$0.15 && 1.87$\pm$0.18 && 1.78$\pm$0.13 && 1.66$\pm$0.15 && 1.74$\pm$0.15 \\

N$_{dBB}$ \footnote{Normalization of the DiskBB model.} & 50.10$\pm$23.2 && 32.71$\pm$9.42 && 40.32$\pm$16.12 && 58.56$\pm$19.6 && 71.34$\pm$31.12 && 63.51$\pm$24.31 \\
R$_{in}$ (i=$28^{\circ})$\footnote{Inner disk radii from DiskBB normalization}&9.79 km &&7.91 km &&8.78 km&& 10.59 km&&11.68 km&&11.02 km\\

R$_{eff}$(i=$28^{\circ})$\footnote{Effective radius obtained by using DiskBB Normalizaton and the spectral corrections 1.18--1.64.}&11.55 -- 16.06 km && 9.34 -- 12.97 km&&10.37 -- 14.41 km &&12.49--17.36 km&& 13.79--19.16 km&&13.01--18.08 km\\
kT$_{BB}$ (keV) \footnote{Temperature of the BB model.} &2.84$\pm$0.07 && 3.04$\pm$0.10 && 2.84$\pm$0.10 && 2.80$\pm$0.08 && 2.68$\pm$0.07 && 2.73$\pm$0.08 \\

N$_{BB}$ \footnote{Normalization of the BB model.} & 0.083$\pm$0.006 && 0.074$\pm$0.006 && 0.069$\pm$0.007 && 0.075$\pm$0.006 && 0.080$\pm$0.007 && 0.075$\pm$0.007 \\

$\Gamma$$_{pl}$ \footnote{Power-law index.} & 2.93$\pm$0.11 && 3.42$\pm$0.22 && 3.01$\pm$0.10 && 3.11$\pm$0.14 && 2.89$\pm$0.12 && 2.98$\pm$0.13 \\

N$_{pl}$ \footnote{Normalization of the PL model.} & 6.80$\pm$2.17 && 12.02$\pm$3.81 && 7.65$\pm$2.17 && 7.01$\pm$2.62 && 6.07$\pm$2.36 && 6.20$\pm$2.40 \\

$E_{Fe}$ (keV) \footnote{Line Energy of the Gaussian model for Iron line.} & 6.7 && 6.7 && 6.7 && 6.7&&6.7 && 6.7\\

$\sigma_{Fe}$ (keV)\footnote{Line width of the Gaussian model for Iron line.} & 0.91$\pm$0.40 && 0.86$\pm$0.48 && 0.56$\pm$0.53 && 0.75$\pm$0.47 && 0.84$\pm$0.45 && 0.84$\pm$0.52\\
 
$N_{Fe}$ \footnote{Normalization of the Gaussian model for Iron line.} & 0.021$\pm$0.011 && 0.015$\pm$0.008 && 0.014$\pm$0.008 && 0.016$\pm$0.008 && 0.020$\pm$0.011 && 0.018$\pm$0.010 \\

DiskBB flux & 0.50$\pm$0.06&&  0.66$\pm$0.09&& 0.56$\pm$0.08&& 0.65$\pm$0.06&&0.55$\pm$0.06&&0.63$\pm$0.07\\
bbody flux & 0.66$\pm$0.06&&0.59$\pm$0.07 && 0.55$\pm$0.06&& 0.60$\pm$0.06 &&0.64$\pm$0.05 &&0.59$\pm$0.06\\	
Powerlaw flux& 0.36$\pm$0.06&& 0.25$\pm$0.06&&0.36$\pm$0.03&&0.26$\pm$0.05&&0.35$\pm$0.06&&0.30$\pm$0.07 \\
Total flux& 1.55$\pm$0.01 && 1.53$\pm$0.02 && 1.48$\pm$0.01 && 1.53$\pm$0.01 && 1.57$\pm$0.01 && 1.55$\pm$0.01 \\
L$_{3-50}$ & 3.12 $\pm$ 0.02 && 3.08 $\pm$0.04 && 2.98$\pm$0.02 && 3.08 $\pm$ 0.02 && 3.16$\pm$0.02 && 3.12 $\pm$0.02 \\
$\dot{m}$& 1.67 $\pm$ 0.01 && 1.65$\pm$0.02 && 1.60$\pm$0.01 && 1.65 $\pm$0.01 && 1.69$\pm$0.01 && 1.67$\pm$0.01 \\
$\chi^{2}$/dof& 29.72/37 && 47.54/37 && 28.31/37 &&  31.96/37 && 38.02/37 && 34.98/37 \\
$\chi^{2}$/dof (without power-law) & 93.69/39 && 112.21/39 && 76.06/39 &&  67.96/39 && 87.22/39 && 211.60/39 \\
Lag observed && {\bf 422 $\pm$ 85 s} &&&& {\bf 369 $\pm$ 30 s} &&&& {\bf 306 $\pm$ 70 s} \\
H$_{c}$ \footnote{ Coronal height estimated from lag }&& {\bf 98 -- 489 km} &&&& {\bf 78 -- 392 km} &&&& {\bf 58 -- 292 km} \\

F-test(Gaussian) prob. & 1.39$\times$10$^{-6}$ && 1.37$\times$10$^{-3}$ && 1.02$\times$10$^{-5}$ && 2.16$\times$10$^{-5}$ && 1.57$\times$10$^{-5}$  && 4.346$\times$10$^{-5}$ \\
F-test(powerlaw) prob. & 5.95$\times$10$^{-10}$ && 1.25$\times$10$^{-7}$ && 1.14$\times$10$^{-8}$ && 8.68$\times$10$^{-7}$ && 2.13$\times$10$^{-7}$ && 3.45$\times$10$^{-15}$ \\

\hline\hline

Parameter && D  \\
          & a && b \\
\hline
$k$T$_{in}$ (keV) &1.66$\pm$0.12&&1.80$\pm$0.09\\
N$_{dBB}$ &90.04$\pm$33.12&&72.15$\pm$15.23\\
R$_{in}$ (i=$28^{\circ})$&13.12 km&&11.75 km\\
$R_{eff}$(i=$28^{\circ}$)&15.49--21.52 km&& 13.87--19.27 km\\
$k$T$_{BB}$ (keV) &2.60$\pm$0.07&&2.79$\pm$0.08\\
N$_{BB}$ &0.072$\pm$0.007&&0.063$\pm$0.005\\
$\Gamma$$_{pl}$ &2.88$\pm$0.21&&3.24$\pm$0.20\\
N$_{pl}$ &3.96$\pm$2.47&&3.06$\pm$1.95\\
$E_{Fe}$ (keV) &6.7 &&6.7  \\
$\sigma_{Fe}$ (keV) &0.78$\pm$0.63&&0.78$\pm$0.54\\ 
$N_{Fe}$ &0.017$\pm$0.011&&0.014$\pm$0.009 \\
DiskBB flux & 0.70$\pm$0.07&&  0.83$\pm$0.11 \\
bbody flux &	 0.57$\pm$0.01&&0.50$\pm$0.04 \\	
Powerlaw flux & 0.23$\pm$0.08&& 0.26$\pm$0.04 \\
Total flux & 1.52$\pm$ 0.01&& 1.46$\pm$0.01\\
L$_{3-50}$& 3.06$\pm$0.02 && 2.94$\pm$0.02\\
$\dot{m}$& 1.64$\pm$0.01 && 1.57$\pm$0.01\\
$\chi^{2}$/dof& 30.31/37 && 32.98/37 \\
$\chi^{2}$/dof (without power-law) & 70.30/39 && 64.97/39 \\
Lag observed && {\bf -571 $\pm$ 37 s} \\
H$_{c}$  && {\bf 104 -- 519 km} \\

F-test(Gaussian) prob. & 1.28$\times$10$^{-6}$ && 2.17$\times$10$^{-3}$ \\
F-test(power-law) prob. & 1.74$\times$10$^{-7}$ && 3.56$\times$10$^{-6}$ \\

\hline
\end{tabular}
\end{minipage}
\end{table*}

\begin{table*}[htb]
\begin{minipage}[t]{\columnwidth}
\scriptsize
\caption{Best-fit spectral parameters for sections exhibiting lags using
  nthComp + Gaussian + Power-law model. 
The flux in units of 10$^{-8}$ ergs cm$^{-2}$ s$^{-1}$ is calculated in the energy band 3--50 keV. Errors are quoted at a 90\% confidence level.
Luminosity is in units of 10$^{38}$ erg s$^{-1}.$ assuming the distance 13 kpc for GX 17+2. Mass accretion rate is in units of 10$^{18}$ g s$^{-1}$.} 
\begin{tabular}{ccccccccccccccccccccccccccccccc}
\hline
Parameter && A &&&& B &&&& C  \\
          & a && b && a && b&& a && b  \\
\hline

$\Gamma$$_{pl}$\footnote{Powerlaw index.} &  3.0  && 3.0 && 3.0 && 3.0 && 3.0 && 3.0 \\

N$_{pl}$\footnote{Normalization of the PL model.} &  7.27$\pm$1.02 && 6.52$\pm$0.95 && 6.45$\pm$1.02 && 3.42$\pm$0.99 && 7.73$\pm$0.90 && 5.59$\pm$0.91 \\

$E_{Fe}$ (keV) \footnote{Line Energy of the Gaussian model for Iron line.} &  6.7 && 6.7 && 6.7 && 6.7 && 6.7 && 6.7 \\

$\sigma_{Fe}$ (keV)\footnote{Line width of the Gaussian model for Iron line.} & 0.77$\pm$0.30 && 0.95$\pm$0.29  && 0.80$\pm$0.32 && 0.76$\pm$0.29 && 0.67$\pm$0.31 && 0.81$\pm$0.35 \\

$N_{Fe}$ \footnote{Normalization of the Gaussian model for Iron line.} &  0.019$\pm$0.007 && 0.022$\pm$0.006  && 0.018$\pm$0.007 && 0.019$\pm$0.006 && 0.018$\pm$0.007 &&  0.20$\pm$0.007 \\

$\Gamma$$_{Nth}$ \footnote{nthComp power-law index.} &  1.80$\pm$0.04 && 1.94$\pm$0.04  && 1.88$\pm$0.04 && 1.98$\pm$0.03 && 1.87$\pm$0.04 && 1.95$\pm$0.04 \\

$k$T$_{e}$ (keV)\footnote{Electron temperature ( nthComp ).} &  3.22$\pm$0.07 && 3.45$\pm$0.07  && 3.19$\pm$0.08 && 3.25$\pm$0.07 && 3.07$\pm$0.07 && 3.13$\pm$0.07\\

$k$T$_{bb}$ (keV)\footnote{Seed photon temperature ( nthComp )}&0.64$\pm$0.09 && 0.42$\pm$0.07 && 0.66$\pm$0.08 && 0.60$\pm$0.06 && 0.69$\pm$0.07 && 0.66$\pm$0.06\\



N$_{Nth}$\footnote{Normalization of the nthComp model.} &  0.64$\pm$0.12 && 0.72$\pm$0.18  && 0.68$\pm$0.16 && 1.09$\pm$0.22 && 0.64$\pm$0.13 && 0.82$\pm$0.15\\

nthComp flux        & 1.19$\pm$0.05 &&1.17$\pm$0.04 && 1.16$\pm$0.04 &&1.35$\pm$0.04 &&1.18$\pm$0.04 &&1.27$\pm$0.04 \\
Powerlaw flux       & 0.34$\pm$0.05 &&0.28$\pm$0.05 && 0.30$\pm$0.05 &&0.16$\pm$0.05 &&0.36$\pm$0.04 &&0.26$\pm$0.04 \\
Total flux          & 1.55$\pm$0.01 &&1.52$\pm$0.01 && 1.48$\pm$0.01 &&1.53$\pm$0.01 &&1.56$\pm$0.01 &&1.55$\pm$0.01 \\
L$_{3-50}$ & 3.12 $\pm$ 0.02 && 3.06 $\pm$0.02 && 2.98$\pm$0.02 && 3.08 $\pm$ 0.02 && 3.14$\pm$0.02 && 3.12 $\pm$0.02 \\
$\dot{m}$& 1.67 $\pm$ 0.01 && 1.64$\pm$0.01 && 1.60$\pm$0.01 && 1.65 $\pm$0.01 && 1.68$\pm$0.01 && 1.67$\pm$0.01 \\
$\chi^{2}$/dof      & 35.64/38 && 44.47/38 && 26.18/38 && 30.04/38 && 42.52/38 && 34.40/38  \\
$\chi^{2}$/dof (without power-law) & 139.46/39 && 48.04/39 && 107.35/39 && 57.43/39  && 188.0/39 && 114.27/39  \\

F-test(Gaussian) prob. & 9.91$\times$10$^{-7}$ && 1.23$\times$10$^{-6}$ && 1.79$\times$10$^{-7}$ && 1.16$\times$10$^{-7}$ && 1.17$\times$10$^{-5}$  && 1.95$\times$10$^{-6}$ \\
F-test(powerlaw) prob. & 8.16$\times$10$^{-13}$&& 8.87$\times$10$^{-2}$ && 3.33$\times$10$^{-13}$ && 8.15$\times$10$^{-7}$  && 7.87$\times$10$^{-14}$ && 1.88$\times$10$^{-11}$ \\

\hline\hline

Parameter && D  \\
          & a && b \\
\hline

$\Gamma$$_{pl}$    &  3.0  && 3.0  \\

N$_{pl}$           & 5.20$\pm$0.77  && 5.36$\pm$0.86 \\

$E_{Fe}$ (keV)     &  6.7 && 6.7  \\

$\sigma_{Fe}$ (keV) &  0.74$\pm$0.37 && 0.97$\pm$0.21   \\

$N_{Fe}$            &  0.019$\pm$0.008 && 0.027$\pm$0.006  \\

$\Gamma$$_{Nthcomp}$    &  2.04$\pm$0.04 && 2.15$\pm$0.02 \\

$k$T$_{e}$ (keV)    &  2.98$\pm$0.07 && 3.21$\pm$0.04 \\

$k$T$_{bb}$ (keV)   &  0.70$\pm$0.05 && 0.59$\pm$0.04 \\



N$_{Nthcomp}$           &  0.82$\pm$0.12 && 0.92$\pm$0.22 \\

nthComp flux        & 1.26$\pm$0.03 &&1.21$\pm$0.04 \\
Powerlaw flux       & 0.24$\pm$0.04 &&0.28$\pm$0.05  \\
Total flux          & 1.52$\pm$0.01 &&1.46$\pm$0.01  \\
L$_{3-50}$& 3.06$\pm$0.02 && 2.94$\pm$0.02\\
$\dot{m}$& 1.64$\pm$0.01 && 1.57$\pm$0.01\\
$\chi^{2}$/dof      & 35.67/38 && 31.50/38   \\ 
$\chi^{2}$/dof (without power-law) & 131.06/39 && 101.30/39 \\

F-test(Gaussian) prob. & 2.67$\times$10$^{-6}$ && 8.78$\times$10$^{-9}$ \\
F-test(Powerlaw) prob. & 2.72$\times$10$^{-12}$ && 3.52$\times$10$^{-11}$ \\

\hline\hline

\end{tabular}
\end{minipage}
\end{table*}

\begin{table*}[htb]
\begin{minipage}[t]{\columnwidth}
\scriptsize
\caption{Best-fit spectral parameters for sections which exhibited HBOs using
 DiskBB +  Gaussian + bbody + Power-law model. The subscript BB represents the bbody model and dBB represents DiskBB model.
The flux in units of 10$^{-8}$ ergs cm$^{-2}$ s$^{-1}$ is calculated in the energy band 3--50 keV. Errors are quoted at a 90\% confidence level.
Luminosity is in units of 10$^{38}$ erg s$^{-1}.$ assuming the distance 13 kpc for GX 17+2. Mass accretion rate is in units of 10$^{18}$ g s$^{-1}$.} 
\begin{tabular}{ccccccccccccccccccccccccccccccc}
\topline
Parameter &For HBO 1 & For HBO 2 & For HBO 3  \\
\hline
$k$T$_{in}$ (keV)\footnote{Temperature of the DiskBB model.} & 2.10$\pm$0.21 & 1.92$\pm$0.14 & 1.87$\pm$0.13 \\
N$_{dBB}$\footnote{Normalization of the DiskBB model.} & 20.53$\pm$7.89 & 33.52$\pm$11.32 & 39.54$\pm$12.75  \\
$R_{eff}$(i=$28^{\circ})$\footnote{Effective radius obtained by using DiskBB Normalizaton and the spectral corrections 1.18--1.64.} & 7.39 -- 10.28 km & 9.45 -- 13.13 km & 10.26 -- 14.27 km \\
$k$T$_{BB}$ (keV) \footnote{Temperature of the BB model.} & 3.05$\pm$0.08 & 2.93$\pm$0.07 & 2.91$\pm$0.06\\
N$_{BB}$ \footnote{Normalization of the BB model.} & 0.077$\pm$0.007 & 0.077$\pm$0.005 & 0.077$\pm$0.005 \\
$\Gamma$$_{pl}$\footnote{Powerlaw index.} &  2.96$\pm$0.29 & 2.97$\pm$0.31 & 3.09$\pm$0.08  \\
N$_{pl}$ \footnote{Normalization of the PL model.}& 5.99$\pm$4.01 & 4.33$\pm$2.12 & 8.53$\pm$2.13  \\
$E_{Fe}$ (keV) \footnote{Line Energy of the Gaussian model for Iron line.}&6.7 &6.7 &6.7 \\
$\sigma_{Fe}$ (keV) \footnote{Line width of the Gaussian model for Iron line.}& 0.86$\pm$0.29 & 0.82$\pm$0.40 & 0.88$\pm$0.39 \\
$N_{Fe}$\footnote{Normalization of the Gaussian model for Iron line.}& 0.02$\pm$0.007 & 0.019$\pm$0.008 & 0.019$\pm$0.009 \\

DiskBB flux & 0.50$\pm$0.07 & 0.56$\pm$0.06& 0.56$\pm$0.06\\
bbody flux & 0.62$\pm$0.06 & 0.61$\pm$0.05 & 0.62$\pm$0.04 \\
Powerlaw flux & 0.39$\pm$0.04 & 0.33$\pm$0.04 & 0.33$\pm$0.04 \\
Total flux & 1.54$\pm$0.01 & 1.53$\pm$0.01 & 1.53$\pm$0.01\\
L$_{3-50}$ & 3.10 $\pm$ 0.02 & 3.08 $\pm$ 0.02 & 3.08$\pm$ 0.02\\
$\dot{m}$& 1.66 $\pm$ 0.01 & 1.65 $\pm$ 0.01 & 1.65 $\pm$ 0.01\\
$\chi^{2}$/dof& 47.44/37 & 42.60/37 & 41.77/37 \\
$\chi^{2}$/dof (without Gaussian) & 79.12/39 & 71.40/39 & 73.59/39 \\
$\chi^{2}$/dof (without power-law) & 155.14/39 & 118.78/39 &  324.97/39 \\
F-test(Gaussian) prob.& 7.76$\times$10$^{-5}$ & 7.09$\times$10$^{-5}$ & 2.81$\times$10$^{-5}$ \\
F-test(Powerlaw) prob.& 3.02$\times$10$^{-10}$ & 5.77$\times$10$^{-9}$ & 3.28$\times$10$^{-17}$ \\

\hline\hline
\end{tabular}
\end{minipage}
\end{table*}

\begin{table*}[htb]
\begin{minipage}[t]{\columnwidth}
\scriptsize
\caption{Same model as used to obtain Table 4 and Best-fit spectral parameters for sections which exhibited HBOs using
 nthComp +  Gaussian  + Power-law model.
The flux in units of 10$^{-8}$ ergs cm$^{-2}$ s$^{-1}$ is calculated in the energy band 3--50 keV. Errors are quoted at a 90\% confidence level.
Luminosity is in units of 10$^{38}$ erg s$^{-1}.$ assuming the distance 13 kpc for GX 17+2. Mass accretion rate is in units of 10$^{18}$ g s$^{-1}$.} 
\begin{tabular}{ccccccccccccccccccccccccccccccc}

\topline

Parameter &For HBO 1 & For HBO 2 & For HBO 3  \\
\hline
$\Gamma$$_{pl}$\footnote{power-law index.} & 3.0 & 3.0 & 3.0 \\
N$_{pl}$ \footnote{Normalization of the PL model.}& 6.96$\pm$0.89 & 4.88$\pm$0.77 & 4.93$\pm$0.74  \\
$E_{Fe}$ (keV) \footnote{Line Energy of the Gaussian model for Iron line.}&6.7 &6.7 &6.7 \\
$\sigma_{Fe}$ (keV) \footnote{Line width of the Gaussian model for Iron line.}& 0.88$\pm$0.27 & 0.82$\pm$0.26 & 0.84$\pm$0.25 \\
$N_{Fe}$\footnote{Normalization of the Gaussian model for Iron line.}& 0.021$\pm$0.006 & 0.020$\pm$0.006& 0.020$\pm$0.004 \\
$\Gamma$$_{Nth}$\footnote{nthComp power-law index.} & 1.72$\pm$0.03 & 1.85$\pm$0.02& 1.86$\pm$0.02 \\
$k$T$_{e}$ (keV)\footnote{Electron temperature ( nthComp ). nthComp model.} & 3.31$\pm$0.06 & 3.30$\pm$0.05 & 3.28$\pm$0.05 \\
$k$T$_{bb}$ (keV)\footnote{Seed photon temperature ( nthComp).} &0.55$\pm$0.11 & 0.53$\pm$0.09 & 0.54$\pm$0.08 \\
N$_{Nth}$ \footnote{Normalization of the nthComp model.} & 0.70$\pm$0.23 & 1.03$\pm$0.3 & 1.03$\pm$0.3\\
nthComp flux &1.19$\pm$0.05 &1.28$\pm$0.04 &1.28$\pm$0.04 \\
Powerlaw flux & 0.32$\pm$0.05 & 0.22$\pm$0.05 & 0.23$\pm$0.05 \\
Total flux & 1.54$\pm$0.01 & 1.53$\pm$0.01 & 1.53$\pm$0.01\\
L$_{3-50}$ & 3.10 $\pm$ 0.02 & 3.08 $\pm$ 0.02 & 3.08$\pm$ 0.02\\
$\dot{m}$& 1.66 $\pm$ 0.01 & 1.65 $\pm$ 0.01 & 1.65 $\pm$ 0.01\\
$\chi^{2}$/dof& 47/38 & 43/38 & 44/38 \\

F-test(Gaussian) prob.& 3.14$\times$10$^{-6}$ & 1.19$\times$10$^{-6}$ & 6.74$\times$10$^{-7}$ \\
F-test(Powerlaw) prob.& 1.46$\times$10$^{-12}$ & 7.70$\times$10$^{-11}$ & 2.39$\times$10$^{-11}$ \\

\hline\hline
\end{tabular}
\end{minipage}
\end{table*}

\vspace{-2em}




\begin{figure}[!ht]
\caption{ The LAXPC 10 spectral fit for the the {\it DiskBB +  Gaussian + bbody + Powerlaw} model. 
The top panel gives the unfolded spectra (thick line) with the
component models (dashed lines) and the bottom panel gives the residuals obtained from the fit. Here in the top panel,
the light green colour line gives the Gaussian model component, cyan gives the Powerlaw component, red  gives the diskBB component and
dark blue gives the bbody component.}
\begin{subfigure}[b]{0.7\columnwidth}
\includegraphics[width=0.7\columnwidth, angle=270]{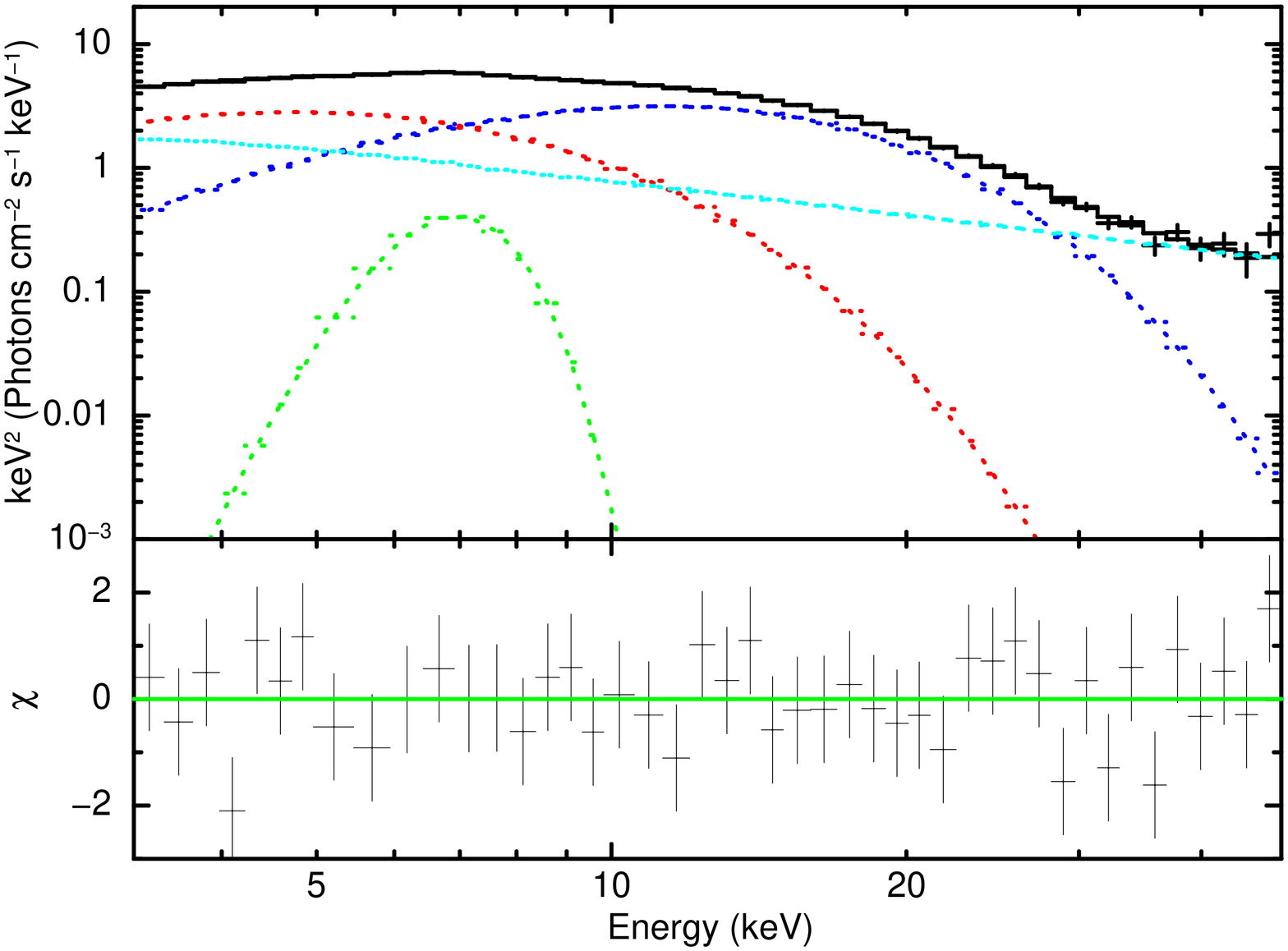}
\caption*{A(a)}
\end{subfigure}

\begin{subfigure}[b]{0.7\columnwidth}
\includegraphics[width=0.7\columnwidth, angle=270]{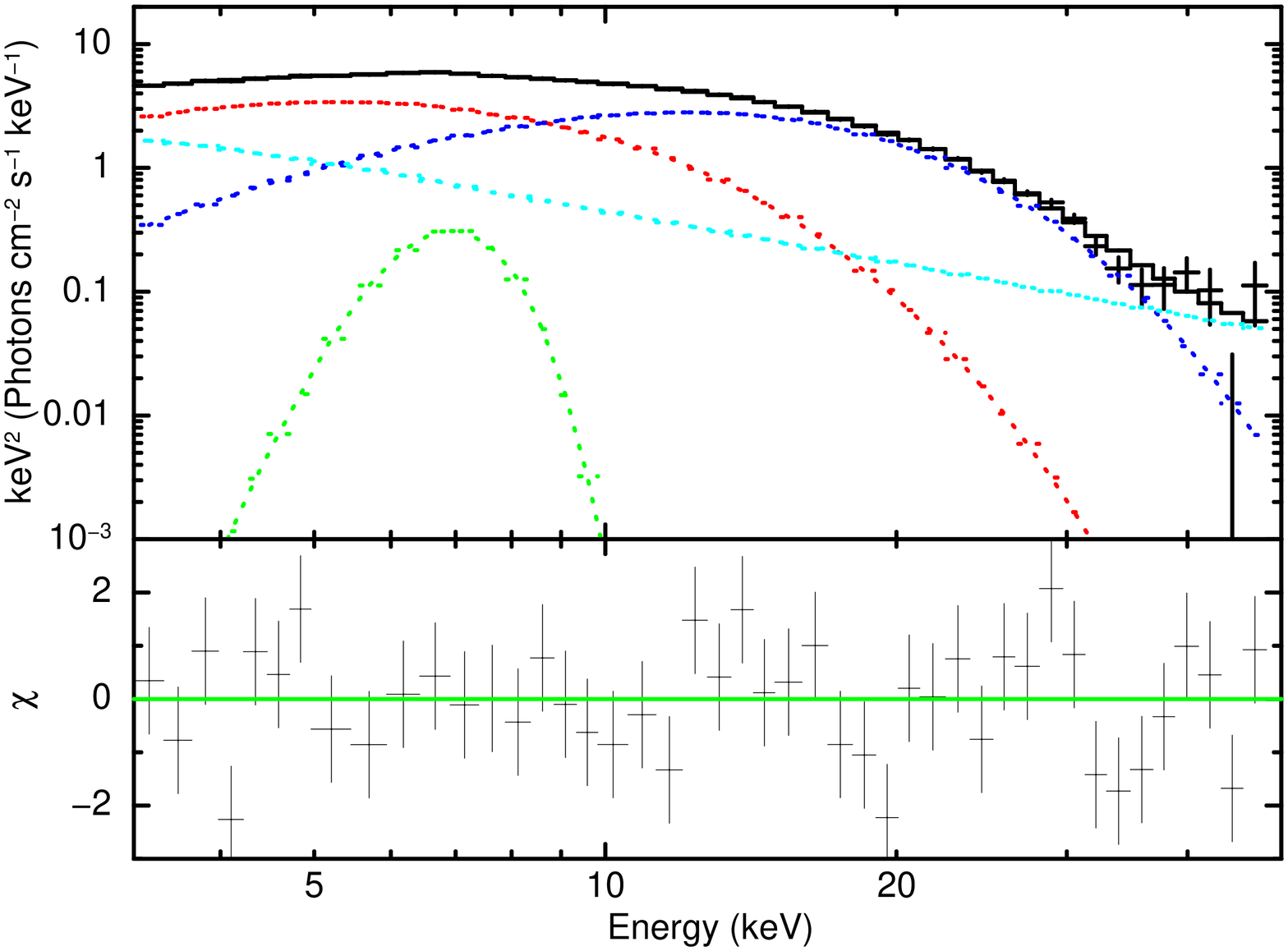}
\caption*{A(b)}
\end{subfigure}

\begin{subfigure}[b]{0.7\columnwidth}
\includegraphics[width=0.7\columnwidth, angle=270]{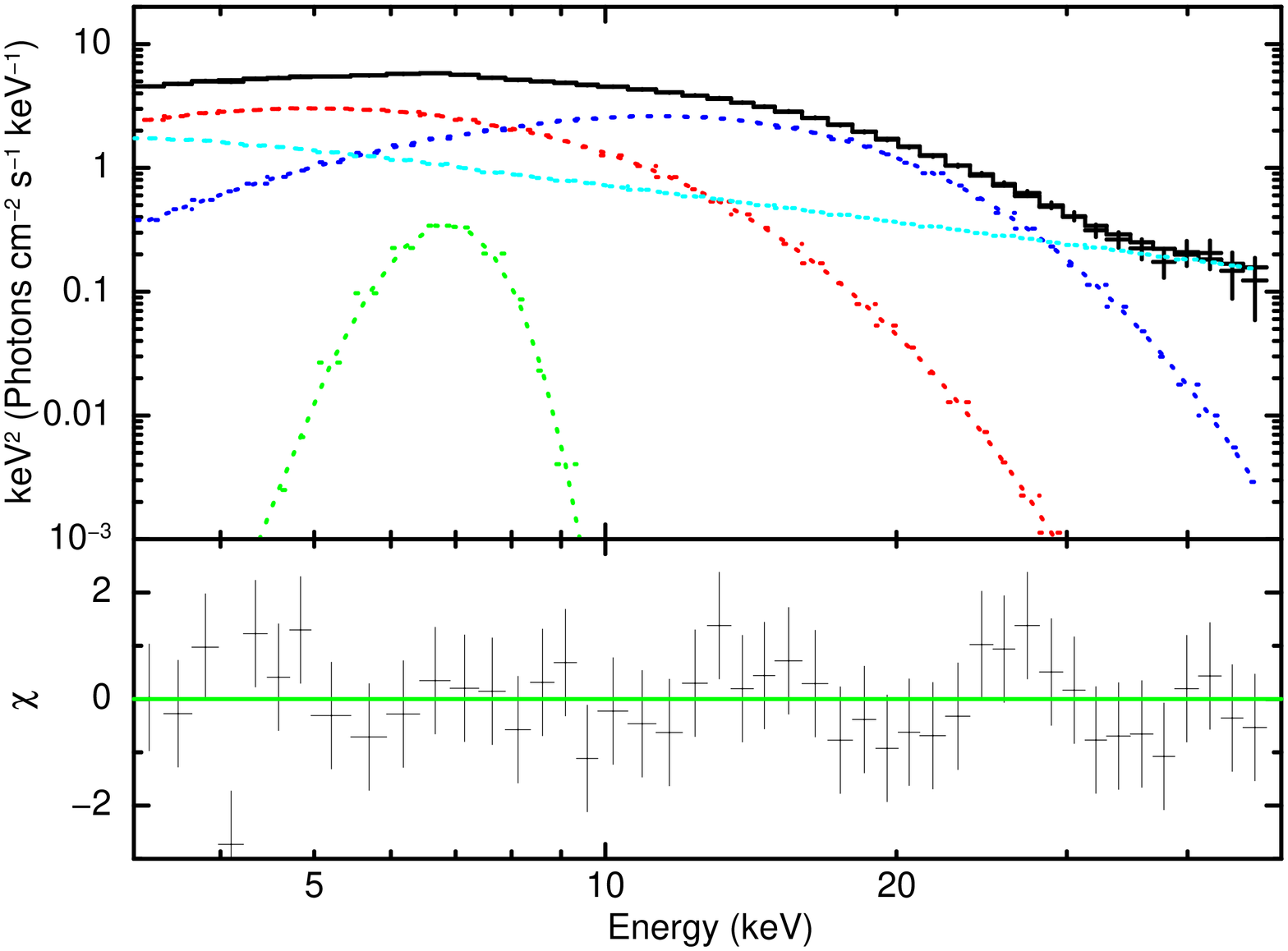}
\caption*{B(a)}
\end{subfigure}

\begin{subfigure}[b]{0.7\columnwidth}
\includegraphics[width=0.7\columnwidth, angle=270]{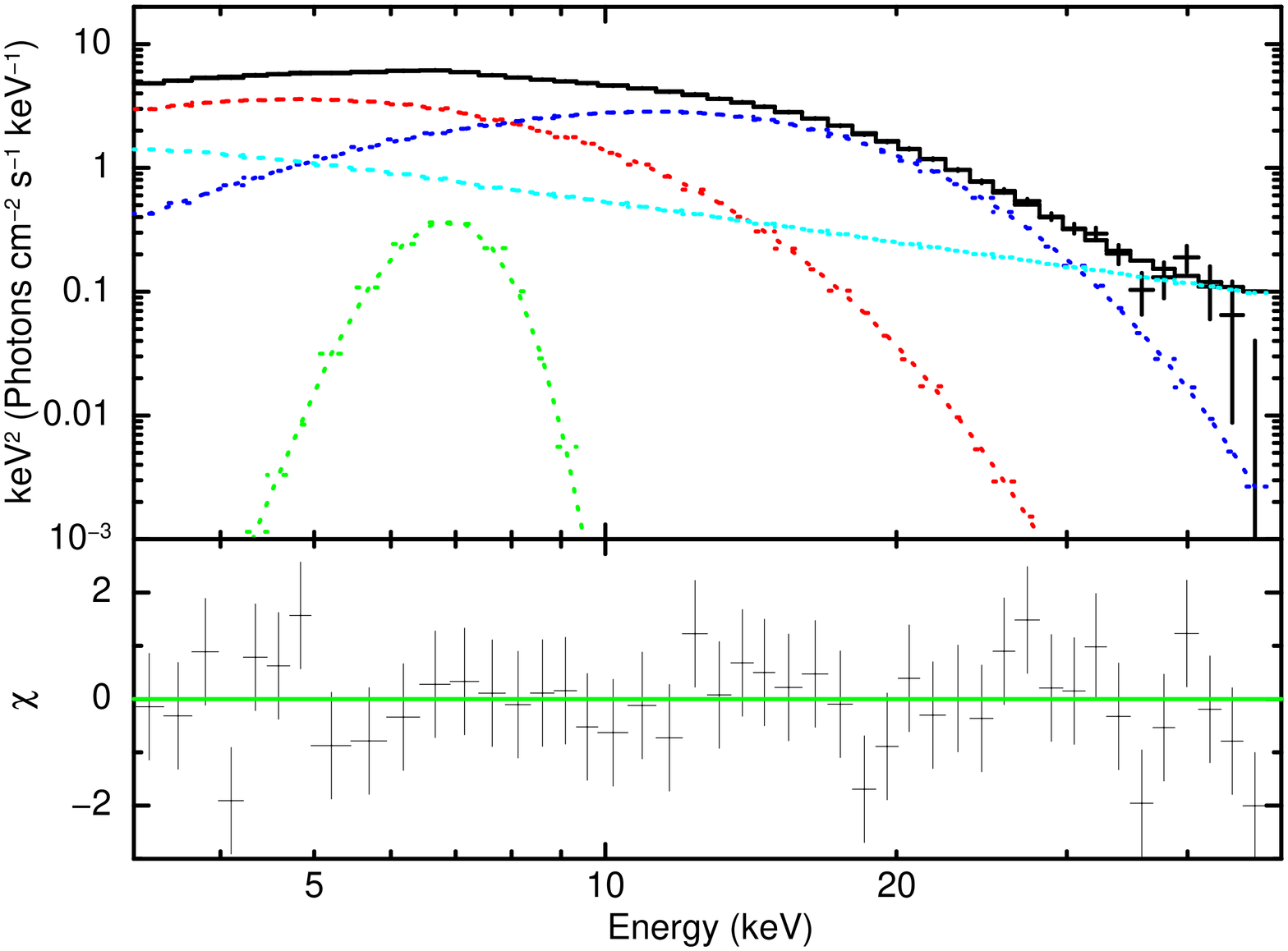}
\caption*{B(b)}
\end{subfigure}
\end{figure}
\begin{figure}[!ht]\ContinuedFloat
\begin{subfigure}[b]{0.7\columnwidth}
\includegraphics[width=0.7\columnwidth, angle=270]{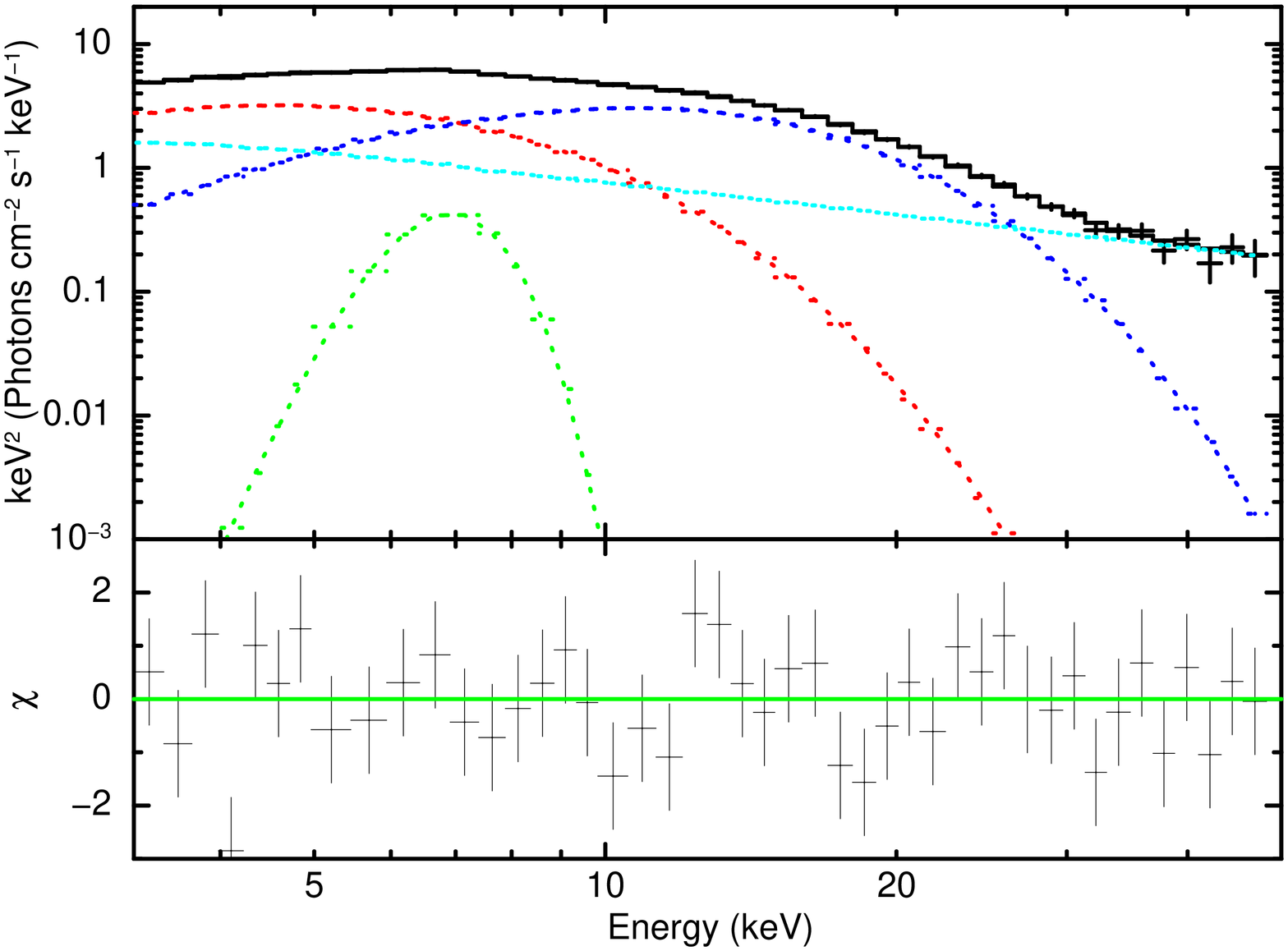}
\caption*{C(a)}
\end{subfigure}

\begin{subfigure}[b]{0.7\columnwidth}
\includegraphics[width=0.7\columnwidth, angle=270]{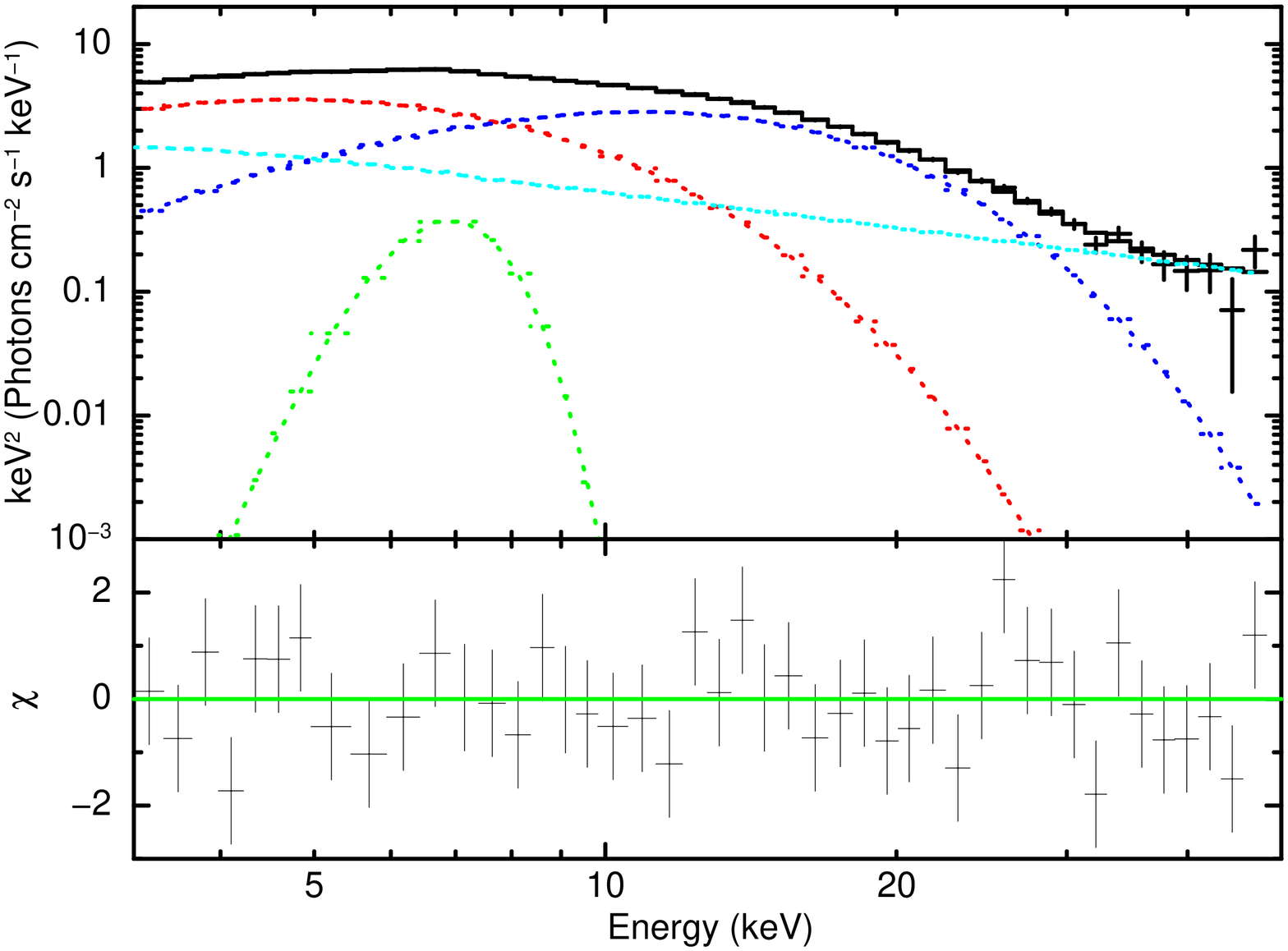}
\caption*{C(b)}
\end{subfigure}

\begin{subfigure}[b]{0.7\columnwidth}
\includegraphics[width=0.7\columnwidth, angle=270]{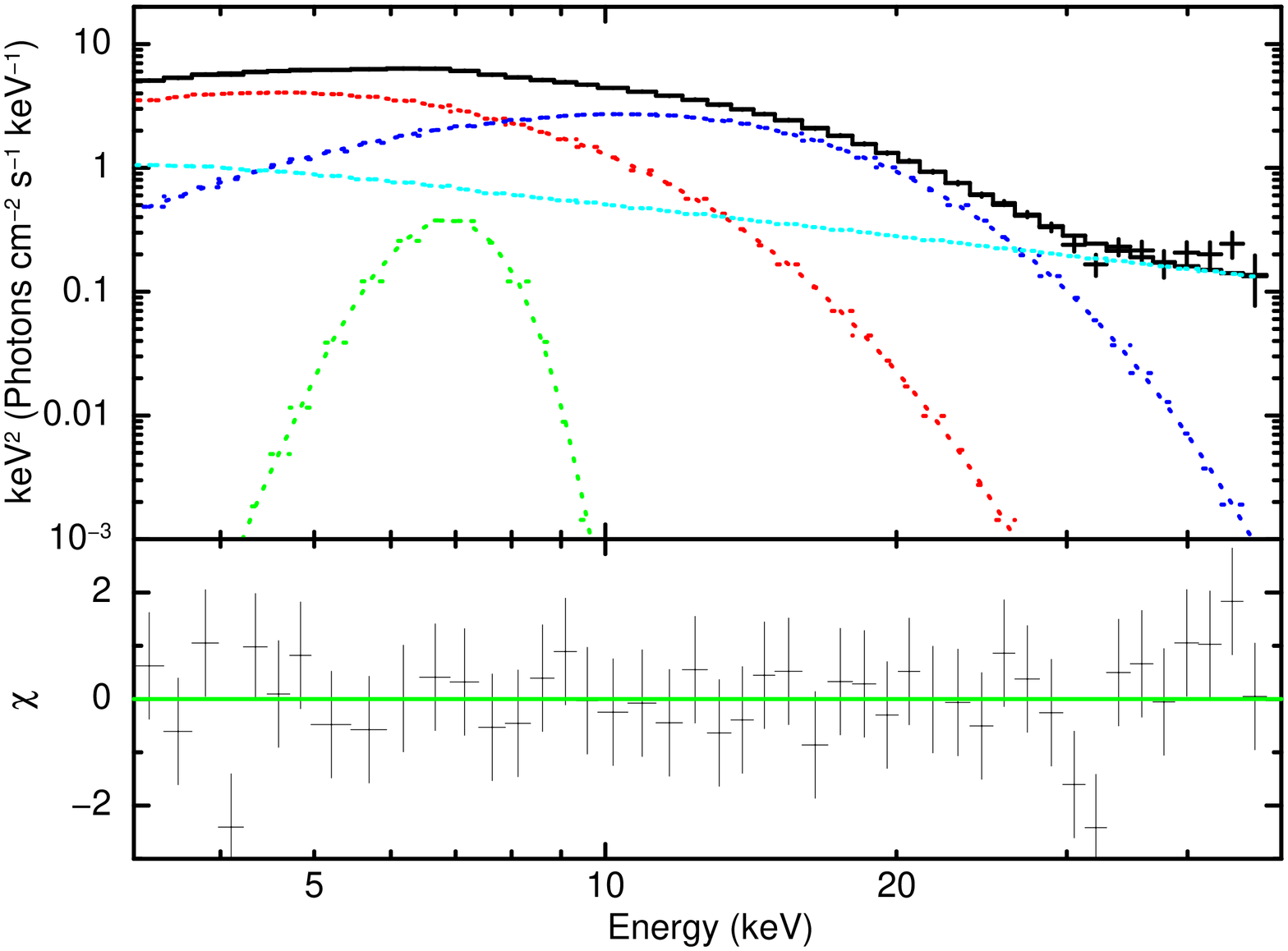}
\caption*{D(a)}
\end{subfigure}

\begin{subfigure}[b]{0.7\columnwidth}
\includegraphics[width=0.7\columnwidth, angle=270]{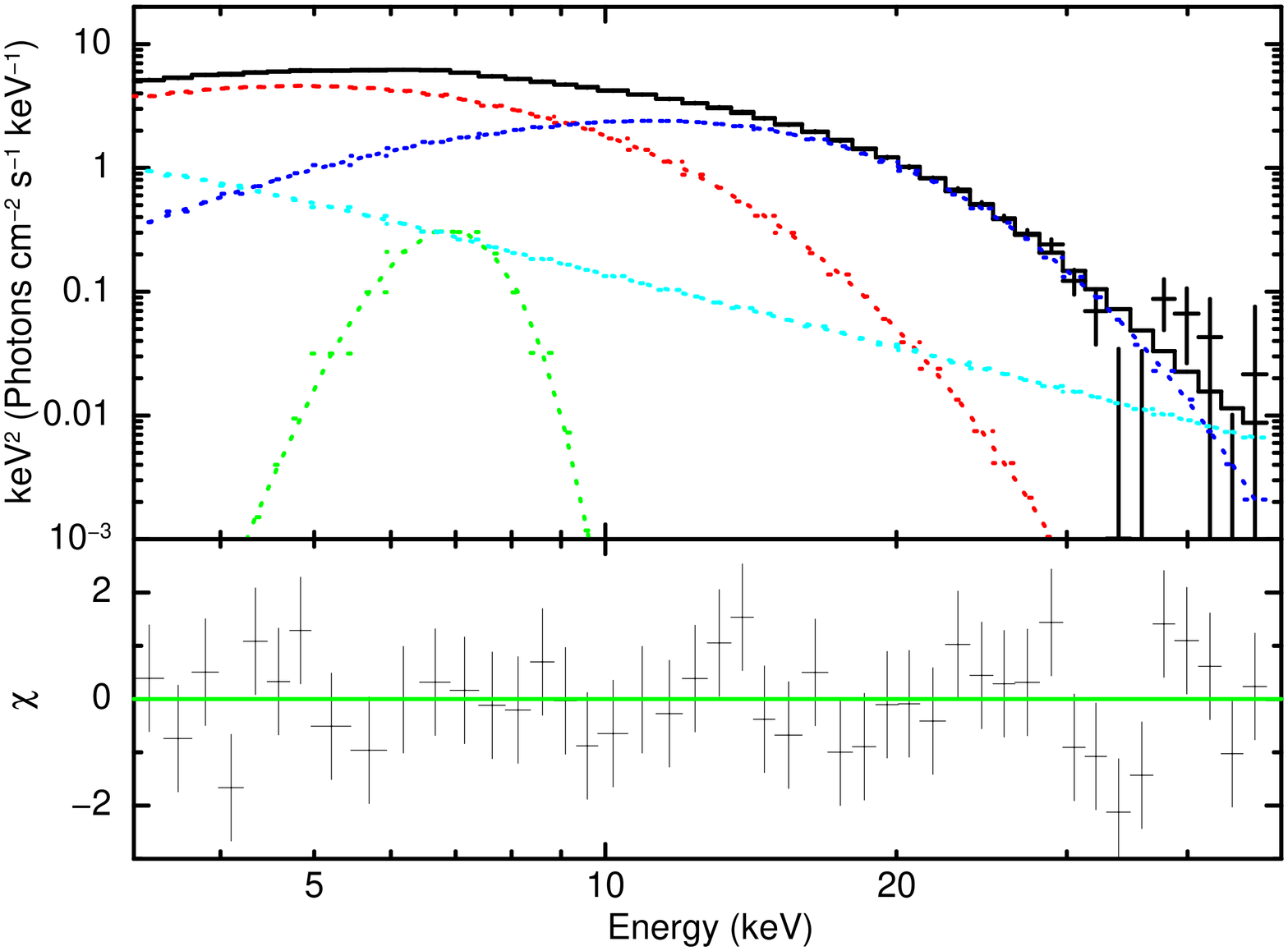}
\caption*{D(b)}
\end{subfigure}
\end{figure}

\begin{figure}[!ht]
\caption{ The LAXPC 10 spectral fit for the {\it Nthcomp +  Gaussian  + Power-law} model. 
The top panel gives the unfolded spectra (thick line) with the
component models (dashed lines) and the bottom panel gives the residuals obtained from the fit. Here in the top panel,
the light green colour line gives the Gaussian model component, red  gives the Power-law component and
dark blue gives the Nthcomp component.}
\begin{subfigure}[b]{0.7\columnwidth}
\includegraphics[width=0.7\columnwidth, angle=270]{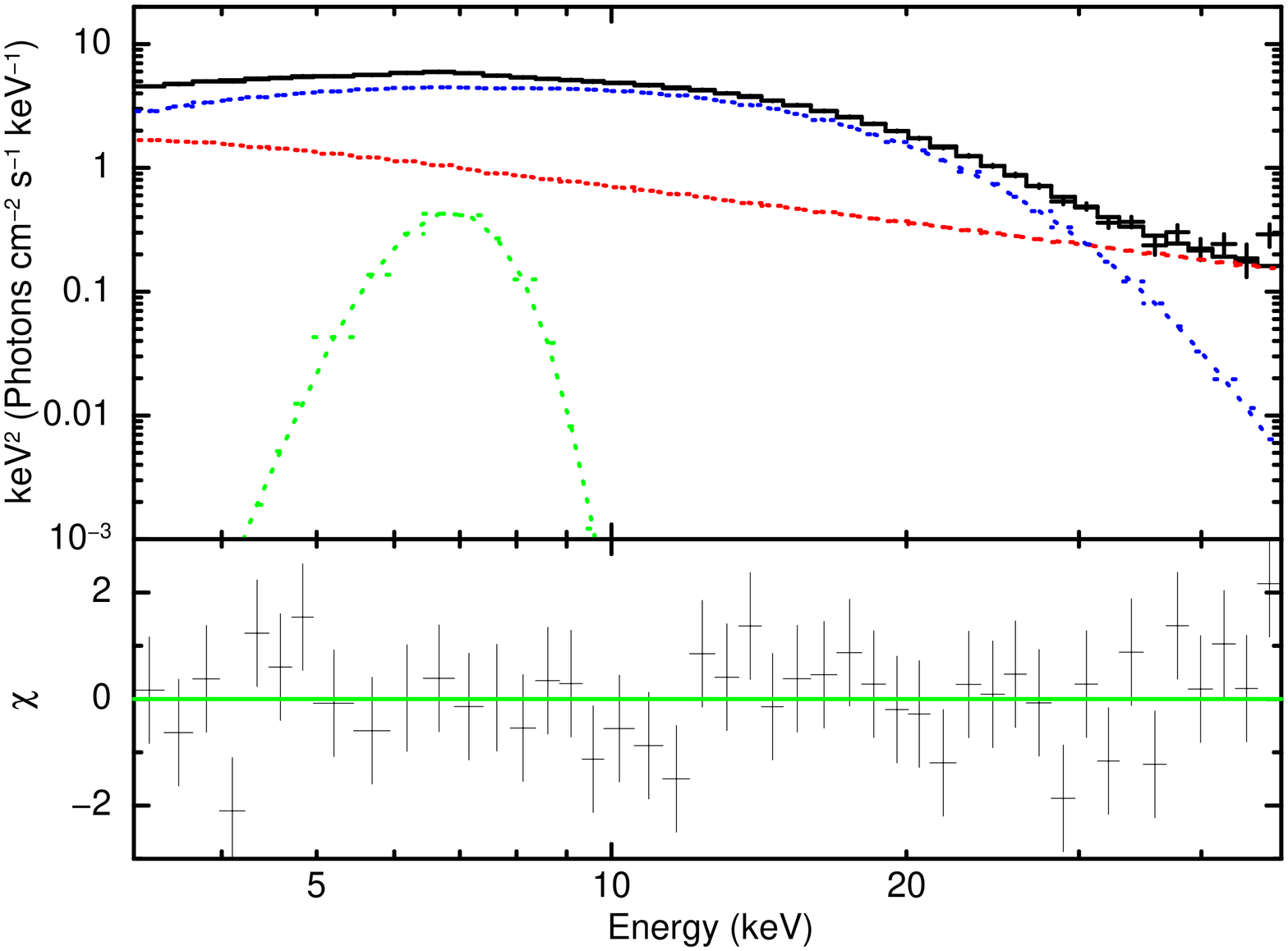}
\caption*{A(a)}
\end{subfigure}

\begin{subfigure}[b]{0.7\columnwidth}
\includegraphics[width=0.7\columnwidth, angle=270]{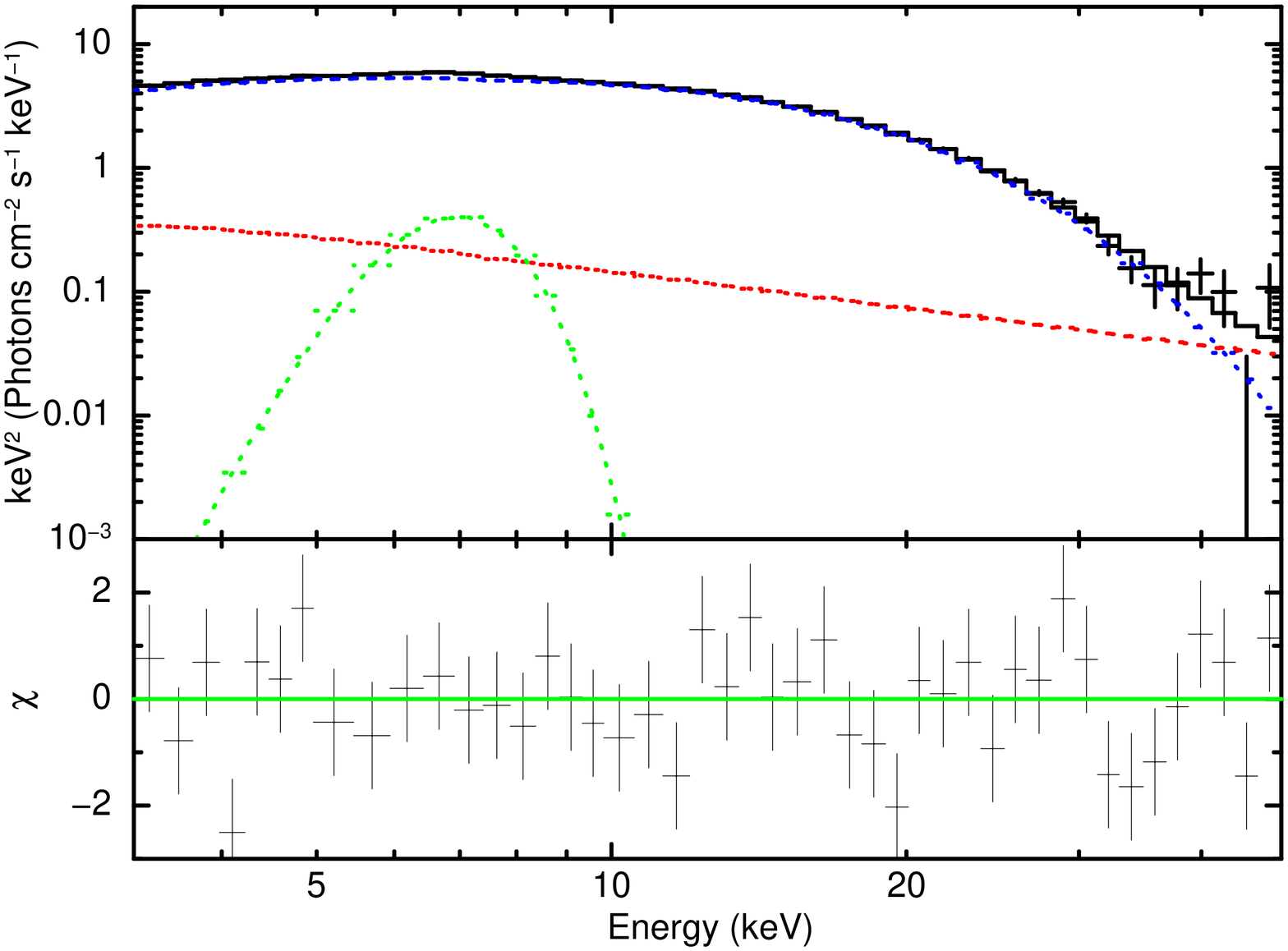}
\caption*{A(b)}
\end{subfigure}

\begin{subfigure}[b]{0.7\columnwidth}
\includegraphics[width=0.7\columnwidth, angle=270]{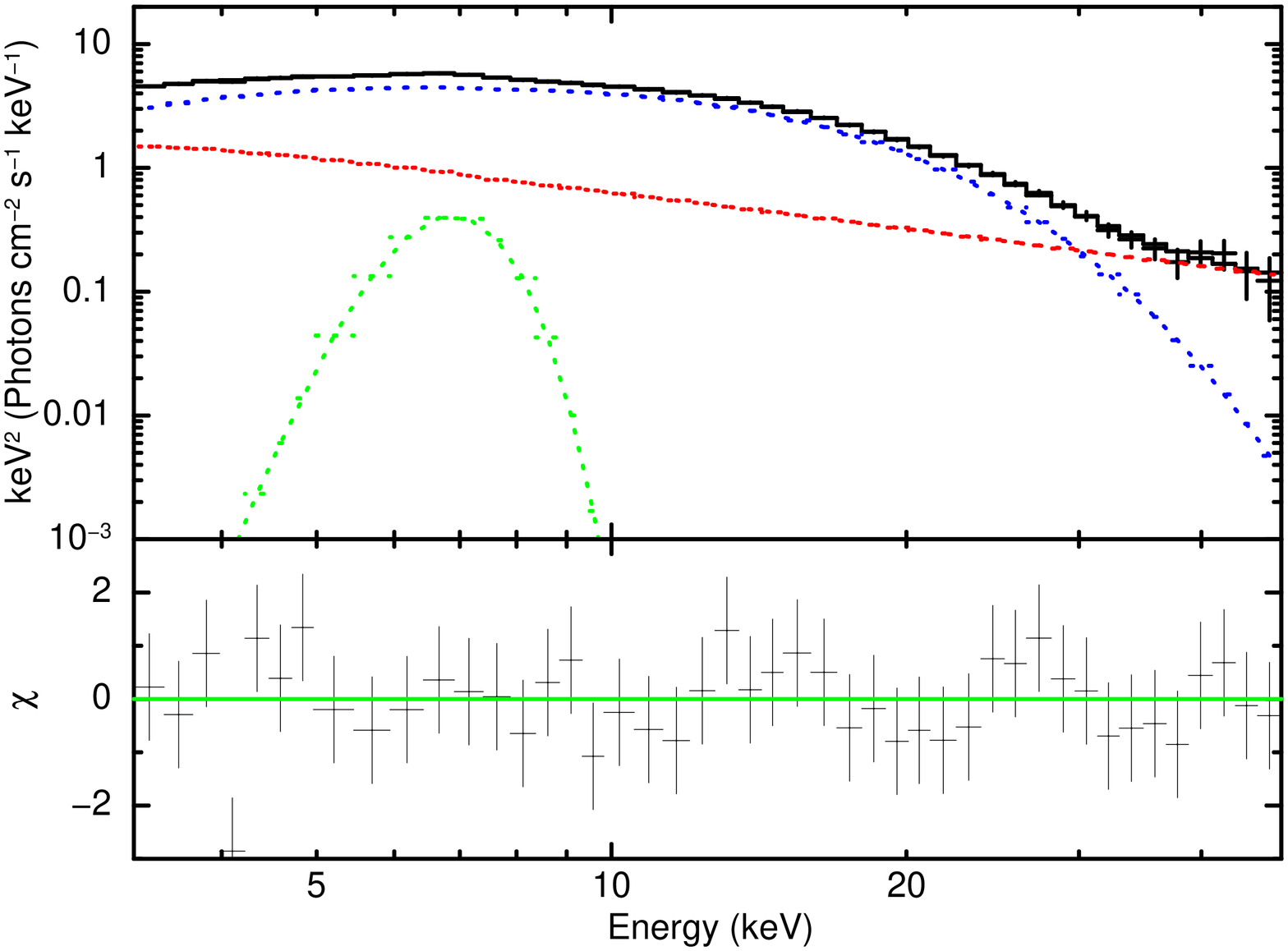}
\caption*{B(a)}
\end{subfigure}

\begin{subfigure}[b]{0.7\columnwidth}
\includegraphics[width=0.7\columnwidth, angle=270]{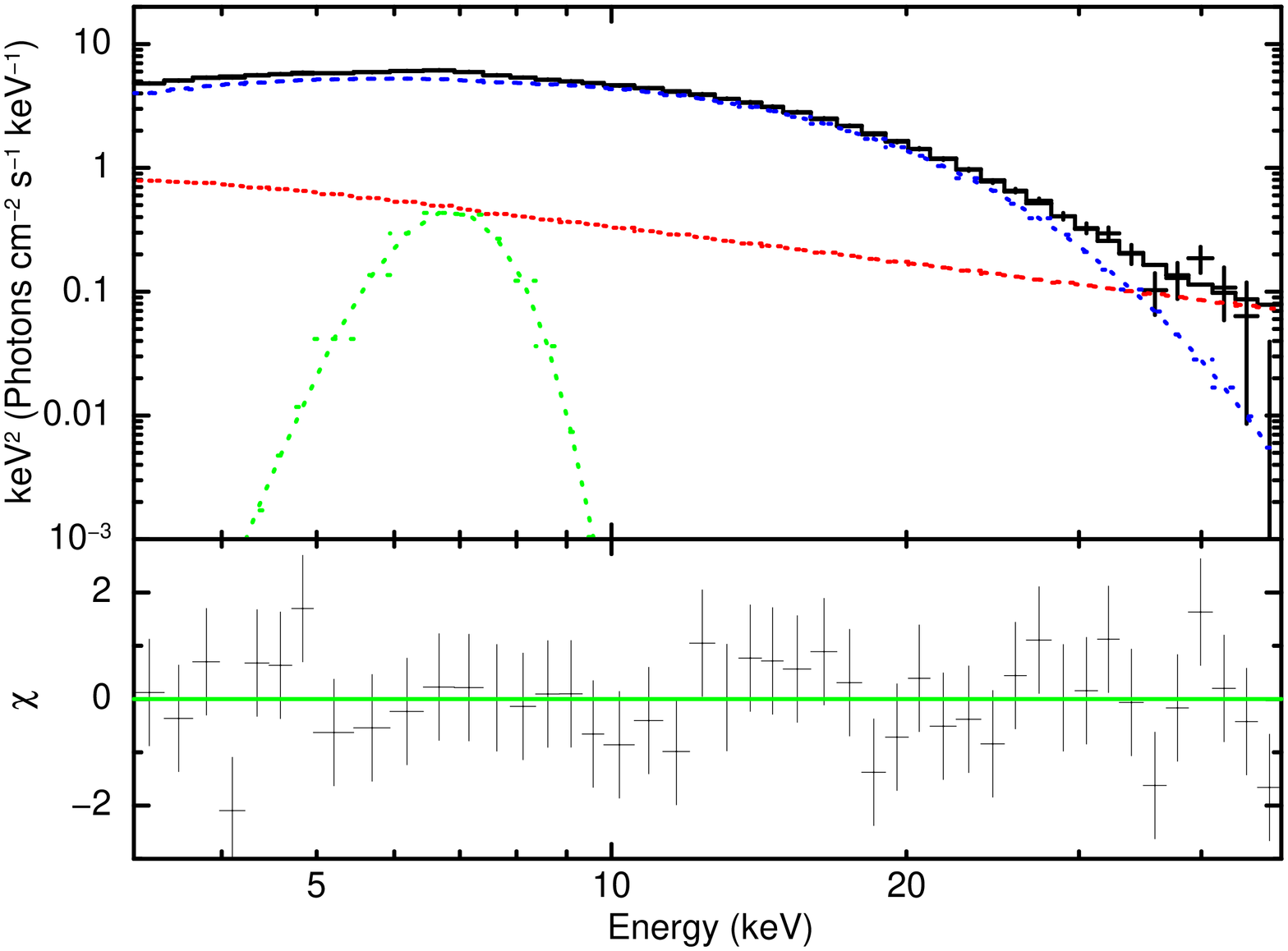}
\caption*{B(b)}
\end{subfigure}
\end{figure}
\begin{figure}[!ht]\ContinuedFloat
\begin{subfigure}[b]{0.7\columnwidth}
\includegraphics[width=0.7\columnwidth, angle=270]{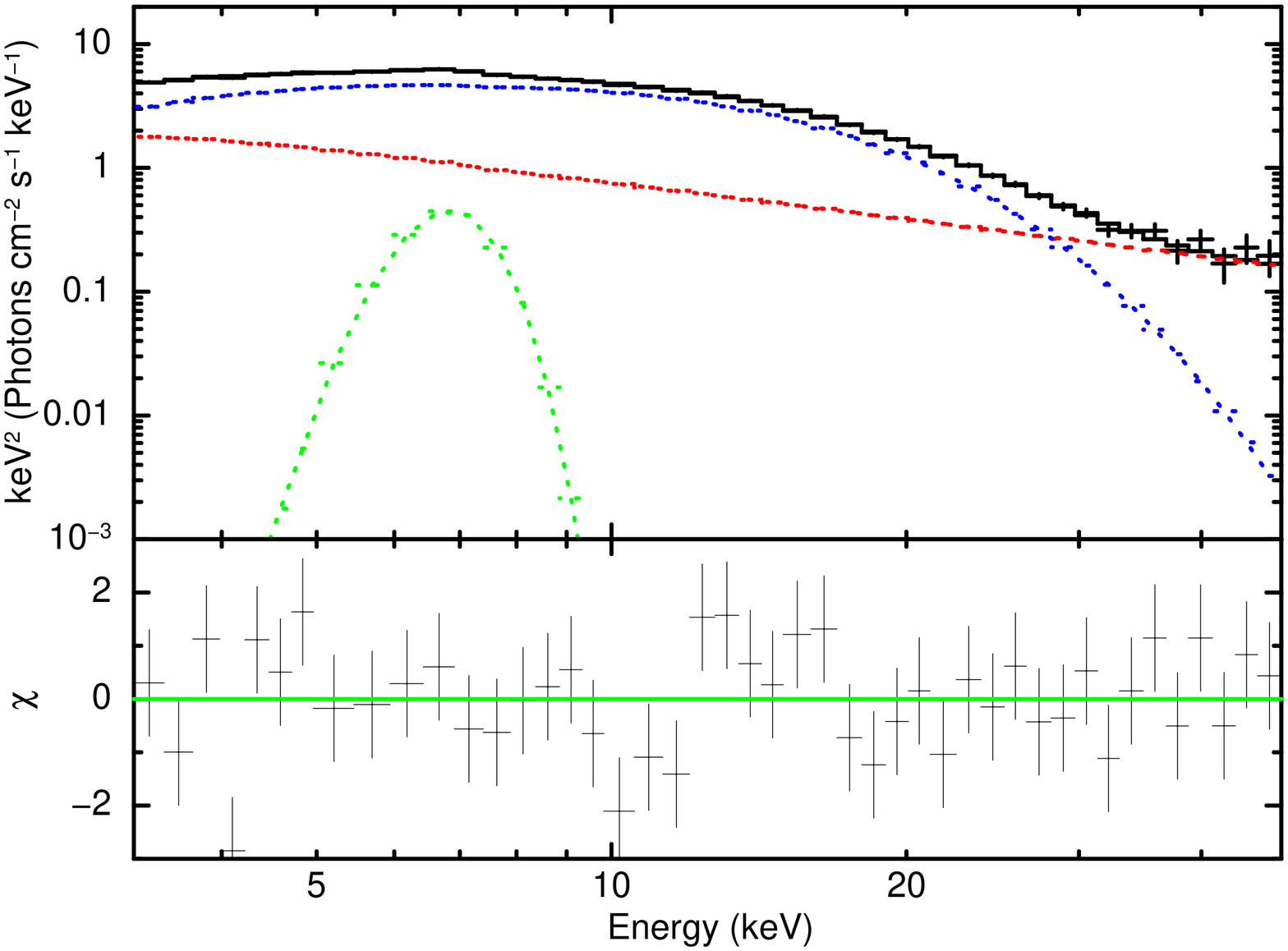}
\caption*{C(a)}
\end{subfigure}

\begin{subfigure}[b]{0.7\columnwidth}
\includegraphics[width=0.7\columnwidth, angle=270]{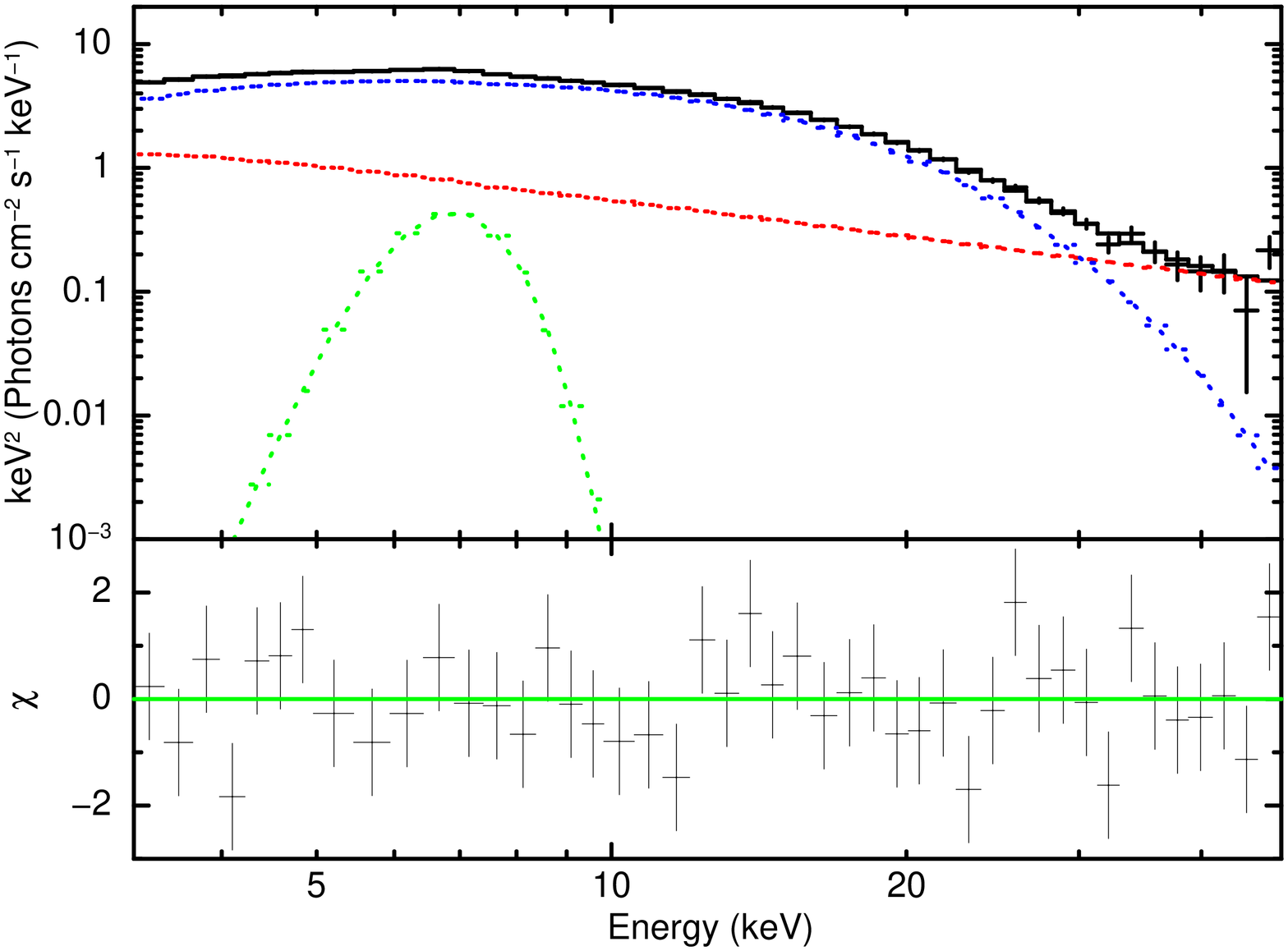}
\caption*{C(b)}
\end{subfigure}

\begin{subfigure}[b]{0.7\columnwidth}
\includegraphics[width=0.7\columnwidth, angle=270]{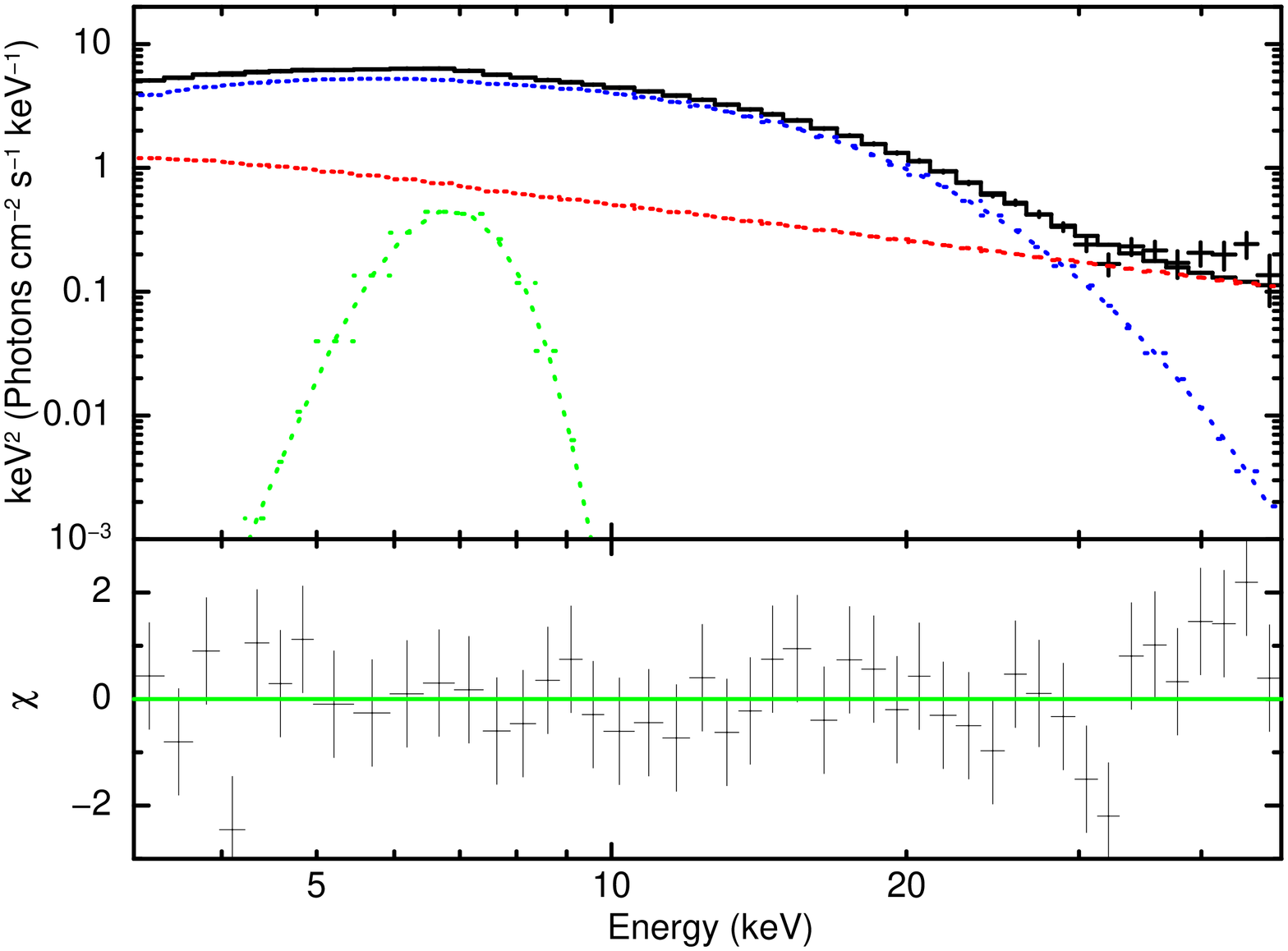}
\caption*{D(a)}
\end{subfigure}

\begin{subfigure}[b]{0.7\columnwidth}
\includegraphics[width=0.7\columnwidth, angle=270]{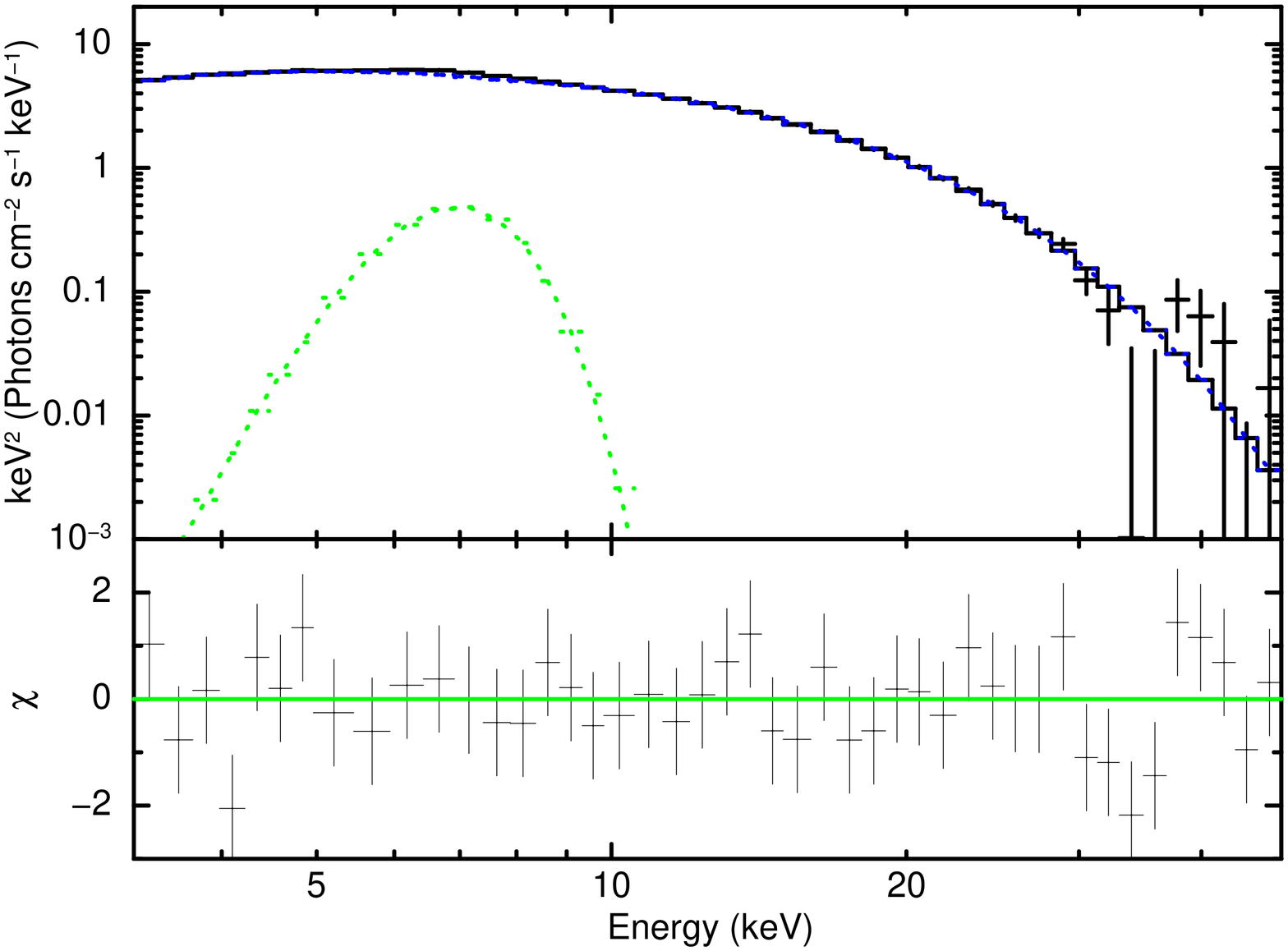}
\caption*{D(b)}
\end{subfigure}
\end{figure}

\begin{figure}[!ht]
\caption{ The LAXPC 10 spectral fit for the segments exhibiting HBOs using the {\it DiskBB +  Gaussian + bbody + Powerlaw} model. 
The top panel gives the unfolded spectra (thick line) with the
component models (dashed lines) and the bottom panel gives the residuals obtained from the fit. Here in the top panel,
the light green colour line gives the Gaussian model component, cyan gives the power-law component, red  gives the diskBB component and
dark blue gives the bbody component.}
\begin{subfigure}[b]{0.7\columnwidth}
\includegraphics[width=0.7\columnwidth, angle=270]{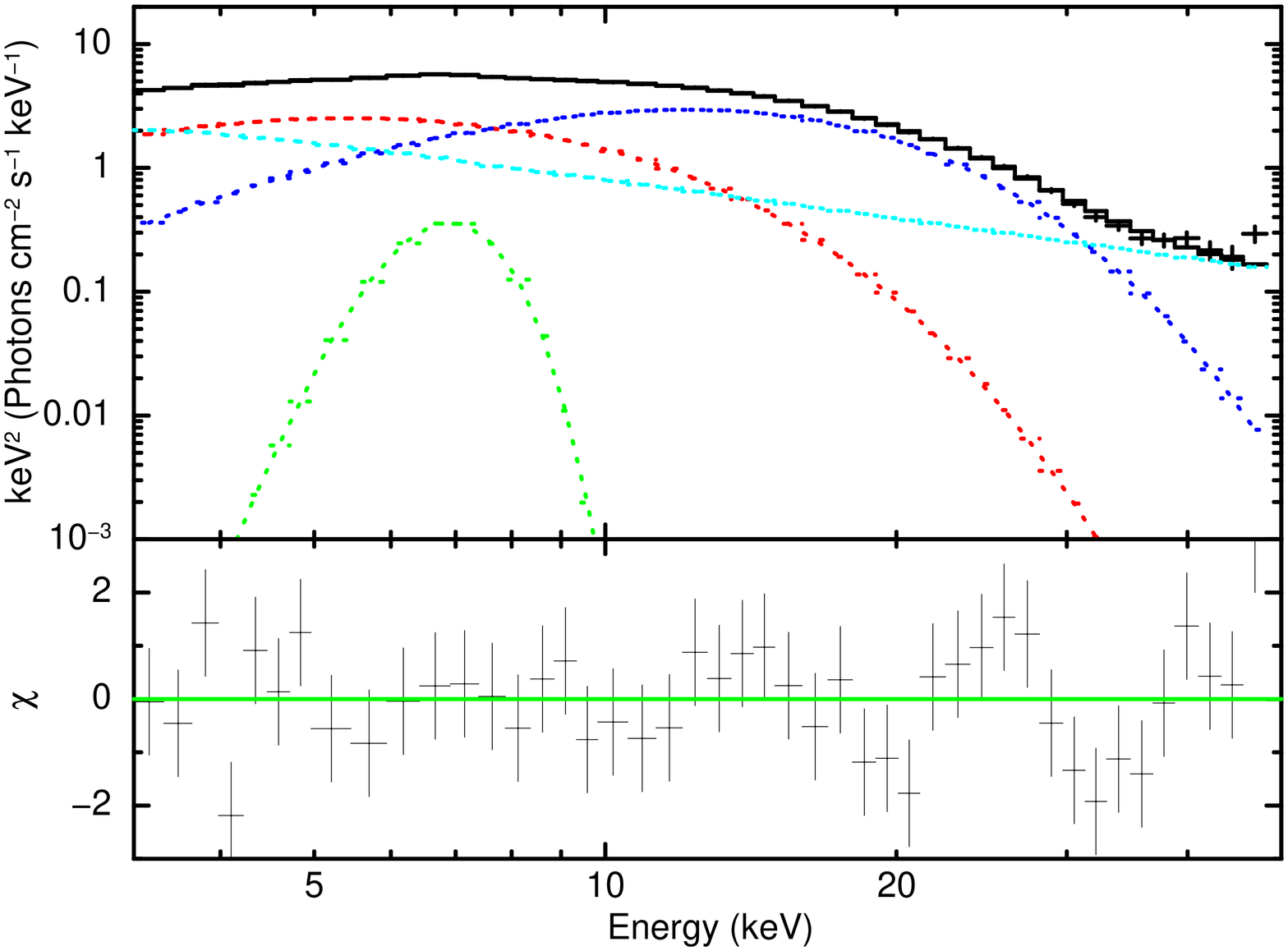}
\caption*{For HBO 1}
\end{subfigure}

\begin{subfigure}[b]{0.7\columnwidth}
\includegraphics[width=0.7\columnwidth, angle=270]{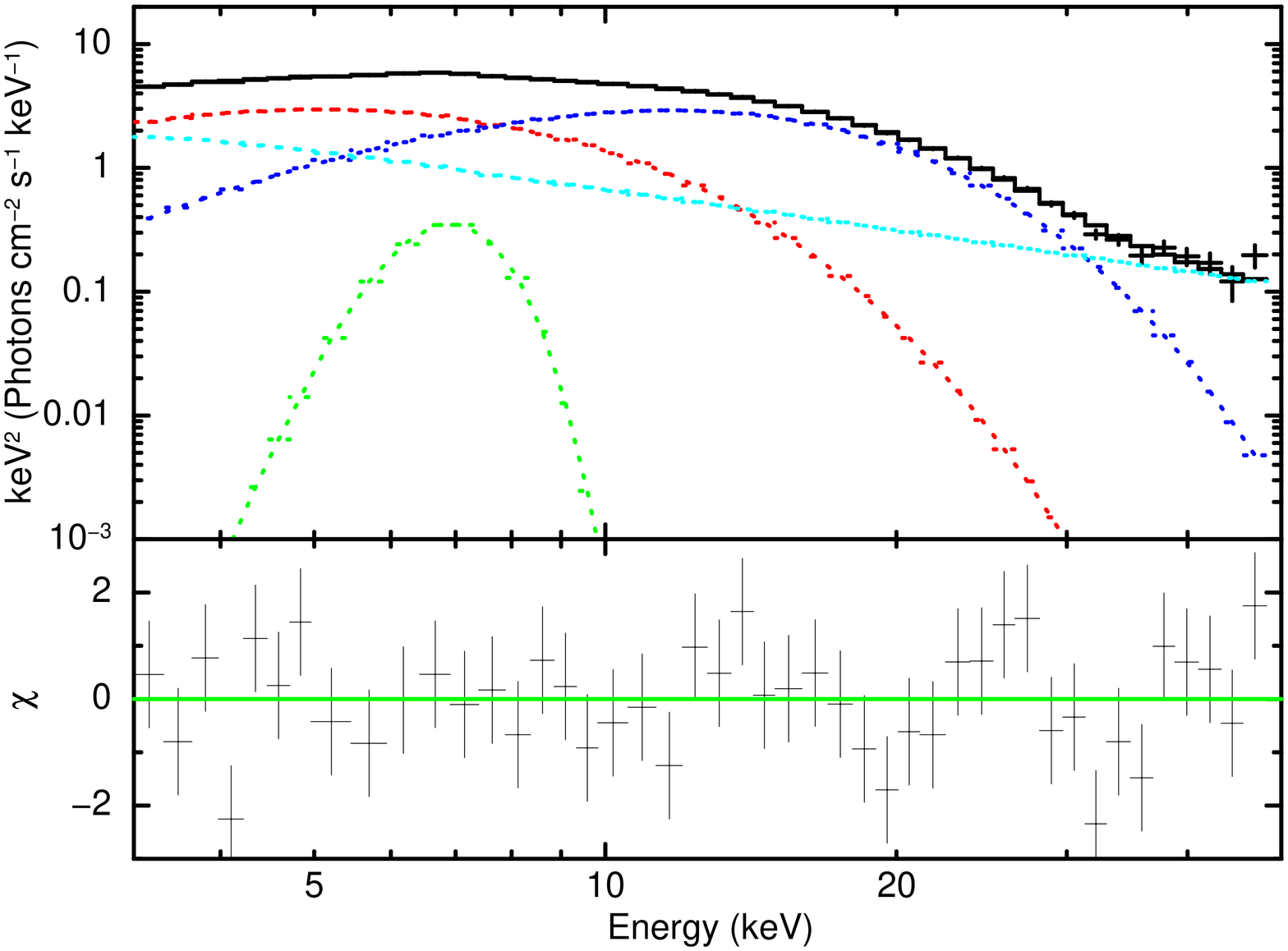}
\caption*{For HBO 2}
\end{subfigure}

\begin{subfigure}[b]{0.7\columnwidth}
\includegraphics[width=0.7\columnwidth, angle=270]{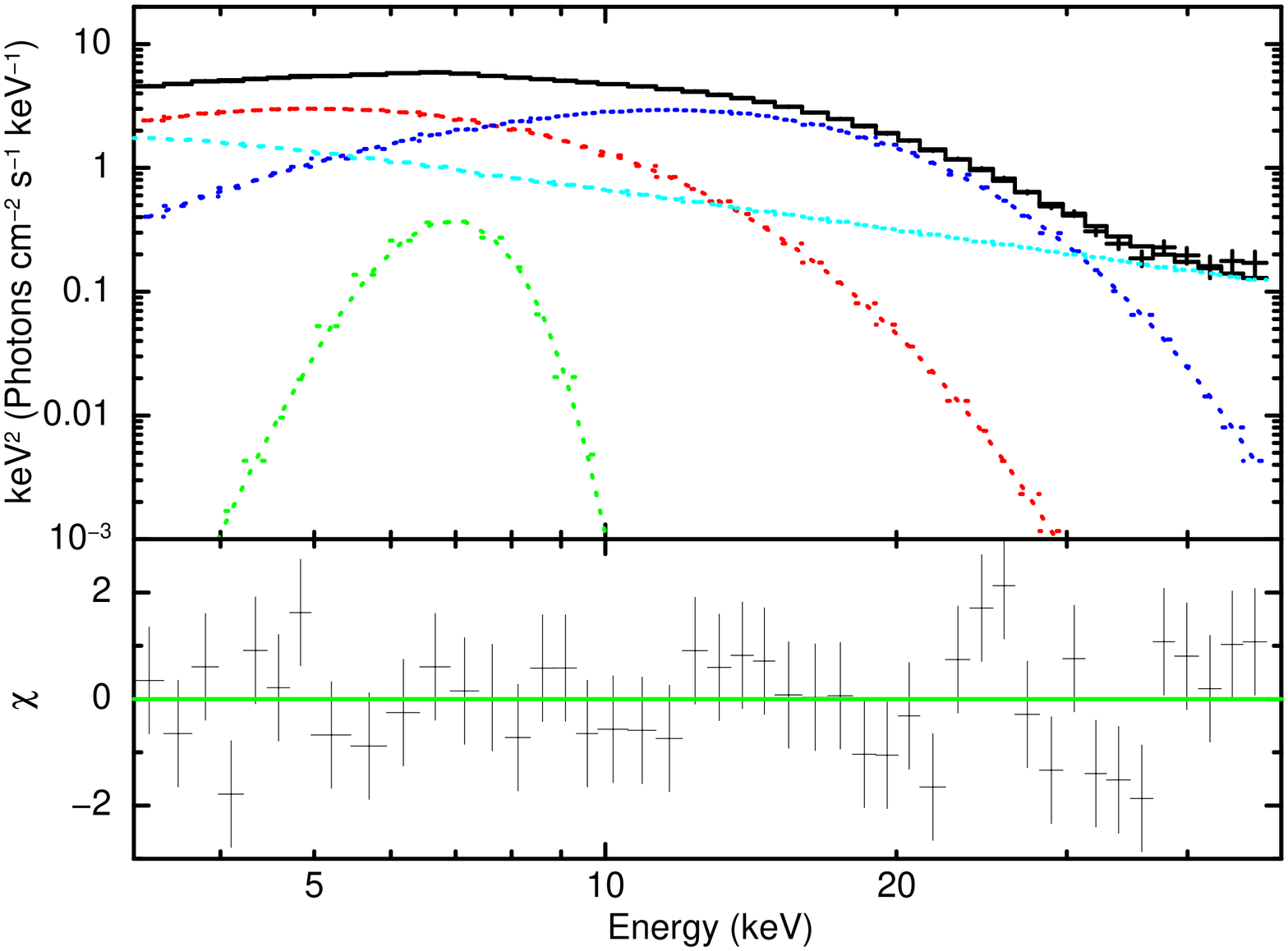}
\caption*{For HBO 3}
\end{subfigure}
\end{figure}

\begin{figure}[!ht]
\caption{ The LAXPC 10 spectral fit for the segments exhibiting HBOs using the {\it Nthcomp +  Gaussian + Powerlaw} model. 
The top panel gives the unfolded spectra (thick line) with the
component models (dashed lines) and the bottom panel gives the residuals obtained from the fit. Here in the top panel,
the light green colour line gives the Gaussian model component, red  gives the Powerlaw component and
dark blue gives the Nthcomp component.}
\begin{subfigure}[b]{0.7\columnwidth}
\includegraphics[width=0.7\columnwidth, angle=270]{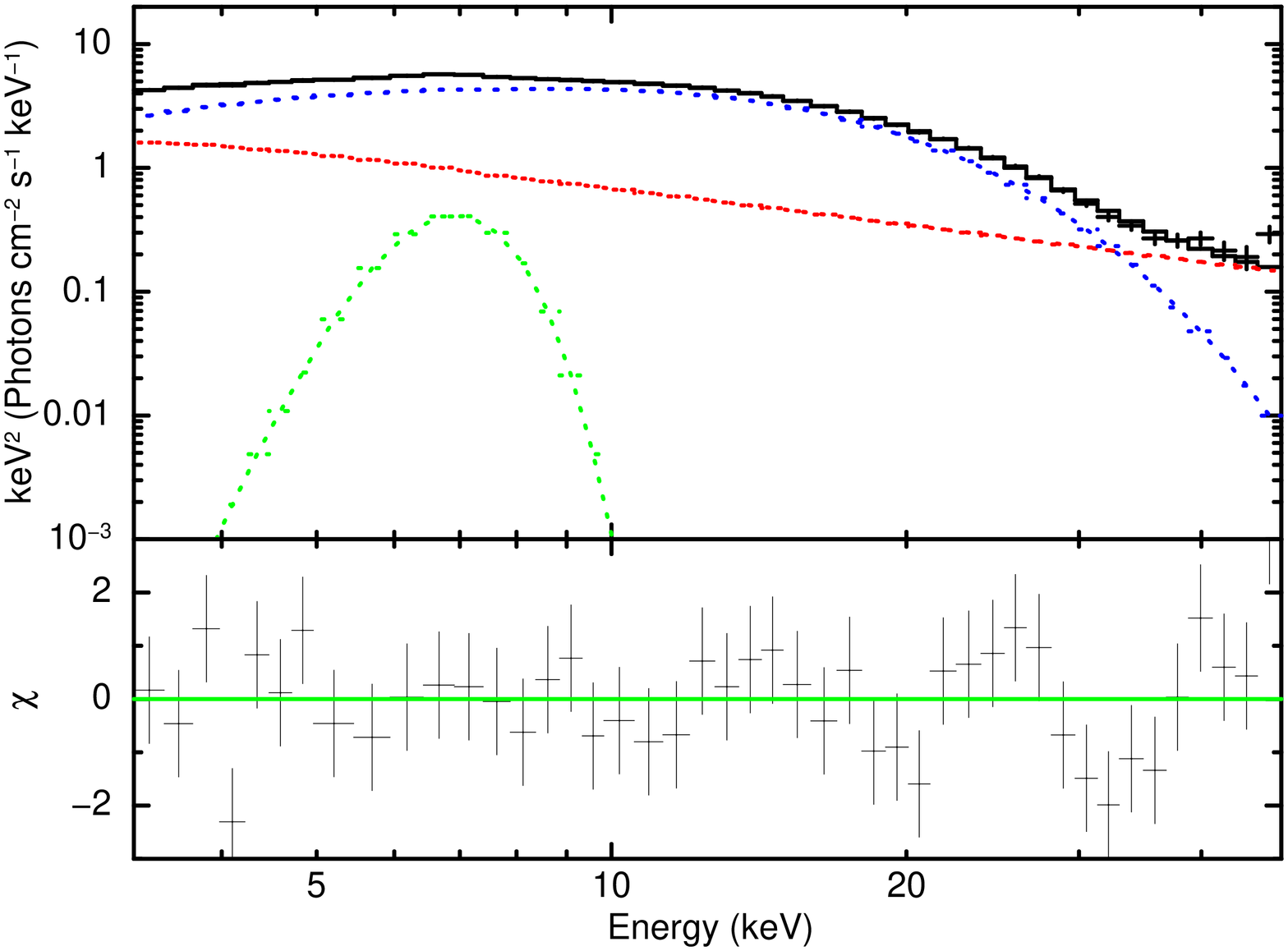}
\caption*{For HBO 1}
\end{subfigure}

\begin{subfigure}[b]{0.7\columnwidth}
\includegraphics[width=0.7\columnwidth, angle=270]{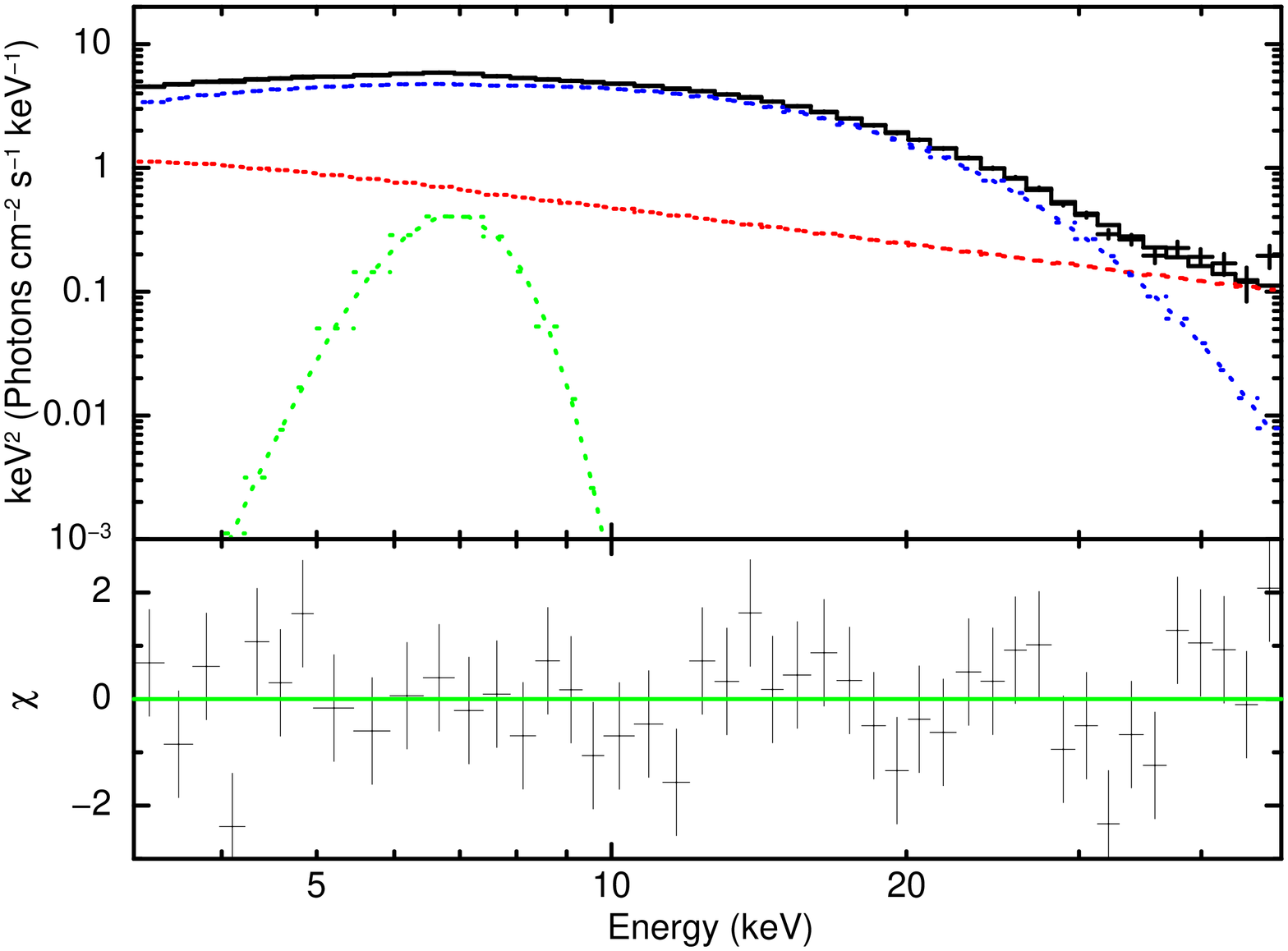}
\caption*{For HBO 2}
\end{subfigure}

\begin{subfigure}[b]{0.7\columnwidth}
\includegraphics[width=0.7\columnwidth, angle=270]{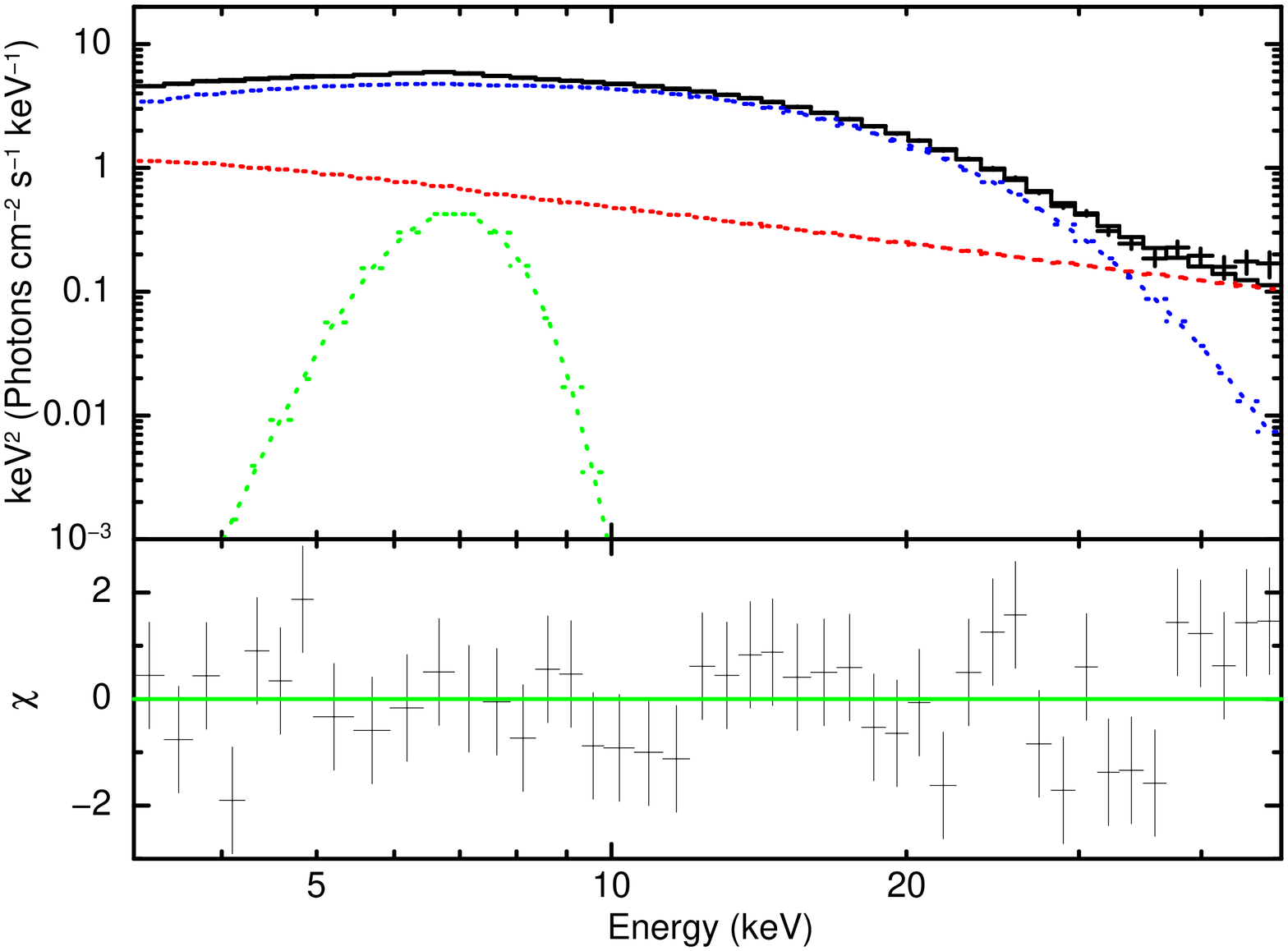}
\caption*{For HBO 3}
\end{subfigure}
\end{figure}

\clearpage

\vspace{-1em}
\section{RESULTS AND DISCUSSION}
\subsection{Constraining the inner disk radius}
Based on the PDS analysis, we found HBOs with a centroid frequency $\sim$ 25 Hz and 33 Hz.
Based on the relativistic precession model (RPM), higher frequency kHz QPOs,  lower frequency kHz QPOs 
and  HBOs correspond to the Keplerian, periastron precession and nodal precession frequencies, respectively Stella \& Vietri 1999; 
Stella et al. 1999).

Based on the equation given by Di Matteo \& Psaltis (1999) we can estimate an upper limit to the inner disk radius. 
It should be noted that the below relation is based on the empirical correlation between  HBOs and the 
upper kHz QPO $\nu_{HBO}$ $=$ 63 $\times$ ($\nu_{upp}$ / 1 kHz)$^{1.9}$ reported by Psaltis et al. (1999). 
This relation should be verified for GX 17+2 by considering all the available detections of QPOs.

\begin{equation}
\frac{R_{in}}{R_g} \le 27\nu^{-0.35} \Bigg[{\frac{M}{2 M_{\odot}}}\Bigg]^{-2/3}
\end{equation} 

With $\nu$=25 Hz, we found R$_{in}$ to be $\sim$ 23 km and with $\nu$=33 Hz, we found R$_{in}$ to be $\sim$ 21 km.

\subsection{Constraining the inner disk radius: TL Model}

We constrained the inner disk radius using the transition layer (TL) model (Titarchuk \& Osherovich 1999; Osherovich \& Titarchuk 1999a,b). 
Using the correlation equation between HBO and kHz QPO in Z sources given by Wu (2001) and the relation between
khz QPO and inner disk radius given by Matteo \& Psaltis (1999), Sriram et al. (2019) arrived at an equation connecting
HBO frequency and the inner disk radius as below,

\begin{equation}
\frac{R_{in}} {R_{g}} = 220\Bigg[\bigg(\frac{\Omega}{\pi}\bigg)^{-2/3} \bigg((\frac{\Omega\ \ sin \delta}{\nu_{HBO}\ \ \pi})^{2} - 1\bigg)^{1/3}\Bigg] \Bigg[{\frac{M}{10 M_{\odot}}}\Bigg]^{-2/3}
\end{equation}

where $\delta$ is the angle between rotational angular velocity ($\Omega$) and normal to the Keplerian oscillation ($\delta$=6.30 for GX 17+2; Wu 2001). 

We found R$_{in}$ to be $\sim$ 21 R$_{g}$ ($\sim$ 44 km ; Rg = GM/c$^2$) for 25 Hz HBO.

For $\nu$ $\sim$ 25 Hz, we found R$_{in}$ to be $\sim$ 22 Rg ($\sim$ 46 km) and for $\nu$ $\sim$ 33 Hz, R$_{in}$ $\sim$
18 Rg {$\sim$ 38 km).

\subsection{Inner disk radii from spectral modeling}

Based on the DiskBB model normalization (Mitsuda et al. 1984), N$_{dBB}$ is proportional to the inner disk radius. Using the equation, R$_{in}$ (km) = $\sqrt{(N / cos i)}$ $\times$ D / 10 kpc, and taking i = 28$^\circ$ (Malu et al. 2020), 
we estimated the inner disk radii R$_{in}$ to be 9.79 km  (A(a)),7.91 km  (A(b)), 8.78 km (B(a)), 10.59 km (B(b)), 11.68 km (C(a)), 11.02 km (C(b)) , 13.12 km ( D(a)), 11.75 km  (D(b)).

Since these radii needs to be corrected for spectral hardening using the relation,
R$_{eff}$ = $\kappa^{2}$ $\xi$ R$_{in}$ (Kubota et al. 2001) where $\kappa$ $\sim$ 1.7--2.0 (Shimura \& Takahara 1995)
and $\xi$ = 0.41 (Kubota et al. 1998), we estimate the R$_{in}$ to be 11.55 -- 16.06 km (A(a)),9.34 -- 12.97 km (A(b)), 10.37 -- 14.41 km  (B(a)), 
12.49--17.36 km (B(b)), 13.79--19.16 km  (C(a)), 13.01--18.08 km (C(b)), 15.49--21.52 km (D(a)) and 13.87--19.27 km (D(b)), which
is close to that estimated by Cackett et al. (2010), Ludlam et al. (2017), Agrawal et al. (2020) and Malu et al. (2020) for GX 17+2. The derived R$_{in}$ values suggest that disk front is found to be near or almost at the last stable orbit.
Along the HID track we find no change in the inner radius, hence suggesting that it could be the coronal structure which is responsible for the source traversing along the HID and not the mass accretion rate. 
This is similar to the results by Lin et al. (2012) and Homan et al. (2002).

\subsection{ Coronal height estimation from CCF lags}
An extensive timing and spectral study using the archival data of AstroSat LAXPC for GX 17+2 is reported here.
Based on the CCF study between LAXPC soft (3-5 keV) and hard (16-40 keV) light curves, we detect correlated and anti-correlated lags of the order of few hundred seconds (see Table 1).
This is in accordance with previously observed results for this source (Sriram et al. 2019, Malu et al. 2020).
The hard component could be arising from a coronal structure or a hot boundary layer component.
Previous studies have shown that the causative factor for these large lags could only be the readjustment 
of the coronal structure, especially since here the inner disk radius is found to be almost at the last stable orbit at all times along
the HID (for a detail spectral modeling see Agrawal et al. (2020)), ruling it out as the agent for change causing the large CCF lags. 
All the lags were found in the HB and NB.
Sriram et al. (2019), derived the equation (shown below) for determining the height of the coronal structure based on the condition
that the CCF lags are the readjustment timescale of the corona. 

\begin{equation}
H_{corona}=\Bigg[\frac{t_{lag} \dot{m}}{2 \pi R_{disk} H_{disk} \rho}-R_{disk}\Bigg] \times \beta \; cm
\end{equation}

Here H$_{disk}$ = 10$^{8}$ $\alpha^{-1/10}$ $\dot{m}_{16}^{3/20} R_{10}^{9/8} f^{3/20} $ cm, 
$\rho$ = 7 $\times$ 10$^{-8}$ $\alpha^{-7/10}$ $\dot{m}^{11/20}$ $R^{-15/8}$ $f^{11/20}$  g cm$^{-3}$, f = (1-(R$_s$/R)$^{1/2}$)$^{1/4}$ 
and $\beta$ = v$_{corona}$/v$_{disk}$ (Shakura \& Sunyaev 1973, Sriram et al. 2019).

For the sections A, B, C and  D the coronal height was constrained
to be  98 km, 78 km, 58 km and 104 km (for $\beta$ = 0.1) and  489 km, 392 km, 292 km and 519 km for $\beta$ =0.5,
for $\beta$ = 0.1 -- 0.5 (Manmoto et al. 1997, Pen et al. 2003, McKinney et al. 2012), an average R$_{disk}$ estimated from DiskBB normalization for each segment
and $\dot{m}$ estimated from the luminosity considering the equation L =  GM$\dot{m}$/R. Here $\alpha$ was taken to be 0.1.

The lack of spectral variation among the sections and the presence of lags, along with the detection of inner disk radius almost at 
the last stable orbit, indicate towards the possibility of a non varying disk front and varying coronal structure, which leads to the conclusion
that the mass accretion rate is not the factor causing the source to move along the HID but the varying comptonized corona. This was independently reported by Homan et al. (2002) and Lin et al. (2012), 
where upon studying GX 17+2 they conclude the same to address the evolution of the source across the HID. 
Hard lags suggest that the corona is decreasing in size, supported by most of the observations that hard count rates are often found to be decreasing in the light curves (Fig. 2 \& 3 ) and vice-verse for soft lags i.e. corona is increasing in height. 
However as it can be seen from CCFs that it is not always the scenario and more studies are required in this direction. 
The decreasing corona size should result in narrow equivalent width of iron line (Kara et al. 2019) 
but similar studies require better spectral resolution. 

\vspace{-0.5em}

\section{Conclusion}
1. Based on energy dependent CCF studies, we detect correlated and anti-correlated soft and hard lags of the order of few tens to 
few hundred seconds in the HB and NB. We interpret these lags as the readjustment timescales of the compact corona close to the neutron star. 
These lags constrain the coronal height to few tens to few hundred kms. 

2. PDS study has led to the detection of HBOs of $\sim$ 25 Hz and $\sim$ 33 Hz, along with their harmonics, 
which are characteristic features of these type of sources. 
Based on the RPM and TLM, we find the inner disk radius to be 10--21 Rg.
 
3. Spectral studies of the segments associated with lags show no significant variation in any of the spectral parameters, 
though we noticed small variations in fluxes for few sections. 
The inner disk radius was found to be close to the last stable orbit along the HB - NB/FB, suggesting that the disk is not truncated.

4. The lack of variation in the disk front suggests that the contributing factor for the source traversing along the HID could be a
variation in the corona or boundary layer. Hence we can conclude that the mass accretion rate is not the primary factor for the movement along
the HID. This result is in congruence with that of Homan et al. (2002) and Lin et al. (2012). 
The readjustment velocity factor of the corona $\beta$ could play a significant role in this regard. 
 
\vspace{-1em}

\section*{Acknowledgements}
We thank the Referee for the comments that has improved the quality of the paper.
K.S and C. P. acknowledge the financial support of ISRO under AstroSat
archival Data utilization program. This publication uses data from the
AstroSat mission of the Indian Space Research Organisation (ISRO),
archived at the Indian Space Science Data Centre (ISSDC). K.S also 
acknowledges the financial support from SERB CRG program. M.S.
acknowledges the financial support from DST-INSPIRE fellowship.
VKA thanks GH SAG; DD PDMSA, and Director URSC for encouragement and 
continuous support to carry out this research.
Authors sincerely acknowledge the contribution of the LAXPC team
toward the development of the LAXPC instrument on-board the AstroSat. 
This work uses data from the LAXPC instruments developed at TIFR, Mumbai, and the LAXPC POC at
TIFR is thanked for verifying and releasing the data via the ISSDC
data archive. Authors thank the AstroSat Science Support Cell hosted
by IUCAA and TIFR for providing the LAXPC software that was
used for LAXPC data analysis.
\vspace{-2em}




\begin{thebibliography}{xx}

\vspace{-1.5em}

\bibitem[]{}Agrawal V. K., Nandi A., Ramadevi M. C., 2020, Ap\&SS, 365, 41
\bibitem[]{}Alpar, M. A., \& Shaham, J., 1985, Nature, 316, 239
\bibitem[]{}Antia H. M., Yadav J. S., Agrawal P. C., et al. 2017, ApJS, 231, 10
\bibitem[]{}Cackett E. M. et al., 2009a, ApJ, 690, 1847
\bibitem[]{}Cackett, E. M., Miller, J. M., Ballantyne, D. R. et al. 2010, ApJ, 720, 205
\bibitem[]{}Davis, S.W., Blaes, O.M., Hubeny, I., Turner, N.J. 2005 ApJ, 621, 372
\bibitem[]{}Di Salvo T.,  D'AÃ­ A., Iaria R., Burderi  L., Dovaiak  M., Karas V., Matt G., Papitto  A., Piraino  S., Riggio  A., Robba  N. R., Santangelo A. 2009, MNRAS, 398, 2022 
\bibitem[]{}Fortner B., Lamb F. K., Miller G. S., 1989, Nature, 342, 775
\bibitem[]{}Galloway, D. K., Muno, M. P., Hartman, J. M., Psaltis, D., \& Chakrabarty, D. 2008, ApJS, 179, 360
\bibitem[]{}Hasinger G., \& van der Klis, M. 1989, A\&A, 225, 79
\bibitem[]{}Homan, J., van der Klis, M., Jonker, P. G. et al. 2002, ApJ, 568, 878 
\bibitem[]{}Lamb F. K., 1989, in Hunt J., Battrick B., eds, Proceedings of the 23rd ESLAB Symposium on Two Topics in X-Ray Astronomy. Vol. 1: X Ray Binaries.
Vol. 2: AGN and the X Ray Background. ESA, Noordwijk, p. 215
\bibitem[]{}Lei, Y. J., Qu, J. L., Song, L. M. et al. 2008, ApJ, 677, 461
\bibitem[]{}Lin, D., Remillard, R. A., Homan, J. 2007, ApJ, 667, 1073
\bibitem[]{}Lin, D., Remillard, R. A., Homan, J., \& Barret, D. 2012, ApJ, 756, 34
\bibitem[]{}Ludlam R. M. et al., 2017a, ApJ, 836, 140
\bibitem[]{}Malu S., Sriram, K. \& Agrawal, V.K 2020, MNRAS, Volume 499, Issue 2 (arXiv:2009.11002v2)
\bibitem[]{}Manmoto T., Mineshige S., Kusunose M., 1997, ApJ, 489, 791
\bibitem[]{}McKinney J. C., Tchekhovskoy A., Blandford R. D., 2012, MNRAS, 423, 3083
\bibitem[]{}Mitsuda, K., Inoue, H., Koyama, K. et al. 1984, PASJ, 36, 741
\bibitem[]{}Pen U.-L., Matzner C. D., Wong S., 2003, ApJ, 596, L207
\bibitem[]{}Penninx, W., Lewin, W. H. G., Zijlstra, A. A. et al. 1988, Nature, 336, 146
\bibitem[]{}Priedhorsky W., Hasinger G., Lewin W. H. G., Middleditch J., Parmar A., Stella L., White N., 1986, ApJ, 306, L9
\bibitem[]{}Psaltis et al. 1999, ApJ, 520, 763
\bibitem[]{}Shakura N. I., Sunyaev R. A., 1973, A\&A, 24:337
\bibitem[]{}Sriram K., Agrawal, V. K., Pendharkar, Jayant K., \& Rao, A. R. 2007, ApJ, 661, 1055
\bibitem[]{}Sriram K., Agrawal, V. K., \& Rao A. R. 2009, RAA, 9, 901
\bibitem[]{}Sriram K., Rao, A. R., Choi, C. S. 2010, ApJ, 725, 1317
\bibitem[]{}Sriram K., Rao, A. R., \& Choi, C. S. 2012, A\&A, 541, A6
\bibitem[]{}Sriram K., Malu S., \& Choi, C. S. 2019, ApJS, 244, 5S
\bibitem[]{}Stella, L., Vietri, M. 1999, Phys. Rev. Lett., 82, 17
\bibitem[]{}Stella, L., Vietri, M., \& Morsink, S.M. 1999, ApJL, 524, 63
\bibitem[]{}Titarchuk, L., Bradshaw, C. F., Geldzahler, B. J., \& Fomalont, E. B. 2001, ApJ, 555, 45
\bibitem[]{}Titarchuk, L.G. \& Osherovich, V.A. 1999, ApJL, 518, 95 
\bibitem[]{}van der Klis M. 2006, in Lewin W., van der Klis M., eds, Compact Stellar X-ray Sources, Cambridge Univ. Press, Cambridge, p. 39
\bibitem[]{}van der Klis M. 1988, in Ögelman H., van den Heuvel E. P. J., eds, NATO Advanced Science Institutes (ASI) Series C Vol.
262, NATO Advanced Science Institutes (ASI) Series C. p. 27
\bibitem[]{}Vaughan, B. A., van der Klis, M., Lewin, W. H. G., van Paradijs, J., Mitsuda, K., \& Dotnai, T. 1999, A\&A, 343, 197
\bibitem[]{}Vrtilek S. D., Raymond J. C., Garcia M. R. et al., 1990, A\&A, 235, 162
\bibitem[]{}Wijnands R. et al., 1997, ApJ, 490, L157
\bibitem[]{}Yadav J. S., Agrawal P. C., Antia H. M. 2016, Proc. SPIE, 9905, 99051D
\bibitem[]{}Zdziarski, A.A., Johnson, W.N., Magdziarz, P. 1996, MNRAS, 283, 193


\end{thebibliography}
\end{document}